\documentclass[a4paper, 11pt]{article}
\usepackage{jheppub}\makeatletter\gdef\@fpheader{}\makeatother 
\usepackage{mathtools, bbm, microtype, tabularx, caption, placeins}
\usepackage[noscale]{youngtab}\Yboxdim 6.5pt\Ylinethick 0.4pt 
\usepackage{dcolumn}\newcolumntype{d}[1]{D{.}{.}{#1}}
\usepackage[dvipsnames]{xcolor}   

\graphicspath{{Figures/}}

\newcommand{\T}{\mathbbm{T}} 
\newcommand{\Z}{\mathbbm{Z}}  
\newcommand{\R}{\mathcal{R}}
\newcommand{\U}{\textrm{U}}
\newcommand{\SU}{\textrm{SU}}
\newcommand{\USp}{\textrm{USp}} 
\newcommand{\SO}{\textrm{SO}}    
 
\author[a]{Mudassar Sabir,}
\author[b]{Adeel Mansha,}
\author[c,d,e]{Tianjun Li,}
\author[a]{Zhi-Wei Wang} 

\affiliation[a]{School of Physics, University of Electronic Science and Technology of China, Chengdu, Sichuan 611731, P. R. China} 
\affiliation[b]{College of Physics and Optoelectronic Engineering, Shenzhen University, Shenzhen 518060, P.R. China}
\affiliation[c]{CAS Key Laboratory of Theoretical Physics, Institute of Theoretical Physics, Chinese Academy of Sciences, Beijing 100190, P. R. China}
\affiliation[d]{School of Physical Sciences, University of Chinese Academy of Sciences, Beijing, P. R. China}
\affiliation[e]{School of Physics, Henan Normal University, Xinxiang 453007, P. R. China}

\emailAdd{mudassar.sabir@uestc.edu.cn}
\emailAdd{adeelmansha@alumni.itp.ac.cn}
\emailAdd{tli@itp.ac.cn} 
\emailAdd{zhiwei.wang@uestc.edu.cn}

\arxivnumber{2410.09093}

\begin{document}

\title{Susy breaking soft terms in the supersymmetric Pati-Salam landscape from Intersecting D6-Branes}

\abstract{We investigate the supersymmetry breaking soft terms for all the viable models in the complete landscape of three-family supersymmetric Pati-Salam models arising from intersecting D6-branes on a $\mathbb{T}^6/(\mathbb{Z}_2\times \mathbb{Z}_2)$ orientifold in type IIA string theory. The calculations are performed in the general scenario of $u$-moduli dominance with the $s$-moduli turned on, where the soft terms remain independent of the Yukawa couplings and the Wilson lines. The results for the trilinear coupling, gaugino-masses, squared-mass parameters of squarks, sleptons and Higgs depend on the brane wrapping numbers and the susy breaking parameters. We find that unlike the Yukawa couplings which remain unchanged for the models dual under the exchange of two SU(2) sectors, the corresponding soft term parameters only match for the trilinear coupling and the mass of the gluino. This can be explained by the internal geometry where the Yukawa interactions depend only on the triangular areas of the worldsheet instantons while the soft terms have an additional dependence on the orientation-angles of D6-branes in the three two-tori. In the special limit of parameter space we find universal masses for the Higgs and the gauginos.}

\maketitle
\flushbottom
\section{Introduction}\label{sec:Intro}
In the standard model, quarks and leptons arise from chiral representations of the gauge group $\SU(3)_C \times \SU(2)_L \times \U(1)_Y$. Intersecting D6-branes in type IIA string theory, filling the four-dimensional spacetime and extending into three compact dimensions, provide a geometric realization of standard model gauge interactions. Open strings stretched between two intersecting D6-branes act as chiral fermions \cite{Berkooz:1996km, Aldazabal:2000cn}. Family replication is readily achieved as branes generically intersect at several points. The four-dimensional gauge couplings are determined by the volumes of the cycles wrapped by D6-branes while the gravitational coupling depends on the total internal volume of the compact dimensions, thus allowing the possibility of having the string scale $M_\mathrm{S}$ parametrically smaller than the Planck scale $M_\mathrm{P}$. Hierarchical Yukawa couplings arise from open world-sheet instantons, which are suppressed by $\exp(-A_{ijk}T)$, where $A_{ijk}$ represents the area of the triangle formed by the intersections $\{i, j, k\}$, and $T$ denotes the string tension.

Since realistic Yukawa textures cannot arise from a single stack of D-branes, this necessitates that the gauge group of the standard model be a direct product of unitary groups rather than a simple unitary group. Statistically, the configurations of the branes tend to favor direct products of unitary groups constructed from an even number of D-branes within the stack. This preference can be explained by the K-theory conditions \cite{Witten:1998cd, Uranga:2000xp}, constrained by mod 4 and thus are easier to satisfy for $\U(2N)$ with $N \in \Z$. For example, in trinification models, there is currently no single consistent three-family model that satisfies the stringent requirements of $\mathcal{N}=1$ supersymmetry, tadpole cancellation, and K-theory constraints \cite{Mansha:2024yqz}. Note that even if the models are consistent under the K-theory, no tadpoles and susy conditions, the viable models with realistic Yukawa textures are extremely difficult to engineer. Consequently, the left-right symmetric Pati-Salam group, $\SU(4)_C\times \SU(2)_L\times \SU(2)_R$, stands out as the most promising option for realistic models.

The model building rules to construct supersymmetric Pati-Salam models on a $\T^6/(\Z_2\times \Z_2)$ orientifold utilizing intersecting D6-branes with the requirement of $\mathcal{N}=1$ supersymmetry, no-tadpole constraints and ensuring K-theory conditions were laid out in \cite{Cvetic:2004ui, Blumenhagen:2006ci, Blumenhagen:2005mu}. A similar approach has been applied in more recent studies \cite{Li:2019nvi, Li:2021pxo, Mansha:2022pnd, Sabir:2022hko, Mansha:2023kwq}. The complete landscape of consistent three-family $\mathcal{N}=1$ supersymmetric Pati-Salam models containing 206,752 models has been derived in \cite{He:2021gug}. There are 33 distinct physical models with distinct gauge coupling relations up to type I and type II T-dualities. In \cite{Sabir:2024cgt}, we have presented the results of Yukawa couplings for all the feasible models in this landscape, and in \cite{Sabir:2024mfv} the most realistic model explaining the fermion masses and mixings has been discussed.  

In this paper, we investigate the supersymmetry breaking soft terms for all the viable three-family supersymmetric Pati-Salam models in the landscape, focusing on the $u$-moduli dominated scenario with the $s$-moduli turned on. In this setup, the soft terms remain independent of the Yukawa couplings and the Wilson lines. The results for the trilinear coupling, gaugino-masses, squared-mass parameters of squarks, sleptons and Higgs are determined by the D-brane wrapping numbers and the supersymmetry breaking parameters, which include the gravitino mass and the goldstino angles.  

For any two models which are dual under the exchange of two SU(2)s, the Yukawa couplings produce identical results, however, the corresponding soft term parameters exhibit notable differences. Specifically, the trilinear coupling and the gluino mass match, whereas the other soft terms, such as bino, wino and squared masses of squarks and sleptons, differ between the dual models. This discrepancy can be understood from the underlying geometry of the compact space. While the Yukawa interactions depend solely on the triangular area of the worldsheet instantons, the soft terms are influenced by additional factors, including the orientation angles of the D6-branes on the three two-tori. As a result, dual models not only exhibit distinct gauge coupling relations but also feature different gaugino masses and soft scalar mass parameters. 

Since the scale of supersymmetry breaking is still unknown, we mainly focus on generic results in the present work. Nonetheless, we have found an interesting regime of parameters where all gaugino-masses become degenerate and the Higgs mass parameter is half the gravitino-mass which determines the susy breaking scale. This special limit corresponds to setting the three goldstino angles $\{\Theta_1,\Theta_2,\Theta_3\}$ equal to 1/2 and setting the dilaton angle $\Theta_s = -1/2$ with all CP-violating phases $\{\gamma_1,\gamma_2,\gamma_3,\gamma_s\}$ set to zero.   

The contents of the paper are organized as follows. In section~\ref{sec:orientifold} we briefly review the rules to construct supersymmetric Pati-Salam models from stacks of intersecting D6-branes on a $\T^6/(\Z_2\times \Z_2)$ orientifold. In section~\ref{sec:EFT} we describe the four-dimensional effective field theory and the soft terms from supersymmetry breaking in the case of $u$-moduli domination with the dilaton modulus $s$ turned on. We outline the computational strategy to calculate the susy breaking soft terms from the given wrapping numbers and the angles of D6-branes. In section~\ref{sec:soft_terms} we systematically compute the soft terms for all viable models in the three-family Pati-Salam landscape. Finally, we conclude in section~\ref{sec:conclusion}.  

\section{Pati-Salam model building on a $\T^6/(\Z_2\times \Z_2)$ orientifold}\label{sec:orientifold}
$\T^6/(\Z_2\times \Z_2)$ is the product of three two-tori $\T^2$ with the orbifold group $\Z_2\times \Z_2$, having the generators $\theta$ and $\omega$, which are respectively associated with the twist vectors $(1/2,-1/2,0)$ and $(0,1/2,-1/2)$ such that their action on the complex coordinates $z_i$ is given as,
\begin{align}
\theta: \quad & (z_1,z_2,z_3) \to (-z_1,-z_2,z_3), \nonumber \\
\omega: \quad & (z_1,z_2,z_3) \to (z_1,-z_2,-z_3). \label{orbifold}
\end{align}
Orientifold projection is the gauged $\Omega \R$ symmetry, where $\Omega$ is world-sheet parity that interchanges the left- and right-moving sectors of a closed string and swaps the two ends of an open string as,
\begin{alignat}{2}
\textrm{Closed}&: \quad & \Omega : (\sigma_1, \sigma_2) & \mapsto (2\pi -\sigma_1, \sigma_2), \nonumber \\
\textrm{Open}&:  \quad & \Omega : (\tau, \sigma) & \mapsto (\tau, \pi - \sigma) ,
\end{alignat}
and $\R$ acts as complex conjugation on coordinates $z_i$. This results in four different kinds of orientifold 6-planes (O6-planes) corresponding to $\Omega \R$, $\Omega \R\theta$, $\Omega \R\omega$, and $\Omega \R\theta\omega$ respectively. These orientifold projections are only consistent with either the rectangular or the tilted complex structures of the factorized two-tori. Denoting the wrapping numbers for the rectangular and tilted tori as $n_a^i[a_i]+m_a^i[b_i]$ and $n_a^i[a'_i]+m_a^i[b_i]$ respectively, where $[a_i']=[a_i]+\frac{1}{2}[b_i]$. Then a generic 1-cycle $(n_a^i,l_a^i)$ satisfies $l_{a}^{i}\equiv m_{a}^{i}$ for the rectangular two-torus and $l_{a}^{i}\equiv 2\tilde{m}_{a}^{i}=2m_{a}^{i}+n_{a}^{i}$ for the tilted two-torus such that $l_a^i-n_a^i$ is even for the tilted tori. 

Note that the two different bases $(n^i,l^i)$ and $(n^i,m^i)$ are related as, 
\begin{align}\label{basis-l-m}
m^i &= 2^{-\beta_i} l^i - \frac{\beta_i}{2} n^i, \quad \beta_i = \begin{cases}
0  & \mathrm{rectangular}~\T^2, \\ 
1  & \mathrm{tilted}~\T^2. \end{cases}
\end{align}
We use the basis $(n^i,l^i)$ to specify the model wrapping numbers in appendix~\ref{appA} while the basis $(n^i,m^i)$ is convenient to sketch the Yukawa textures in section~\ref{sec:soft_terms}.

The homology cycles for a stack $a$ of $N_a$ D6-branes along the cycle $(n_a^i,l_a^i)$ and their $\Omega \R$ images ${a'}$ stack of $N_a$ D6-branes with cycles $(n_a^i,-l_a^i)$ are respectively given as,
\begin{align}
[\Pi_a ]&=\prod_{i=1}^{3}\left(n_{a}^{i}[a_i]+2^{-\beta_i}l_{a}^{i}[b_i]\right), \nonumber \\
[\Pi_{a'}] &=\prod_{i=1}^{3}\left(n_{a}^{i}[a_i]-2^{-\beta_i}l_{a}^{i}[b_i]\right).
\end{align}
The homology three-cycles, which are wrapped by the four O6-planes, are given by
\begin{alignat}{2}
\Omega \R : &\quad& [\Pi_{\Omega \R}] &= 2^3 [a_1]\times[a_2]\times[a_3],  \nonumber\\
\Omega \R\omega : && [\Pi_{\Omega \R\omega}] &=-2^{3-\beta_2-\beta_3}[a_1]\times[b_2]\times[b_3],  \nonumber\\
\Omega \R\theta\omega : && [\Pi_{\Omega \R\theta\omega}] &=-2^{3-\beta_1-\beta_3}[b_1]\times[a_2]\times[b_3], \nonumber\\
\Omega \R\theta : && [\Pi_{\Omega \R \theta}] &=-2^{3-\beta_1-\beta_2}[b_1]\times[b_2]\times[a_3]. \label{orienticycles}
\end{alignat}
The intersection numbers can be calculated in terms of wrapping numbers as,
\begin{align}
I_{ab}&=[\Pi_a][\Pi_b] =2^{-k}\prod_{i=1}^3(n_a^il_b^i-n_b^il_a^i),\nonumber\\
I_{ab'}&=[\Pi_a]\left[\Pi_{b'}\right] =-2^{-k}\prod_{i=1}^3(n_{a}^il_b^i+n_b^il_a^i),\nonumber\\
I_{aa'}&=[\Pi_a]\left[\Pi_{a'}\right] =-2^{3-k}\prod_{i=1}^3(n_a^il_a^i),\nonumber\\
I_{aO6}&=[\Pi_a][\Pi_{O6}] =2^{3-k}(-l_a^1l_a^2l_a^3+l_a^1n_a^2n_a^3+n_a^1l_a^2n_a^3+n_a^1n_a^2l_a^3),\label{intersections}
\end{align}
where $k=\sum_{i=1}^3\beta_i$ and $[\Pi_{O6}]=[\Pi_{\Omega \R}]+[\Pi_{\Omega \R\omega}]+[\Pi_{\Omega \R\theta\omega}]+[\Pi_{\Omega \R\theta}]$.

In order to have three families of the left chiral and right chiral standard model fields, the intersection numbers must satisfy
\begin{align}
I_{ab} + I_{ab'} = 3 , \quad I_{ac} = -3, \quad I_{ac'} = 0. \label{eq:NoG}
\end{align}  
 
\subsection{Constraints from tadpole cancellation and $\mathcal{N}=1$ supersymmetry}\label{subsec:constraints}

Since D6-branes and O6-orientifold planes are the sources of Ramond-Ramond charges they are constrained by the Gauss's law in compact space implying the sum of D-brane and cross-cap RR-charges must vanish \cite{Gimon:1996rq}
\begin{align}\label{RRtadpole}
\sum_a N_a [\Pi_a]+\sum_a N_a \left[\Pi_{a'}\right]-4[\Pi_{O6}] &= 0,
\end{align}
where the last terms arise from the O6-planes, which have $-4$ RR charges in D6-brane charge units. RR tadpole constraint is sufficient to cancel the $\SU(N_a)^3$ cubic non-Abelian anomaly while \U(1) mixed gauge and gravitational anomaly or $[\SU(N_a)]^2 \U(1)$ gauge anomaly can be cancelled by the Green-Schwarz mechanism, mediated by untwisted RR fields \cite{Green:1984sg}.

Let us define the following products of wrapping numbers,
\begin{alignat}{4}
A_a &\equiv -n_a^1n_a^2n_a^3,   &\quad B_a &\equiv n_a^1l_a^2l_a^3,       &\quad     C_a &\equiv l_a^1n_a^2l_a^3,  &\quad   D_a &\equiv l_a^1l_a^2n_a^3, \nonumber\\
\tilde{A}_a &\equiv -l_a^1l_a^2l_a^3, & \tilde{B}_a &\equiv l_a^1n_a^2n_a^3, & \tilde{C}_a &\equiv n_a^1l_a^2n_a^3, & \tilde{D}_a &\equiv n_a^1n_a^2l_a^3.\,\label{variables}
\end{alignat}
Cancellation of RR tadpoles requires introducing a number of orientifold planes also called ``filler branes'' that trivially satisfy the four-dimensional $\mathcal{N}=1$ supersymmetry conditions. The no-tadpole condition is given as,
\begin{align}
 -2^k N^{(1)}+\sum_a N_a A_a&=-2^k N^{(2)}+\sum_a N_a B_a= \nonumber\\
 -2^k N^{(3)}+\sum_a N_a C_a&=-2^k N^{(4)}+\sum_a N_a D_a=-16,\,
\end{align}
where $2 N^{(i)}$ is the number of filler branes wrapping along the $i^\mathrm{th}$ O6-plane. The filler branes belong to the hidden sector USp group and carry the same wrapping numbers as one of the O6-planes as shown in table~\ref{orientifold}. USp group is hence referred with respect to the non-zero $A$, $B$, $C$ or $D$-type.

\begin{table}[h]
\renewcommand{\arraystretch}{1.3}
\centering
\begin{tabular}{|c|c|c|}
\hline
  Orientifold action & O6-plane & $(n^1,l^1)\times (n^2,l^2)\times (n^3,l^3)$\\
\hline
    $\Omega \R$& 1 & $(2^{\beta_1},0)\times (2^{\beta_2},0)\times (2^{\beta_3},0)$ \\
\hline
    $\Omega \R\omega$& 2& $(2^{\beta_1},0)\times (0,-2^{\beta_2})\times (0,2^{\beta_3})$ \\
\hline
    $\Omega \R\theta\omega$& 3 & $(0,-2^{\beta_1})\times (2^{\beta_2},0)\times (0,2^{\beta_3})$ \\
\hline
    $\Omega \R\theta$& 4 & $(0,-2^{\beta_1})\times (0,2^{\beta_2})\times (2^{\beta_3},0)$ \\
\hline
\end{tabular} 
\caption{The wrapping numbers for four O6-planes.}
\label{orientifold}
\end{table}

Preserving $\mathcal{N}=1$ supersymmetry in four dimensions after compactification from ten-dimensions restricts the rotation angle of any D6-brane with respect to the orientifold plane to be an element of $\SU(3)$, i.e.
\begin{equation}
\theta_a^1 + \theta_a^2 + \theta_a^3 = 0 \mathrm{~mod~} 2\pi ,
\end{equation}
where $\theta_a^i = \frac{1}{\pi} \arctan \left(\frac{2^{- \beta_i} l_a^i}{n_a^i}\chi_i\right)$ is the angle between the D6-brane and orientifold-plane and $\chi_i=R^2_i/R^1_i$ are the complex structure moduli of the $i^\mathrm{th}$ two-torus. $\mathcal{N}=1$ supersymmetry conditions are given as,
\begin{align}
x_A\tilde{A}_a+x_B\tilde{B}_a+x_C\tilde{C}_a+x_D\tilde{D}_a &= 0,\nonumber\\
\frac{A_a}{x_A}+\frac{B_a}{x_B}+\frac{C_a}{x_C}+\frac{D_a}{x_D} &< 0, \label{susyconditions}
\end{align}
where $x_A=\lambda,\; x_B=2^{\beta_2+\beta_3}\cdot\lambda /\chi_2\chi_3,\; x_C=2^{\beta_1+\beta_3}\cdot\lambda /\chi_1\chi_3,\; x_D=2^{\beta_1+\beta_2}\cdot\lambda /\chi_1\chi_2$.

Orientifolds also have discrete D-brane RR charges classified by the $\Z_2$ K-theory groups, which are subtle and invisible by the ordinary homology \cite{Witten:1998cd, Cascales:2003zp, Marchesano:2004yq, Marchesano:2004xz}, which should also be taken into account \cite{Uranga:2000xp}. The K-theory conditions are,
\begin{align}
\sum_a \tilde{A}_a  = \sum_a  N_a  \tilde{B}_a = \sum_a  N_a  \tilde{C}_a = \sum_a  N_a \tilde{D}_a &= 0 \mathrm{~mod~} 4 \label{K-charges}.
\end{align}
In our case, we avoid the nonvanishing torsion charges by taking an even number of D-branes, {\it i.e.}, $N_a \in 2 \Z$.
 
\subsection{Particle spectrum}

\begin{table}[th]
\renewcommand{\arraystretch}{1.3}
\centering
\begin{tabular}{|c|c|}
\hline {\bf Sector} & \phantom{---------------} {\bf Representation} \phantom{---------------} \\
\hline 
$aa$   & $\U(N_a/2)$ vector multiplet  \\
       & 3 adjoint chiral multiplets  \\
\hline $ab+ba$   & $ \mathcal{M}(\frac{N_a}{2}, \frac{\overline{N}_b}{2})= I_{ab}(\yng(1)_{a},\overline{\yng(1)}_{b})$ \\
\hline $ab'+b'a$ & $ \mathcal{M}(\frac{N_a}{2}, \frac{N_b}{2})=I_{ab'}(\yng(1)_{a},\yng(1)_{b})$ \\
\hline $aa'+a'a$ &  $\mathcal{M} (a_{\yng(2)})= \frac 12 (I_{aa'} - \frac 12 I_{aO6})$ \\
                 &  $\mathcal{M} (a_{\yng(1,1)_{}})= \frac 12 (I_{aa'} + \frac 12 I_{aO6})$\\
\hline
\end{tabular}
\caption{General spectrum for intersecting D6-branes at generic angles, where $\mathcal{M}$ is the multiplicity, and $a_{\protect\yng(2)}$ and $a_{\protect\yng(1,1)}$ denote respectively the symmetric and antisymmetric representations of $\U(N_a/2)$. Positive intersection numbers in our convention refer to the left-handed chiral supermultiplets.}
\label{tab:spectrum}
\end{table}

To have three families of the SM fermions, we need one torus to be tilted, which is chosen to be
the third torus. So we have  $\beta_1=\beta_2=0$ and $\beta_3=1$. 
Placing the $a'$, $b$ and $c$ stacks of D6-branes on the top of each other on the third two-torus results in additional vector-like particles from $\mathcal{N} = 2$ subsectors \cite{Cvetic:2004ui}. The anomalies from three global U(1)s of $\U(4)_C$, $\U(2)_L$ and $\U(2)_R$ are cancelled by the Green-Schwarz mechanism, and the gauge fields of these U(1)s obtain masses via the linear $B\wedge F$ couplings. Thus, the effective gauge symmetry is $\SU(4)_C\times \SU(2)_L\times \SU(2)_R$.

\section{Supersymmetry breaking and $\mathcal{N}=1$ Effective theory}\label{sec:EFT}
 
\begin{figure*}[h]
\centering
\includegraphics[width=\textwidth]{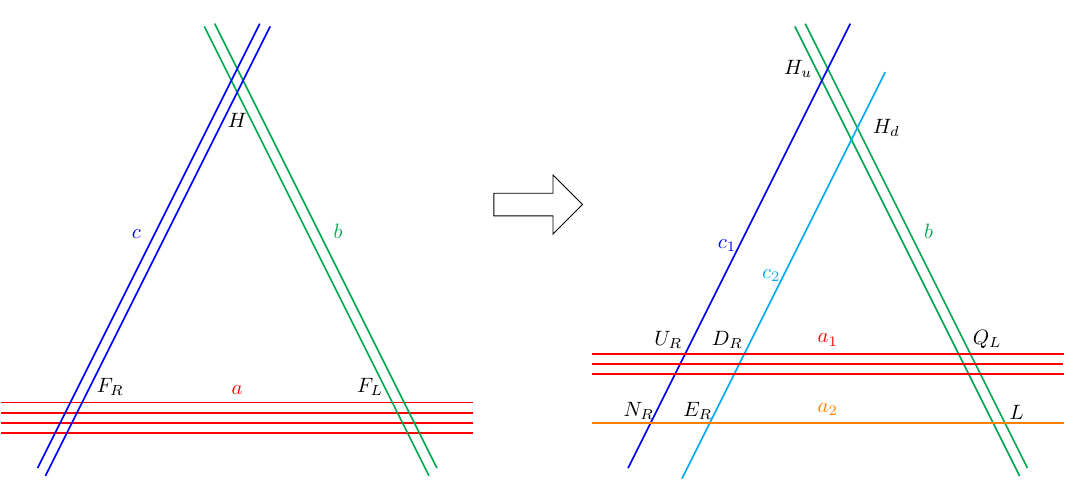}
\caption{Pati-Salam gauge group $\SU(4)_C\times \SU(2)_L\times \SU(2)_R$ is broken down to the standard model gauge group $\SU(3)_C\times \U(2)_L\times \U(1)_{I3R}\times \U(1)_{B-L}$ via the process of brane splitting that corresponds to assigning VEVs to the adjoint scalars, which arise as open-string moduli associated with the positions of stacks $a$ and $c$ in the internal space.} \label{brnsplit}
\end{figure*}

Pati-Salam gauge group $\SU(4)_C\times \SU(2)_L\times \SU(2)_R$ is higgsed down to the standard model gauge group $\SU(3)_C\times \U(2)_L\times \U(1)_{I3R}\times \U(1)_{B-L}$ by assigning vacuum expectation values to the adjoint scalars which arise as open-string moduli associated to the stacks $a$ and $c$, see figure~\ref{brnsplit},
\begin{eqnarray}
\textcolor{red}{a} &\rightarrow & \textcolor{red}{a_1} + \textcolor{orange}{a_2} , \nonumber \\
\textcolor{blue}{c} &\rightarrow & \textcolor{blue}{c_1} + \textcolor{cyan}{c_2} .
\end{eqnarray}
Moreover, the $\U(1)_{I_{3R}}\times \U(1)_{B-L}$ gauge symmetry may be broken to $\U(1)_Y$ by giving vacuum expectation values (VEVs) to the vector-like particles with the quantum numbers $({\bf 1, 1, 1/2, -1})$ and $({\bf 1, 1, -1/2, 1})$ under the $\SU(3)_C\times \SU(2)_L\times \U(1)_{I_{3R}} \times \U(1)_{B-L} $
gauge symmetry from $a_2 c_1'$ intersections \cite{Cvetic:2004ui,Chen:2006gd}. This brane-splitting results in standard model quarks and leptons as \cite{Cvetic:2004nk},
\begin{eqnarray}
F_L(Q_L, L_L)  &\rightarrow &  Q_L + L , \nonumber \\
F_R(Q_R, L_R)  &\rightarrow &  U_R + D_R + E_R + N.
\end{eqnarray}
Three-point Yukawa couplings for the quarks and the charged leptons can be read from the following superpotential,
\begin{align}\label{eq:WY3} 
\mathcal{W}_3 \sim  Y^u_{ijk} Q_i  U^c_j H^u_k + Y^d_{ijk} Q_i D^c_j H^d_k + Y^\nu_{ijk} L_i N^c_j H^u_k + Y^e_{ijk} L_i  E^{c}_j H^d_k  ,
\end{align}
where $Y^u_{ijk}$, $Y^d_{ijk}$, $Y^\nu_{ijk}$ and $Y^e_{ijk}$ are Yukawa couplings, and $Q_i$, $U^c_i$, $D^c_i$, $L_i$, $N^c_i$, and $E^c_i$ are the left-handed quark doublet, right-handed up-type quarks, right-handed down-type quarks, left-handed lepton doublet, right-handed neutrinos, and right-handed leptons, respectively. The superpotential including the four-point interactions is
\begin{align}\label{eq:WY4}
\mathcal{W}_4 \sim {1\over {M_\mathrm{S}}} \left( Y^{\prime u}_{ijkl} Q_i U^c_j H^{\prime u}_k S^L_l + Y^{\prime d}_{ijkl} Q_i D^{c}_j H^{\prime d}_k S^L_l 
+ Y^{\prime \nu}_{ijkl} L_i N^c_j H^{\prime u}_k S^L_l + Y^{\prime e}_{ijkl} L_i E^{c}_j H^{\prime d}_k S^L_l \right),
\end{align}
where $Y^{\prime u}_{ijkl}$, $Y^{\prime d}_{ijkl}$, $Y^{\prime \nu}_{ijkl}$, and $Y^{\prime e}_{ijkl}$ are Yukawa couplings of the four-point functions, and $M_\mathrm{S}$ is the string scale.

The additional exotic particles must be made superheavy to ensure gauge coupling unification at the string scale. Similar to refs. \cite{Cvetic:2007ku, Chen:2007zu} we can decouple the additional exotic particles except the charged chiral multiplets under $\SU(4)_C$ anti-symmetric representation. These chiral multiplets can be decoupled via instanton effects in principle \cite{Blumenhagen:2006xt, Haack:2006cy, Florea:2006si}.

We now turn our attention toward the four dimensional low energy effective field theory. $\mathcal{N}=1$ supergravity action is encoded by three functions viz. the gauge kinetic function $f_{x}$, the K\"{a}hler potential $K$ and the superpotential $W$ \cite{Cremmer:1982en}. Each of these functions in turn depend on a dilaton modulus $S$, three K\"{a}hler moduli $T^i$ and three complex structure moduli $U^i$.

The complex structure moduli $U^i$ are defined as,
\begin{align}\label{U-moduli}
U^j & = \frac{4i \chi_j+2\beta_j\chi_j^2}{4+\beta_j\chi_j^2}, \qquad \because \chi_j \equiv \frac{R_2^j}{R_1^j}.
\end{align}
These upper case moduli in string theory basis can be transformed to lower-case $s$, $t^i$ and $u^i$ moduli in field theory basis as \cite{Lust:2004cx},
\begin{align}
\mathrm{Re}\,(s) &= \frac{e^{-{\phi}_4}}{2\pi}\,\left(\frac{\sqrt{\mathrm{Im}\,U^{1}\, \mathrm{Im}\,U^{2}\,\mathrm{Im}\,U^3}}{|U^1U^2U^3|}\right) , \nonumber \\
\mathrm{Re}\,(u^j) &= \frac{e^{-{\phi}_4}}{2\pi}\left(\sqrt{\frac{\mathrm{Im}\,U^{j}}{\mathrm{Im}\,U^{k}\,\mathrm{Im}\,U^l}}\right)\; \left|\frac{U^k\,U^l}{U^j}\right|, \quad (j,k,l)=(\overline{1,2,3}), \nonumber \\
\mathrm{Re}(t^j) &= \frac{i\alpha'}{T^j} , \label{eq:moduli}
\end{align}
where $j$ denotes the $j^\mathrm{th}$ two-torus, and $\phi_4$ is the four dimensional dilaton which is related to the supergravity moduli as  
\begin{equation}
2\pi e^{\phi_4}=\Big(\mathrm{Re}(s)\,\mathrm{Re}(u^1)\,\mathrm{Re}(u^2)\,\mathrm{Re}(u^3)\Big)^{-1/4}.
\end{equation}
Inverting the above formulas we can solve for $U^i$ moduli in string theory basis in terms of $s$ and $u^i$ as,
\begin{equation}
\frac{|U^j|^2}{\mathrm{Im}\,(U^j)} = \sqrt{\frac{\mathrm{Re}\,(u^k)\,\mathrm{Re}\,(u^l)}{\mathrm{Re}\,(u^j)\mathrm{Re}\,(s)}}, \quad  (j,k,l)=(\overline{1,2,3}) . \label{eq:b}
\end{equation}
The holomorphic gauge kinetic function for any D6-brane stack $x$ wrapping a calibrated 3-cycle is given as \cite{Blumenhagen:2006ci},
\begin{equation}
f_x = \frac{1}{2\pi \ell_s^3}\left[e^{-\phi}\int_{\Pi_x} \mathrm{Re}(e^{-i\theta_x}\Omega_3)-i\int_{\Pi_x}C_3\right],
\end{equation}
where the integral involving 3-form $\Omega_3$ gives,
\begin{equation}
\int_{\Pi_x}\Omega_3 = \frac{1}{4}\prod_{i=1}^3(n_x^iR_1^i + 2^{-\beta_i}il_x^iR_2^i).
\end{equation}
It can then be shown that,
\begin{align}
f_x &= \frac{1}{4 k_x}(n_x^1\,n_x^2\,n_x^3\,s-\frac{n_x^1\,l_x^2\,l_x^3\,u^1}{2^{(\beta_2+\beta_3)}}-\frac{l_x^1\,n_x^2\,l_x^3\,u^2}{2^{(\beta_1+\beta_3)}}-
\frac{l_x^1\,l_x^2\,n_x^3\,u^3}{2^{(\beta_1+\beta_2)}}), \label{kingauagefun}
\end{align}
where the factor $k_x$ is related to the difference between the gauge couplings for $\U(N_x)$ and $\USp(2N_x),\SO(2N_x)$ such that $k_x =1$ for $\U(N_x)$ and $k_x =2$ for $\USp(2N_x)$ or $\SO(2N_x)$ \cite{Klebanov:2003my, Blumenhagen:2003jy}. 
Since, the standard model hypercharge $\U(1)_Y$ is a linear combination of several U(1)s,
\begin{equation}
Q_Y=\frac{1}{6}Q_{a_1}+\frac{1}{2}Q_{a_2}-\frac{1}{2}Q_{c_1}-\frac{1}{2}Q_{c_2},
\end{equation}
similarly, the holomorphic gauge kinetic function for the hypercharge is also taken as a linear combination of the kinetic gauge functions from several stacks \cite{Blumenhagen:2003jy, Ibanez:2001nd},
\begin{equation}\label{fY}
f_Y=\frac{1}{6}f_{a_1}+\frac{1}{2}f_{a_2}+\frac{1}{2}f_{c_1}+\frac{1}{2}f_{c_2}.
\end{equation}
The K\"{a}hler potential to the second order for the moduli $M$ and open string matter fields, $C_i, C_\theta$ has the structure,
\begin{align}
K(M,\bar{M},C,\bar{C}) &= \hat{K}(M,\bar{M}) +  \sum_{\mathrm{untwisted}\,i,j} \tilde{K}_{C_{i}\bar{C}_j}(M,\bar{M}) C_i\bar{C}_j \nonumber\\
& \quad  + \sum_{\mathrm{twisted}, \, \theta} \tilde{K}_{C_{\theta}\bar{C}_{\theta}}(M,\bar{M}) C_{\theta}\bar{C}_{\theta} ,
\shortintertext{where}
\kappa^2 \hat{K}(M,\bar{M})&= -\log (S+\bar{S}) - \sum_{i=1}^3 \log (T^i+\bar{T}^i) - \sum_{i=1}^3 \log (U^i+\bar{U}^i) , 
\end{align}
where $\kappa^2 =8\pi G_N $, $C_i$ correspond to the D-brane positions and the Wilson lines moduli arising from strings having both ends on the same stack while $C_{\theta}$ correspond to strings stretching between different stacks comprising $1/4$ BPS branes. The untwisted moduli fields $C_{i},\bar{C}_j$ are not present in MSSM and must become heavy via higher dimensional operators\footnote{D-branes wrapping rigid cycles can freeze such open string moduli \cite{Blumenhagen:2005tn}, however such rigid cycles without discrete torsion are not present in $\T^6/(\Z_2 \times \Z_2)$.}.

Let us determine the K\"{a}hler metric $\tilde{K}_{C_\theta\bar{C}_\theta}(M,\bar{M})$ for the twisted moduli. 
The angles $\theta^j_{x}$ made by the cycle wrapped by stack of D6-branes on each of the three two-tori are related to the $s$ and $u^j$ moduli as,
\begin{equation}\label{eq:angle}
\tan(\pi\theta^j_x)=\frac{2^{-\beta_j}l^j_x}{n^j_x} \chi_j=\frac{2^{-\beta_j}l^j_x}{n^j_x} \sqrt{\frac{\mathrm{Re}u^k\,\mathrm{Re}u^l}{\mathrm{Re}u^j\,\mathrm{Re}s}}, \quad (j,k,l)=(\overline{1,2,3}).
\end{equation}
We denote the K\"{a}hler potential arising from strings stretching between stacks $x$ and $y$ as $\tilde{K}_{xy}$ and $\theta^j_{xy}\equiv \theta^j_y-\theta^j_x$ denotes the angle difference between the cycles wrapped by the branes $x$ and $y$ on the $j^\mathrm{th}$ two-torus with the constraint $\sum_j\theta^j_{xy}=0$. 

Following \cite{Font:2004cx,Cvetic:2003ch,Lust:2004cx}, we find two cases for the K\"{a}hler metric for 1/4 BPS brane configurations in type IIA theory:
\begin{itemize}
\item $\theta^j_{xy}<0$, $\theta^k_{xy}>0$, $\theta^l_{xy}>0$
\begin{align}
\tilde{K}_{xy} &= e^{\phi_4} e^{\gamma_E (2-\sum_{j = 1}^3 \theta^j_{xy}) } \sqrt{\frac{\Gamma(\theta^j_{xy})}{\Gamma(1+\theta^j_{xy})}} \sqrt{\frac{\Gamma(1-\theta^k_{xy})}{\Gamma(\theta^k_{xy})}} \sqrt{\frac{\Gamma(1-\theta^l_{xy})}{\Gamma(\theta^l_{xy})}} \nonumber \\ 
&\quad (t^j + \bar{t}^j)^{\theta^j_{xy}} (t^k + \bar{t}^k)^{-1+\theta^k_{xy}} (t^l + \bar{t}^l)^{-1+\theta^l_{xy}}. \label{eq:kahler1}
\end{align}
\item $\theta^j_{xy}<0$, $\theta^k_{xy}<0$, $\theta^l_{xy}>0$
\begin{align}
\tilde{K}_{xy} &= e^{\phi_4} e^{\gamma_E (2+\sum_{j = 1}^3 \theta^j_{xy}) } \sqrt{\frac{\Gamma(1+\theta^j_{xy})}{\Gamma(-\theta^j_{xy})}}
\sqrt{\frac{\Gamma(1+\theta^k_{xy})}{\Gamma(-\theta^k_{xy})}} \sqrt{\frac{\Gamma(\theta^l_{xy})}{\Gamma(1-\theta^l_{xy})}} \nonumber \\ 
&\quad (t^j + \bar{t}^j)^{-1-\theta^j_{xy}} (t^k + \bar{t}^k)^{-1-\theta^k_{xy}} (t^l + \bar{t}^l)^{-\theta^l_{xy}}. \label{eq:kahler2}
\end{align}
\end{itemize}
The K\"{a}hler metric for 1/2 BPS brane configurations, which give rise to non-chiral matter in bifundamental representations like the Higgs from the $\mathcal{N}=2$ sector, is given as,
\begin{equation}
\tilde{K}_\mathrm{Higgs} =\left[(s+\bar{s})(t^1+\bar{t}^1)(t^2+\bar{t}^2)(u^3+\bar{u}^3)\right]^{-1/2}.  \label{nonChiralK}
\end{equation}
The holomorphic superpotential is given as,
\begin{align}
W(M,C) &= \hat{W}(M) + \frac{1}{2} {\mu}_{\alpha \beta}(M)\, C^{\alpha}\,C^{\beta}+ \frac{1}{6}\,Y_{\alpha\beta\gamma}(M)\,C^{\alpha}\,C^{\beta}\,C^{\gamma}+... \label{eq:W}
\end{align}
and the minimum of the tree-level F-term scalar potential of the supergravity is given by\footnote{In our analysis we assume that D-terms do not affect the soft terms \cite{Kawamura:1996ex, Komargodski:2009pc}.}
\begin{align}
V(M,\bar{M}) &= e^{\kappa^2 \hat{K}} \left[\hat{K}^{\bar{I}J}(D_{I}\hat{W})^*(D_J \hat{W}) -3\,\kappa^2 \lvert \hat{W}\rvert ^2\right] \nonumber\\
 &= \hat{K}_{I\bar{J}} F^I \bar{F}^{\bar{J}} -3\,e^{\kappa^2 \hat{K}}\kappa^2 \lvert\hat{W}\rvert^2,
\end{align}
where $D_I=\partial_I + \kappa^2 \partial_I \hat{K}$, $\hat{K}_{\bar{I}J}=\partial_{\bar{I}} \partial_J \hat{K}$, $\hat{K}^{\bar{I}J}$ is the inverse K\"{a}hler metric, and the auxiliary fields $\bar{F}^{\bar{I}}$ are,
\begin{equation}
\bar{F}^{\bar{I}} = \kappa^2 e^{\kappa^2 \hat{K}/2} \hat{K}^{\bar{I}J} D_J \hat{W}, \label{aux}
\end{equation}
where the indices $I$, $J$ run over the dilaton $S$, the complex structure moduli $U^i$ and the K\"{a}hler moduli $T^i$. Thus supersymmetry is broken via F-terms from some of the hidden sector fields $M$ acquiring VEVs, thereby generating soft terms in the observable sector \cite{Font:2004cx, Kane:2004hm, Chen:2007zu}. Gravitino gets massive by absorbing goldstino via the superhiggs mechanism.
\begin{equation}
m_{3/2}=e^{\kappa^2 \hat{K}/2}\kappa^2|\hat{W}|.
\end{equation}
The \emph{normalized} soft parameters viz. the gaugino mass, squared scalar mass and trilinear parameters are given by \cite{Brignole:1997dp},
\begin{align}
M_x &= \frac{1}{2\,\mathrm{Re}\,f_x}\, (F^I\,\partial_I\,f_x), \nonumber \\
m_{xy}^2 &= m_{3/2}^2 + V_0 - \sum_{\bar{I},J}\, \bar{F}^{\bar{I}}F^J\,{\partial}_{\bar{I}}\,{\partial}_{J} \log({\tilde{K}}_{xy}), \nonumber \\
A_{xyz} &= F^I\left[\hat{K}_I+{\partial}_I\,\log(Y_{xyz})-{\partial}_{I} \log(\tilde{K}_{xy}\tilde{K}_{yz}\tilde{K}_{zx})\right], \label{softterms}
\end{align}
where $V_0$ is the VEV of the scalar potential.

Although it appears that soft terms may depend on the Yukawa couplings via the superpotential, however these are \emph{not} the physical Yukawa couplings which exponentially depend on the worldsheet area. Both are related by the following relation,
\begin{equation}
Y^{\mathrm{phys}}_{xyz} = Y_{xyz}\, \frac{\hat{W}^*}{|\hat{W}|}\,e^{\kappa^2\hat{K}/2}\, (\tilde{K}_{xy}\tilde{K}_{yz}\tilde{K}_{zx})^{-1/2}.
\end{equation}

To calculate the soft terms from supersymmetry breaking we ignore the cosmological constant $V_0$ and introduce the following VEVs for the auxiliary fields \eqref{aux} for the $s$, $t^i$ and $u^i$ moduli \cite{Brignole:1993dj},
\begin{align}\label{eq:aux}
F^s &= 2\sqrt{3}C m_{3/2} \mathrm{Re}(s) \Theta_s e^{-i\gamma_s}, \nonumber \\
F^{\{u,t\}^i} &= 2\sqrt{3}C m_{3/2}\left( \mathrm{Re}  ({u}^i) \Theta_i^u e^{-i\gamma^u_i}+  \mathrm{Re} (t^i) \Theta_i^t e^{-i\gamma^t_i}\right),
\end{align}
Here, the factors $\gamma_s$ and $\gamma_i$ denote the CP-violating phases of the moduli. The constant $C$ is given by the gravitino-mass $m_{3/2}$ and the cosmological constant $V_0$ as $C^2 = 1+ \frac{V_0}{3 m^2_{3/2}}$. $\Theta_s$ and $\Theta^{t,u}_i$ are the goldstino angles which determine the degree to which supersymmetry breaking is being dominated by any of the dilaton $s$, complex structure ($u^i$) and K\"ahler ($t^i$) moduli constrained by the relation,
\begin{align}\label{constraint}
\sum_{i=1}^3 (|\Theta_i^u|^2 + |\Theta_i^t|^2) + |\Theta_s|^2 =1.
\end{align}
Unlike the $s$ or $u$-moduli dominant supersymmetry breaking, the case of $t$-moduli dominant susy breaking depends on the physical Yukawa couplings via the area of the triangles. Accordingly we shall concentrate on the general scenario with the $u$-moduli dominated supersymmetry breaking with dilaton-modulus $s$ turned on, $F^s \neq 0$.
We set the cosmological constant, $V_0$ to be zero.

\subsection{Supersymmetry breaking via $u$-moduli and dilaton $s$}
Supersymmetry breaking via $u$-moduli including a non-zero VEV for the dilaton $s$ results in the following auxiliary fields \eqref{eq:aux},
\begin{equation}\label{auxfields_su}
F^{s,u^i} = \sqrt{3}m_{3/2}[(s + \bar{s})\Theta_s e^{-i\gamma_s} + (u^i + \bar{u}^i)\Theta_i e^{-i\gamma_i}].
\end{equation}

To calculate the soft terms, we need to know the derivatives of the K\"{a}hler potential with respect to $u^i$. Defining $\tilde{K}_{xy}\equiv e^{{\phi}_4}\,\tilde{K}^0_{xy}$ and using (\ref{eq:kahler1}) and \eqref{eq:kahler2}, we compute the derivatives with respect to $u^i$ as,
\begin{align}
\frac{\partial \log{\tilde{K}_{xy}}}{\partial u^i} &= \sum_{j=1}^3\frac {\partial \log{\tilde{K}^0_{xy}}}{\partial\theta^j_{xy}}
\frac{\partial\theta^j_{xy}}{\partial u^i} - \frac{1}{4(u^i+\bar{u}^i)}, \\
\frac{\partial^2 \log{\tilde{K}_{xy}}}{\partial u^i\partial\bar ,u^j}&= \sum_{k=1}^3\left(\frac{\partial \log{\tilde{K}^0_{xy}}}{\partial\theta^k_{xy}} \frac{\partial^2\theta^k_{xy}}{\partial u^i\partial\bar u^j} +\frac{\partial^2\log \tilde{K}^0_{xy}}{\partial (\theta^k_{xy})^2} \frac{\partial\theta^k_{xy}}{\partial u^i} \frac{\partial\theta^k_{xy}}{\partial \bar u^j} +\frac{{\delta}_{ij}}{4\,(u^i+\bar{u}^i)^2}\right). \label{eq:derivative-Kahler}
\end{align}
From the K\"{a}hler potential in \eqref{eq:kahler2}, we have
\begin{align}
\Psi(\theta^j_{xy}) &\equiv \frac {\partial \log{\tilde{K}^0_{xy}}}{\partial\theta^j_{xy}}= \gamma_E\!+\!\frac{1}{2}\frac{d}{d\theta^j_{xy}}\log{\Gamma(1-\theta^j_{xy})} -\frac{1}{2}\frac{d}{d\theta^j_{xy}}\log{\Gamma(\theta^j_{xy})}-\log(t^j+\bar t^j), \label{eq:psi}\\
\Psi'(\theta^j_{xy}) &\equiv \frac{\partial^2\log\tilde{K}^0_{xy}} {\partial(\theta^j_{xy})^2}= \frac{d\Psi(\theta^j_{xy})}{d \theta^j_{xy}}.  \label{eq:dpsi}
\end{align}
The first derivative of the angle differences $\theta^j_{xy}$ are defined as,
\begin{equation}
\theta^{j,k}_{xy} \equiv (u^k+\bar u^k)\,\frac{\partial \theta^j_{xy}}{\partial u^k}= \left\{\begin{array}{l} \left[-\frac{1}{4\pi} \sin(2\pi\theta^j) \right]^x_y , \quad j=k  \quad \\
\left[\frac{1}{4\pi}\sin(2\pi\theta^j) \right]^x_y , \quad j\neq k \end{array}\right.\label{eq:dthdu}
\end{equation}
\begin{equation}
\theta^{j,s}_{xy} \equiv (s+\bar s)\,\frac{\partial \theta^j_{xy}}{\partial s}= -\frac{1}{4\pi}\left[ \sin(2\pi\theta^j) \right]^x_y , \label{eq:dthdus2}
\end{equation}
where $[f(\theta^j)]^x_y=f(\theta^j_x)-f(\theta^j_y)$.

And the second order derivatives of the angle differences are,
\begin{equation}
\theta^{j,k\bar{l}}_{xy} \equiv (u^k+\bar u^k)(u^l+\bar u^l)\,\frac{\partial^2 \theta^j_{xy}}{\partial u^k\partial\bar u^l}= \left\{\begin{array}{l}
\frac{1}{16\pi}  \left[ \sin(4\pi\theta^j)+4\sin(2\pi\theta^j) \right]^x_y , \quad j=k=l  \\
 \frac{1}{16\pi}  \left[ \sin(4\pi\theta^j)-4\sin(2\pi\theta^j) \right]^x_y , \quad j\neq k=l \\
 -\frac{1}{16\pi}\left[ \sin(4\pi\theta^j) \right]^x_y , \quad j=k\neq l \mathrm{~or~} j=l\neq k \\
 \frac{1}{16\pi}\left[ \sin(4\pi\theta^j) \right]^x_y , \quad j\neq k\neq l\neq j\end{array}\right.\label{eq:d2thdu2}
\end{equation}
where $k,l\neq s$. While the terms associated with the dilaton $s$ are given as,
\begin{equation}
\theta^{j,k\bar s}_{xy} \equiv (u^k+\bar u^k)(s+\bar s)\,\frac{\partial^2 \theta^j_{xy}}{\partial u^k\partial\bar s}= \left\{\begin{array}{l}
 \frac{1}{16\pi}\left[\sin{4\pi\theta^j}\right]^x_y , \quad  j=k \quad \\
  -\frac{1}{16\pi}\left[\sin{4\pi\theta^j}\right]^x_y , \quad  j\neq k, \end{array}\right.\label{eq:dth2duds}
\end{equation}
and
\begin{equation}
\theta^{j,s\bar s}_{xy} \equiv (s+\bar s)(s+\bar s)\,\frac{\partial^2 \theta^j_{xy}}{\partial s\partial\bar s}= \frac{1}{16\pi}\left[\sin{4\pi\theta^j} 
+ 4\sin(2\pi\theta^j) \right]^x_y. \label{eq:dth2dss}
\end{equation}

\subsection{Soft parameters}
Substituting above parametrizations \eqref{auxfields_su}-\eqref{eq:dpsi} in the general formulas \eqref{softterms}, the soft parameters are found as follows:
\begin{itemize}
\item Gaugino mass parameters:
\begin{gather}
M_x=\frac{-\sqrt{3}m_{3/2}}{4\mathrm{Re} f_x}\Bigg[\sum_{j=1}^3 \mathrm{Re} (u^j)\,\Theta_j\, e^{-i\gamma_j}\, 2^{-(\beta_k +\beta_l)}n^j_x l^k_x l^l_x  +\Theta_s\mathrm{Re}(s) e^{-i\gamma_0}n_x^1\,n_x^2\,n_x^3\, \Bigg], \nonumber\\
                      (j,k,l)=(\overline{1,2,3}).\label{gaugino-masses}
\end{gather}
Bino mass parameter is then related to the linear combination of the gaugino masses for each stack as,
\begin{equation}\label{Bino-mass}
M_Y = \frac{1}{f_Y}\sum_x c_x f_x M_x ,
\end{equation}
where the coefficients $c_x$ correspond to the linear combination of U(1) factors which define the hypercharge, $\U(1)_Y = \sum c_x \U(1)_x$, cf. \eqref{fY}.
\item Trilinear parameters:
\begin{align}\label{tri-coupling}
A_{xyz}&=-\sqrt{3}m_{3/2}\sum_{j=1}^4 \left[\Theta_je^{-i\gamma_j}\left(\frac{1}{2}+\sum_{k=1}^3 \theta_{xy}^{k,j}\Psi(\theta^k_{xy})+\sum_{k=1}^3  \theta_{zx}^{k,j} \Psi(\theta^k_{zx})\right)\right] \nonumber \\
&\quad +\frac{\sqrt{3}}{2}m_{3/2}\left({\Theta}_{3}e^{-i{\gamma}_3} + \Theta_s e^{-i{\gamma}_s}\right),
\end{align}
where $j=4$ corresponds to $\Theta_s$ and $x$, $y$, and $z$ label those stacks of branes whose intersections define the corresponding fields present in the trilinear coupling. Since the differences of the angles may be negative $\theta_{xy} = \theta_y - \theta_x$, it is useful to define the sign parameter,
\begin{equation}\label{eq:etaxy}
\eta_{xy} = \prod_{i=1}^3 (-1)^{1-H(\theta_{xy}^{i})},\quad H(x)= \begin{cases}
      0, & x < 0 \\
      1, & x\geq 0
    \end{cases}
\end{equation}
where the value $\eta_{xy} = -1$ indicates that only one of the angle differences is negative while $\eta_{xy} = +1$ indicates that two of the angle differences are negative.
\item Squarks and sleptons mass-squared (1/4 BPS scalars):
\begin{align}\label{slepton-mass}
m^2_{xy}&= m_{3/2}^2\left[1-3\sum_{m,n=1}^4
\Theta_m\Theta_ne^{-i(\gamma_m-\gamma_n)}\left(
\frac{{\delta}_{mn}}{4}+ \sum_{j=1}^3 \left(\theta^{j,m\bar
n}_{xy}\Psi(\theta^j_{xy})+
 \theta^{j,m}_{xy}\theta^{j,\bar n}_{xy}\Psi'(\theta^j_{xy})\right)\right)
\right],\nonumber\\
\end{align}
where $\Theta_4 \equiv \Theta_s$ and the functions $\Psi(\theta_{xy})=\frac{\partial \log (e^{-\phi_4}\tilde{K}_{xy})}{\partial \theta_{xy}}$ in the case of $\eta_{xy}=-1$ are
\begin{align}
\mathrm{if} \ \theta_{xy} < 0&: \nonumber\\
\Psi(\theta^j_{xy})&=
-\gamma_E+\frac{1}{2}\frac{d}{d\theta^j_{xy}}\,\log{\Gamma(-\theta^j_{xy})}-
\frac{1}{2}\frac{d}{d\theta^j_{xy}}\,\log{\Gamma(1+\theta^j_{xy})}+\log(t^j+\bar t^j)\nonumber\\
\mathrm{if} \ \theta_{xy} > 0&: \label{eqn:Psi1}\\
\Psi(\theta^j_{xy})&=
-\gamma_E+\frac{1}{2}\frac{d}{d\theta^j_{xy}}\,\log{\Gamma(1-\theta^j_{xy})}-
\frac{1}{2}\frac{d}{d\theta^j_{xy}}\,\log{\Gamma(\theta^j_{xy})}+\log(t^j+\bar t^j),\nonumber
\end{align}
and in the case of $\eta_{xy}=+1$ are
\begin{align}
\mathrm{if} \ \theta_{xy} < 0&:  \nonumber\\
\Psi(\theta^j_{xy})&=
\gamma_E+\frac{1}{2}\frac{d}{d\theta^j_{xy}}\,\log{\Gamma(1+\theta^j_{xy})}-
\frac{1}{2}\frac{d}{d\theta^j_{xy}}\,\log{\Gamma(-\theta^j_{xy})}-\log(t^j+\bar t^j)  \nonumber\\
\mathrm{if} \ \theta_{xy} > 0&:  \label{eqn:Psi2}\\
\Psi(\theta^j_{xy})&=
\gamma_E+\frac{1}{2}\frac{d}{d\theta^j_{xy}}\,\log{\Gamma(\theta^j_{xy})}-
\frac{1}{2}\frac{d}{d\theta^j_{xy}}\,\log{\Gamma(1-\theta^j_{xy})}-\log(t^j+\bar t^j) ,\nonumber
\end{align}
and $\Psi'(\theta_{xy})$ is just the derivative $\Psi'(\theta^j_{xy}) =\frac{d\Psi(\theta^j_{xy})}{d \theta^j_{xy}}$.

\item Higgs mass-squared (1/2 BPS scalar):
\begin{align}
m^2_{H} &= m^2_{3/2}\left[1-\frac{3}{2}\left(\left|\Theta_3\right|^2+\left|\Theta_s\right|^2\right)\right].
\end{align}
\end{itemize}

\subsection{Computational strategy} 
For any given model we first analyze the structure of three-point Yukawa couplings in that model by considering the triplets $(a,\,b,\,c)_i$ on each of the two-tori $i={1,2,3}$. Not all models can incorporate the standard model fermion-mass hierarchies due to the generic rank-one problem. Only the models containing the required number of Higgs pairs from either from bulk or from $\mathcal{N}=2$ subsector on a single two-torus are viable to incorporate the Yukawa couplings \cite{Sabir:2024cgt}. 

Once a model is viable to generate fermion masses, we consider it to compute the relevant soft terms from supersymmetry breaking in the general case of $u$-moduli dominance with dilaton modulus $s$ turned on. 

We first need to calculate the complex structure moduli $U^i$ \eqref{U-moduli} in string theory basis and the corresponding $u^i$-moduli and $s$-modulus in the supergravity basis from \eqref{eq:moduli}. We also need to compute the gauge kinetic functions $\{f_a,f_b,f_c\}$ \eqref{kingauagefun} for each of three D6-brane stacks to calculate the bino mass \eqref{Bino-mass}. Finally, using the computed $\{M_a,M_b,M_c\}$ for the relevant triplet intersection of the particular model, we can obtain the gaugino masses, $\{M_Y, M_b, M_a\} \equiv \{M_{\tilde{B}}, M_{\tilde{W}}, M_{\tilde{g}}\}$, for the respective gauge groups $\U(1)_Y$, $\SU(2)_L$ and $\SU(3)_C$. 

Next, to calculate the trilinear coupling we first tabulate the various angles \eqref{eq:angle} with respect to the orientifold plane made by the cycle wrapped by each stack of D6-brane on each of the three two-tori. Care needs to be taken in evaluating the multivalued $\arctan$ function to avoid infinities. Furthermore, unlike the gauge kinetic function \eqref{kingauagefun}, that is even under the orientifold action for any D-brane stack, the angles (and the differences of the angles $\theta_{xy}^{i}= \theta_{y}^{i} -\theta_{x}^{i} $) change signs which makes the computation quite involved. For example, the three-point Yukawa couplings for Model~\hyperref[model22]{22} and Model~\hyperref[model22.5]{22-dual} are completely different even though the models are dual under the exchange of $b$ and $c$ stacks. 

In order to account for the negative angle differences it is convenient to define the sign function $\sigma_{xy}^{i}$ which was first introduced in \cite{Sabir:2022hko}, which is $-1$ only for negative angle difference and $+1$ otherwise,
\begin{align}\label{sigmaK}
\sigma_{xy}^{i} \equiv (-1)^{1-H(\theta_{xy}^{i})},
\end{align}
where $H(x)$ is the unit step function. And the function $\eta_{xy}$ \eqref{eq:etaxy} can thus be defined by taking the product on the torus index $i$ as,
\begin{equation}\label{eta}
\eta_{xy}\equiv  \prod_{i=1}^3 \sigma_{xy}^i.
\end{equation}
Using the above defined $\sigma_{xy}^{i}$ and $\eta_{xy}$, we can readily write the four cases of functions $\Psi(\theta_{xy})$ defined in \eqref{eqn:Psi1} and \eqref{eqn:Psi2} into a single expression as,
\begin{align}\label{Psi}
\Psi(\theta^j_{xy}) &= \eta_{xy}\left( \frac{1}{2} \psi^{(0)}(\sigma_{xy}^{i}\theta^j_{xy})+\frac{1}{2}\psi^{(0)}(1-\sigma_{xy}^{i}\theta^j_{xy})+\gamma_E-\log(t^j+\bar t^j)\right) ,
\end{align}
where $\psi^{(0)}(z)$ is called the digamma function defined as the derivative of the logarithm of the gamma function. The successive derivatives of the $\log\Gamma (z)$ yield the polygamma function $\psi^{(n)}(z)$ as,
\begin{align}\label{polygamma}
\psi^{(n-1)}(z) = \frac{d^{(n)}}{dz^{(n)}}\log\Gamma(z) ,
\end{align}
with the following properties,
\begin{align}\label{eq:property}
\frac{d }{dz}\psi^{(0)}(\pm z) &= \pm \psi^{(1)}(\pm z), \nonumber\\
\frac{d }{dz}\psi^{(0)}(1 \pm z) &= \pm \psi^{(1)}(1 \pm z).
\end{align}
Similarly, the derivative $\Psi'(\theta^j_{xy}) = \frac{d\Psi(\theta^j_{xy})}{d \theta^j_{xy}}$ can be expressed succinctly as,
\begin{align}\label{DPsi}
\Psi'(\theta^j_{xy}) & = \eta_{xy}\sigma_{xy}^{i} \left( \frac{1}{2} \psi^{(1)}(\sigma_{xy}^{i}\theta^j_{xy})+\frac{1}{2}\psi^{(1)}(1-\sigma_{xy}^{i}\theta^j_{xy})\right),
\end{align}
where we have utilized the property \eqref{eq:property} and have neglected the contribution of the $t$-moduli.

Lastly, by making use of appropriate Kronecker deltas and defining $u^4\equiv s$, we can express the various cases of the first and second derivatives of the angles as,
\begin{gather}\label{derivative-angles1}
\theta^{i,m}_{xy} \equiv (u^m+\bar u^m)\,\frac{\partial \theta^i_{xy}}{\partial u^m} = (-1)^{\delta _{m,4}} (-1)^{\delta _{i,j}}\frac{\sin (2 \pi  \theta^i)}{4\pi } \bigg|^x_y , \nonumber \\
 i={1,2,3}; \quad m={1,2,3,4.}
\end{gather}
\begin{align}\label{derivative-angles2}
\theta^{i,mn}_{xy} &\equiv (u^m+\bar u^m)(u^n+\bar u^n)\,\frac{\partial^2 \theta^i_{xy}}{\partial u^m\partial\bar u^n} \nonumber \\
&= \delta _{m,n}\frac{\sin (4 \pi  \theta^i) +(-1)^{(1-\delta _{4,m}) (1-\delta _{i,m})}4 \sin (2 \pi  \theta^i)}{16\pi } \bigg|^x_y \nonumber \\
&\quad + (1-\delta _{m,n}) (-1)^{(1-\delta _{4,m}) (1-\delta _{4,n}) (\delta _{i,m}+\delta _{i,n})} (-1)^{1-\delta _{i,m}-\delta _{i,n}} \frac{\sin (4 \pi  \theta^i)}{16\pi } \bigg|^x_y \nonumber \\
& \qquad\qquad\qquad\qquad  i={1,2,3}; \quad m,n={1,2,3,4}.
\end{align}
Utilizing these results the trilinear coupling \eqref{tri-coupling} and the sleptons and the squarks mass-squared parameters \eqref{slepton-mass} are computed while ignoring the CP-violating phases $\{\gamma_1,\gamma_2,\gamma_3,\gamma_s\}$.  

\section{Susy-breaking soft terms in the Pati-Salam landscape}\label{sec:soft_terms}
In appendix~\ref{appA} we tabulate all 33 independent three-family supersymmetric Pati-Salam models with distinct allowed gauge coupling relations. The complete perturbative particle spectra for all models in the landscape have been listed in the appendix B of ref. \cite{Sabir:2024cgt}. 

We find that the first sixteen models viz. \hyperref[model1]{1}, \hyperref[model1.5]{1-dual}, \hyperref[model2]{2}, \hyperref[model3]{3}, \hyperref[model3.5]{3-dual}, \hyperref[model4]{4}, \hyperref[model5]{5}, \hyperref[model6]{6}, \hyperref[model7]{7}, \hyperref[model8]{8}, \hyperref[model9]{9}, \hyperref[model9.5]{9-dual}, \hyperref[model10]{10}, \hyperref[model11]{11}, \hyperref[model11.5]{11-dual}, and \hyperref[model12]{12} do not possess the correct form of three-point Yukawa textures to generate the fermion masses on a single two-torus. Therefore, we will only focus on the remaining models where viable three-point Yukawa interactions are possible.

\subsection{Model 13}\label{sec:model-13}
\begin{figure}[htb]
\centering
\includegraphics[width=\textwidth]{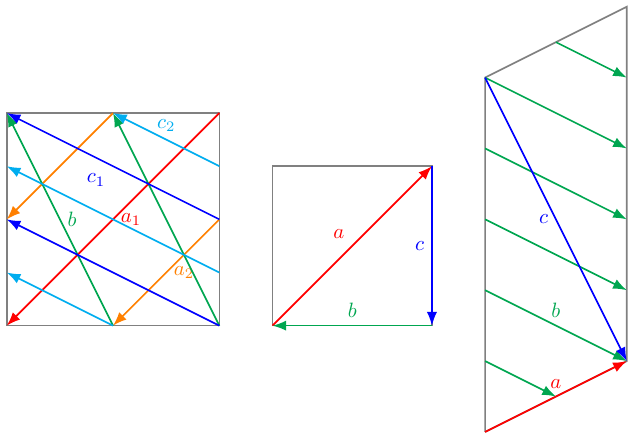}
\caption{Brane configuration for the three two-tori in Model~\hyperref[model13]{13} where the third two-torus is tilted. Fermion mass hierarchies result from the intersections on the first two-torus.}  \label{Fig.13}
\end{figure}

In Model~\hyperref[model13]{13} the three-point Yukawa couplings arise from the triplet intersections from the branes $a$, $b$ and $c$ on the first two-torus with 3 pairs of Higgs from the bulk\footnote{This is the only model in the landscape with 3 adjoint scalars from the bulk. Unlike the Higgs from the $\mathcal{N}=2$ sector which are insensitive to the bulk moduli, the tiny Yukawa couplings from the bulk Higgs are argued to be related to the infinite distance limit~\cite{Lee:2019wij} in the moduli space where a light of tower states, dubbed gonions~\cite{Aldazabal:2000cn}, appears signalling the decompactification of one or two compact dimensions \cite{Casas:2024ttx}.}.
Yukawa matrices for the Model~\hyperref[model13]{13} are of rank 3 and the three intersections required to form the disk diagrams for the Yukawa couplings all occur on the first torus as shown in figure~\ref{Fig.13}.  
\begin{table}[htb]\footnotesize\centering
\renewcommand{\arraystretch}{1.8}
$\begin{array}{|c|c|c|c|}\hline
 \text{\hyperref[model13]{13}} & \theta ^1 & \theta ^2 & \theta ^3 \\\hline
 a & \tan ^{-1}\left(\frac{1}{\sqrt{5}}\right) & -\tan ^{-1}\left(\frac{2}{\sqrt{5}}\right) & -\tan ^{-1}\left(\frac{1}{2 \sqrt{5}}\right) \\\hline
 b & \tan ^{-1}\left(\frac{11}{\sqrt{5}}\right) & 0 & \frac{\pi }{2} \\\hline
 c & \tan ^{-1}\left(2 \sqrt{5}\right) & \tan ^{-1}\left(\frac{2}{\sqrt{5}}\right) & -\tan ^{-1}\left(2 \sqrt{5}\right) \\\hline
\end{array}$
\caption{The angles with respect to the orientifold plane made by the cycle wrapped by stack of D6-branes on each of the three two-tori in Model~\hyperref[model13]{13}.}
\label{Angles13}
\end{table}

The complex structure moduli $U^i$ \eqref{U-moduli} are given as,
\begin{align}\label{eqn:U-moduli13}
\{U^1,U^2,U^3\} &=\left\{\frac{i}{\sqrt{5}},\frac{11 i}{\sqrt{5}},\frac{4}{21} \left(10+i \sqrt{5}\right)\right\},
\end{align}
and the corresponding $u$-moduli and $s$-modulus in supergravity basis from \eqref{eq:moduli} are,
\begin{align}\label{s-u-moduli13}
\{u^1,u^2,u^3\} & = \left\{\frac{\sqrt[4]{5} \sqrt{11} e^{-\phi _4}}{\pi },\frac{\sqrt[4]{5} e^{-\phi _4}}{\sqrt{11} \pi },\frac{\sqrt{11} e^{-\phi _4}}{4\ 5^{3/4} \pi }\right\}, \nonumber\\
s & = \frac{\sqrt[4]{5} e^{-\phi _4}}{4 \sqrt{11} \pi } .
\end{align}
Using \eqref{kingauagefun} and the values from the table~\ref{model13}, the gauge kinetic function becomes,
\begin{align}\label{fx13}
\{f_a,f_b,f_c\} & = \left\{\frac{63 e^{-\phi _4}}{8\ 5^{3/4} \sqrt{11} \pi },\frac{9 \sqrt[4]{5} e^{-\phi _4}}{16 \sqrt{11} \pi },\frac{21 \sqrt{11} e^{-\phi _4}}{16\ 5^{3/4} \pi }\right\},
\end{align}
To calculate the gaugino masses $\{M_Y,M_b,M_a\}$ for the respective gauge groups $\U(1)_Y$, $\SU(2)_L$, and $\SU(3)_C$, we first compute $\{M_a,M_b,M_c\}$ using \eqref{gaugino-masses} as,
\begin{align}\label{Gauginosabc13}
M_a &= \frac{m_{3/2} (110 \Theta _1+10 \Theta _2+11 \Theta _3+5 \Theta _4)}{42 \sqrt{3}},\nonumber \\
M_b &= \frac{m_{3/2} (4 \Theta _2-5 \Theta _4)}{3 \sqrt{3}},\nonumber \\
M_c &= \frac{m_{3/2} (20 \Theta _1+\Theta _3)}{7 \sqrt{3}}.
\end{align} 

Next, to compute the trilinear coupling and the sleptons mass-squared we require the angles, the differences of angles and their first and second order derivatives with respect to the moduli. In table~\ref{Angles13} we show the angles \eqref{eq:angle} made by the cycles wrapped by each stack of D6-branes with respect to the orientifold plane on each two-torus.
The differences of the angles, $\theta_{xy}^{i}= \theta_{y}^{i} -\theta_{x}^{i}$ are,
\begin{align}\label{angle-diff13}\arraycolsep=0pt
\hskip -1em \left[
\begin{array}{ccc}
 \{0.,0.,0.\} & \{0.132923,-0.228657,0.378919\} & \{0.642663,0.0589537,0.298383\} \\
 \{-0.132923,0.228657,-0.378919\} & \{0.,0.,0.\} & \{0.50974,0.287611,-0.080536\} \\
 \{-0.642663,-0.0589537,-0.298383\} & \{-0.50974,-0.287611,0.080536\} & \{0.,0.,0.\} \\
\end{array}
\right]
\end{align}

To account for the negative angle differences we employ the sign function $\sigma_{xy}^{i}$, which is $-1$ only for negative angle difference and $+1$ otherwise,
\begin{align}\label{sigmaK13}
\sigma_{xy}^{i} & = \left(
\begin{array}{ccc}
 \{1,1,1\} & \{1,-1,1\} & \{1,1,1\} \\
 \{-1,1,-1\} & \{1,1,1\} & \{1,1,-1\} \\
 \{-1,-1,-1\} & \{-1,-1,1\} & \{1,1,1\} \\
\end{array}
\right),
\end{align}
and the function $\eta_{xy}$ is evaluated by taking the product on the torus index $i$ as,
\begin{equation}\label{eta13}
\eta_{xy} = \left(
\begin{array}{ccc}
 1 & -1 & 1 \\
 1 & 1 & -1 \\
 -1 & 1 & 1 \\
\end{array}
\right).
\end{equation}
Using the values of $\sigma_{xy}^{i}$ and $\eta_{xy}$ in \eqref{Psi} we can compute the four cases of functions $\Psi(\theta_{xy})$ defined in \eqref{eqn:Psi1} and \eqref{eqn:Psi2}. Similarly we calculate the derivative $\Psi'(\theta^j_{xy}) = \frac{d\Psi(\theta^j_{xy})}{d \theta^j_{xy}}$ using equations \eqref{DPsi}, \eqref{derivative-angles1}, \eqref{derivative-angles2} and the properties of digamma function $\psi^{(0)}(z)$ \eqref{eq:property} while neglecting the contribution of the $t$-moduli. 

Utilizing above results while ignoring the CP-violating phases $\gamma_m$, the gaugino masses; the trilinear coupling \eqref{tri-coupling}; and the squared-masses of squarks and sleptons \eqref{slepton-mass} are obtained as,
\begin{align}\label{GauginosYba123-model13}  
M_{\tilde B} &\equiv M_Y = m_{3/2}\Bigl(\frac{176 \Theta _1+4 \Theta _2+11 \Theta _3+2 \Theta _4}{63 \sqrt{3}}\Bigr), \nonumber\\
M_{\tilde W} &\equiv M_b = m_{3/2}\Bigl(\frac{4 \Theta _2-5 \Theta _4}{3 \sqrt{3}}\Bigr),\nonumber \\
M_{\tilde g} &\equiv M_a = m_{3/2}\Bigl(\frac{110 \Theta _1+10 \Theta _2+11 \Theta _3+5 \Theta _4}{42 \sqrt{3}}\Bigr),\nonumber \\
A_0 \equiv A_{abc} &= m_{3/2}\Bigl(-0.557548 \Theta _1-1.13015 \Theta _2+0.418029 \Theta _3-0.462383 \Theta _4\Bigr),\nonumber\\  
m^2_{L} \equiv m^2_{ab} &= m_{3/2}^2\Bigl(0.185098 \Theta _1{}^2-0.0411881 \Theta _1 \Theta _2-1.44927 \Theta _1 \Theta _3-1.45644 \Theta _1 \Theta _4 \nonumber \\   &\quad +1.06435 \Theta _2{}^2-0.969753 \Theta _2 \Theta _3-0.386382 \Theta _2 \Theta _4-0.290557 \Theta _3{}^2 \nonumber \\  &\quad +0.395625 \Theta _3 \Theta _4-1.0645 \Theta _4{}^2+1\Bigr) ,\nonumber\\  
m^2_{R} \equiv m^2_{ac} &= m_{3/2}^2\Bigl(-2.63181 \Theta _1{}^2+0.77203 \Theta _1 \Theta _2+1.29272 \Theta _1 \Theta _3+1.47197 \Theta _1 \Theta _4 \nonumber \\  &\quad  -2.51929 \Theta _2{}^2+1.29272 \Theta _2 \Theta _3+0.844287 \Theta _2 \Theta _4-2.34591 \Theta _3{}^2 \nonumber \\  &\quad -0.0442256 \Theta _3 \Theta _4-0.0470486 \Theta _4{}^2+1\Bigr). 
\end{align}
All soft terms are subject to the constraint \eqref{constraint}.
\FloatBarrier

\subsection{Model 14}\label{sec:model-14}
\begin{figure}[htb]
\centering
\includegraphics[width=\textwidth]{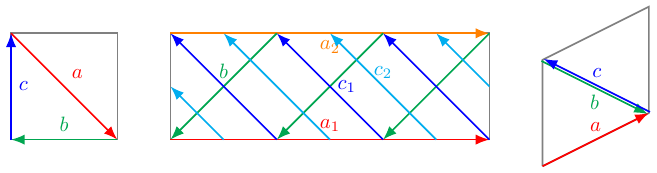}
\caption{Brane configuration for the three two-tori in Model~\hyperref[model14]{14} where the third two-torus is tilted. Fermion mass hierarchies result from the intersections on the second two-torus.}  \label{Fig.14}
\end{figure}

In Model~\hyperref[model14]{14} the three-point Yukawa couplings arise from the triplet intersections from the branes $a$, $b$ and $c$ on the second two-torus with 6 pairs of Higgs from the $\mathcal{N}=2$ sector.
Yukawa matrices for the Model~\hyperref[model14]{14} are of rank 3 and the three intersections required to form the disk diagrams for the Yukawa couplings all occur on the second torus as shown in figure~\ref{Fig.14}.  
\begin{table}[htb]\footnotesize\centering
\renewcommand{\arraystretch}{1.8}
$\begin{array}{|c|c|c|c|}\hline
 \text{\hyperref[model14]{14}} & \theta ^1 & \theta ^2 & \theta ^3 \\\hline
 a & \frac{3 \pi }{4} & 0 & \frac{\pi }{2} \\\hline
 b & 0 & \frac{\pi }{4} & \frac{3 \pi }{4} \\\hline
 c & \frac{\pi }{4} & \frac{3 \pi }{4} & \frac{3 \pi }{4} \\\hline
\end{array}$
\caption{The angles with respect to the orientifold plane made by the cycle wrapped by stack of D6-branes on each of the three two-tori in Model~\hyperref[model14]{14}.}
\label{Angles14}
\end{table}

The complex structure moduli $U^i$ \eqref{U-moduli} are given as,
\begin{align}\label{eqn:U-moduli14}
\{U^1,U^2,U^3\} &=\left\{i,\frac{i}{3},1+i\right\},
\end{align}
and the corresponding $u$-moduli and $s$-modulus in supergravity basis from \eqref{eq:moduli} are,
\begin{align}\label{s-u-moduli14}
\{u^1,u^2,u^3\} & = \left\{\frac{e^{-\phi _4}}{\sqrt{6} \pi },\frac{\sqrt{\frac{3}{2}} e^{-\phi _4}}{\pi },\frac{e^{-\phi _4}}{2 \sqrt{6} \pi }\right\}, \nonumber\\
s & = \frac{\sqrt{\frac{3}{2}} e^{-\phi _4}}{2 \pi } .
\end{align}
Using \eqref{kingauagefun} and the values from the table~\ref{model14}, the gauge kinetic function becomes,
\begin{align}\label{fx14}
\{f_a,f_b,f_c\} & = \left\{\frac{\sqrt{\frac{3}{2}} e^{-\phi _4}}{4 \pi },\frac{\sqrt{\frac{3}{2}} e^{-\phi _4}}{4 \pi },\frac{\sqrt{\frac{3}{2}} e^{-\phi _4}}{4 \pi }\right\},
\end{align}
To calculate the gaugino masses $\{M_Y,M_b,M_a\}$ for the respective gauge groups $\U(1)_Y$, $\SU(2)_L$, and $\SU(3)_C$, we first compute $\{M_a,M_b,M_c\}$ using \eqref{gaugino-masses} as,
\begin{align}\label{Gauginosabc14}
M_a &= \frac{1}{2} \sqrt{3} m_{3/2} (\Theta _2-\Theta _4),\nonumber \\
M_b &= \frac{1}{2} \sqrt{3} m_{3/2} (\Theta _1-\Theta _4),\nonumber \\
M_c &= \frac{1}{2} \sqrt{3} m_{3/2} (\Theta _2+\Theta _3).
\end{align} 

Next, to compute the trilinear coupling and the sleptons mass-squared we require the angles, the differences of angles and their first and second order derivatives with respect to the moduli. In table~\ref{Angles14} we show the angles \eqref{eq:angle} made by the cycles wrapped by each stack of D6-branes with respect to the orientifold plane on each two-torus.
The differences of the angles, $\theta_{xy}^{i}= \theta_{y}^{i} -\theta_{x}^{i}$ are,
\begin{align}\label{angle-diff14}\arraycolsep=0pt
\hskip -1em \left[
\begin{array}{ccc}
 \{0.,0.,0.\} & \{-0.0730092,0.643806,-0.570796\} & \{0.356194,0.356194,-0.429204\} \\
 \{0.0730092,-0.643806,0.570796\} & \{0.,0.,0.\} & \{0.429204,-0.287611,0.141593\} \\
 \{-0.356194,-0.356194,0.429204\} & \{-0.429204,0.287611,-0.141593\} & \{0.,0.,0.\} \\
\end{array}
\right]
\end{align}

To account for the negative angle differences we employ the sign function $\sigma_{xy}^{i}$, which is $-1$ only for negative angle difference and $+1$ otherwise,
\begin{align}\label{sigmaK14}
\sigma_{xy}^{i} & = \left(
\begin{array}{ccc}
 \{1,1,1\} & \{-1,1,-1\} & \{1,1,-1\} \\
 \{1,-1,1\} & \{1,1,1\} & \{1,-1,1\} \\
 \{-1,-1,1\} & \{-1,1,-1\} & \{1,1,1\} \\
\end{array}
\right),
\end{align}
and the function $\eta_{xy}$ is evaluated by taking the product on the torus index $i$ as,
\begin{equation}\label{eta14}
\eta_{xy} = \left(
\begin{array}{ccc}
 1 & 1 & -1 \\
 -1 & 1 & -1 \\
 1 & 1 & 1 \\
\end{array}
\right).
\end{equation}
Using the values of $\sigma_{xy}^{i}$ and $\eta_{xy}$ in \eqref{Psi} we can compute the four cases of functions $\Psi(\theta_{xy})$ defined in \eqref{eqn:Psi1} and \eqref{eqn:Psi2}. Similarly we calculate the derivative $\Psi'(\theta^j_{xy}) = \frac{d\Psi(\theta^j_{xy})}{d \theta^j_{xy}}$ using equations \eqref{DPsi}, \eqref{derivative-angles1}, \eqref{derivative-angles2} and the properties of digamma function $\psi^{(0)}(z)$ \eqref{eq:property} while neglecting the contribution of the $t$-moduli. 

Utilizing above results while ignoring the CP-violating phases $\gamma_m$, the gaugino masses; the trilinear coupling \eqref{tri-coupling}; and the squared-masses of squarks and sleptons \eqref{slepton-mass} are obtained as,
\begin{align}\label{GauginosYba123-model14}  
M_{\tilde B} &\equiv M_Y = m_{3/2}\Bigl(\frac{1}{10} \sqrt{3} (5 \Theta _2+3 \Theta _3-2 \Theta _4)\Bigr), \nonumber\\
M_{\tilde W} &\equiv M_b = m_{3/2}\Bigl(\frac{1}{2} \sqrt{3} (\Theta _1-\Theta _4)\Bigr),\nonumber \\
M_{\tilde g} &\equiv M_a = m_{3/2}\Bigl(\frac{1}{2} \sqrt{3} (\Theta _2-\Theta _4)\Bigr),\nonumber \\
A_0 \equiv A_{abc} &= m_{3/2}\Bigl(-0.444661 \Theta _1-1.36219 \Theta _2+0.260871 \Theta _3-0.186075 \Theta _4\Bigr),\nonumber\\  
m^2_{L} \equiv m^2_{ab} &= m_{3/2}^2\Bigl(-0.0689562 \Theta _1{}^2+1.52807 \Theta _1 \Theta _2-0.25782 \Theta _1 \Theta _3-0.582907 \Theta _1 \Theta _4 \nonumber \\   &\quad -0.127855 \Theta _2{}^2+0.31395 \Theta _2 \Theta _3-0.623439 \Theta _2 \Theta _4-2.04994 \Theta _3{}^2 \nonumber \\  &\quad +1.23441 \Theta _3 \Theta _4-0.809377 \Theta _4{}^2+1\Bigr) ,\nonumber\\  
m^2_{R} \equiv m^2_{ac} &= m_{3/2}^2\Bigl(-2.28519 \Theta _1{}^2-1.00169 \Theta _1 \Theta _2+0.506454 \Theta _1 \Theta _3+1.26634 \Theta _1 \Theta _4 \nonumber \\  &\quad  -0.637087 \Theta _2{}^2+1.13601 \Theta _2 \Theta _3+0.140834 \Theta _2 \Theta _4-0.0262168 \Theta _3{}^2 \nonumber \\  &\quad -1.1866 \Theta _3 \Theta _4-0.492648 \Theta _4{}^2+1\Bigr). 
\end{align}
All soft terms are subject to the constraint \eqref{constraint}.
\FloatBarrier

\subsection{Model 15}\label{sec:model-15}
\begin{figure}[htb]
\centering
\includegraphics[width=\textwidth]{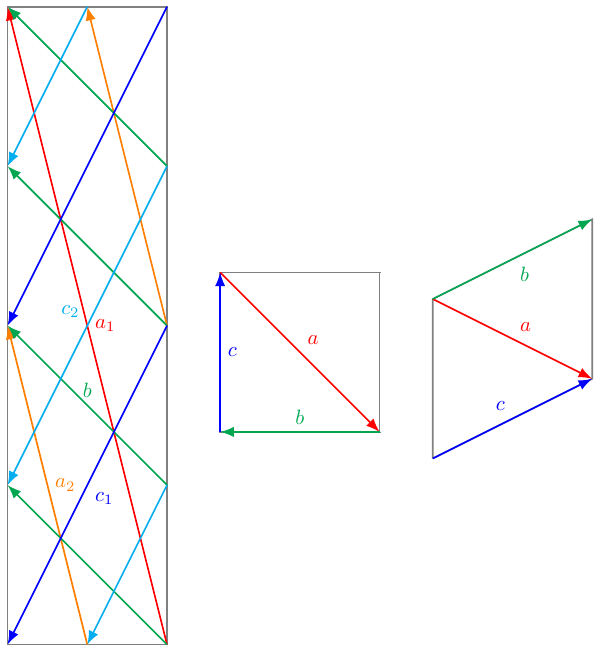}
\caption{Brane configuration for the three two-tori in Model~\hyperref[model15]{15} where the third two-torus is tilted. Fermion mass hierarchies result from the intersections on the first two-torus.}  \label{Fig.15}
\end{figure}

In Model~\hyperref[model15]{15} the three-point Yukawa couplings arise from the triplet intersections from the branes $a$, $b$ and $c$ on the first two-torus with 6 pairs of Higgs from the $\mathcal{N}=2$ sector.
Yukawa matrices for the Model~\hyperref[model15]{15} are of rank 3 and the three intersections required to form the disk diagrams for the Yukawa couplings all occur on the first torus as shown in figure~\ref{Fig.15}.  
\begin{table}[htb]\footnotesize\centering
\renewcommand{\arraystretch}{1.8}
$\begin{array}{|c|c|c|c|}\hline
 \text{\hyperref[model15]{15}} & \theta ^1 & \theta ^2 & \theta ^3 \\\hline
 a & -\tan ^{-1}\left(2 \sqrt{2}\right) & -\tan ^{-1}\left(\frac{1}{\sqrt{2}}\right) & \tan ^{-1}\left(\sqrt{2}\right) \\\hline
 b & -\tan ^{-1}\left(\frac{5}{\sqrt{2}}\right) & 0 & \frac{\pi }{2} \\\hline
 c & -\tan ^{-1}\left(\frac{1}{\sqrt{2}}\right) & \tan ^{-1}\left(\frac{1}{\sqrt{2}}\right) & \tan ^{-1}\left(\frac{1}{\sqrt{2}}\right) \\\hline
\end{array}$
\caption{The angles with respect to the orientifold plane made by the cycle wrapped by stack of D6-branes on each of the three two-tori in Model~\hyperref[model15]{15}.}
\label{Angles15}
\end{table}

The complex structure moduli $U^i$ \eqref{U-moduli} are given as,
\begin{align}\label{eqn:U-moduli15}
\{U^1,U^2,U^3\} &=\left\{2 i \sqrt{2},\frac{5 i}{\sqrt{2}},\frac{2}{3} \left(1+i \sqrt{2}\right)\right\},
\end{align}
and the corresponding $u$-moduli and $s$-modulus in supergravity basis from \eqref{eq:moduli} are,
\begin{align}\label{s-u-moduli15}
\{u^1,u^2,u^3\} & = \left\{\frac{\sqrt{5} e^{-\phi _4}}{2\ 2^{3/4} \pi },\frac{\sqrt[4]{2} e^{-\phi _4}}{\sqrt{5} \pi },\frac{\sqrt{5} e^{-\phi _4}}{2^{3/4} \pi }\right\}, \nonumber\\
s & = \frac{e^{-\phi _4}}{2\ 2^{3/4} \sqrt{5} \pi } .
\end{align}
Using \eqref{kingauagefun} and the values from the table~\ref{model15}, the gauge kinetic function becomes,
\begin{align}\label{fx15}
\{f_a,f_b,f_c\} & = \left\{\frac{27 e^{-\phi _4}}{16\ 2^{3/4} \sqrt{5} \pi },\frac{3 e^{-\phi _4}}{4\ 2^{3/4} \sqrt{5} \pi },\frac{3 \sqrt{5} e^{-\phi _4}}{8\ 2^{3/4} \pi }\right\},
\end{align}
To calculate the gaugino masses $\{M_Y,M_b,M_a\}$ for the respective gauge groups $\U(1)_Y$, $\SU(2)_L$, and $\SU(3)_C$, we first compute $\{M_a,M_b,M_c\}$ using \eqref{gaugino-masses} as,
\begin{align}\label{Gauginosabc15}
M_a &= \frac{m_{3/2} (5 \Theta _1+4 \Theta _2+20 \Theta _3+2 \Theta _4)}{9 \sqrt{3}},\nonumber \\
M_b &= \frac{m_{3/2} (\Theta _2-2 \Theta _4)}{\sqrt{3}},\nonumber \\
M_c &= \frac{m_{3/2} (\Theta _1+2 \Theta _3)}{\sqrt{3}}.
\end{align} 

Next, to compute the trilinear coupling and the sleptons mass-squared we require the angles, the differences of angles and their first and second order derivatives with respect to the moduli. In table~\ref{Angles15} we show the angles \eqref{eq:angle} made by the cycles wrapped by each stack of D6-branes with respect to the orientifold plane on each two-torus.
The differences of the angles, $\theta_{xy}^{i}= \theta_{y}^{i} -\theta_{x}^{i}$ are,
\begin{align}\label{angle-diff15}\arraycolsep=0pt
\hskip -1em \left[
\begin{array}{ccc}
 \{0.,0.,0.\} & \{-0.38452,-0.563254,0.230959\} & \{-0.0969093,-0.13405,0.230959\} \\
 \{0.38452,0.563254,-0.230959\} & \{0.,0.,0.\} & \{0.287611,0.429204,0.\} \\
 \{0.0969093,0.13405,-0.230959\} & \{-0.287611,-0.429204,0.\} & \{0.,0.,0.\} \\
\end{array}
\right]
\end{align}

To account for the negative angle differences we employ the sign function $\sigma_{xy}^{i}$, which is $-1$ only for negative angle difference and $+1$ otherwise,
\begin{align}\label{sigmaK15}
\sigma_{xy}^{i} & = \left(
\begin{array}{ccc}
 \{1,1,1\} & \{-1,-1,1\} & \{-1,-1,1\} \\
 \{1,1,-1\} & \{1,1,1\} & \{1,1,1\} \\
 \{1,1,-1\} & \{-1,-1,1\} & \{1,1,1\} \\
\end{array}
\right),
\end{align}
and the function $\eta_{xy}$ is evaluated by taking the product on the torus index $i$ as,
\begin{equation}\label{eta15}
\eta_{xy} = \left(
\begin{array}{ccc}
 1 & 1 & 1 \\
 -1 & 1 & 1 \\
 -1 & 1 & 1 \\
\end{array}
\right).
\end{equation}
Using the values of $\sigma_{xy}^{i}$ and $\eta_{xy}$ in \eqref{Psi} we can compute the four cases of functions $\Psi(\theta_{xy})$ defined in \eqref{eqn:Psi1} and \eqref{eqn:Psi2}. Similarly we calculate the derivative $\Psi'(\theta^j_{xy}) = \frac{d\Psi(\theta^j_{xy})}{d \theta^j_{xy}}$ using equations \eqref{DPsi}, \eqref{derivative-angles1}, \eqref{derivative-angles2} and the properties of digamma function $\psi^{(0)}(z)$ \eqref{eq:property} while neglecting the contribution of the $t$-moduli. 

Utilizing above results while ignoring the CP-violating phases $\gamma_m$, the gaugino masses; the trilinear coupling \eqref{tri-coupling}; and the squared-masses of squarks and sleptons \eqref{slepton-mass} are obtained as,
\begin{align}\label{GauginosYba123-model15}  
M_{\tilde B} &\equiv M_Y = m_{3/2}\Bigl(\frac{10 \Theta _1+2 \Theta _2+25 \Theta _3+\Theta _4}{12 \sqrt{3}}\Bigr), \nonumber\\
M_{\tilde W} &\equiv M_b = m_{3/2}\Bigl(\frac{\Theta _2-2 \Theta _4}{\sqrt{3}}\Bigr),\nonumber \\
M_{\tilde g} &\equiv M_a = m_{3/2}\Bigl(\frac{5 \Theta _1+4 \Theta _2+20 \Theta _3+2 \Theta _4}{9 \sqrt{3}}\Bigr),\nonumber \\
A_0 \equiv A_{abc} &= m_{3/2}\Bigl(-0.863715 \Theta _1+0.764882 \Theta _2-1.2303 \Theta _3-0.402914 \Theta _4\Bigr),\nonumber\\  
m^2_{L} \equiv m^2_{ab} &= m_{3/2}^2\Bigl(-1.47868 \Theta _1{}^2-0.544215 \Theta _1 \Theta _2+2.13294 \Theta _1 \Theta _3+0.574711 \Theta _1 \Theta _4 \nonumber \\   &\quad -2.39151 \Theta _2{}^2+1.01424 \Theta _2 \Theta _3+1.98343 \Theta _2 \Theta _4+0.200897 \Theta _3{}^2 \nonumber \\  &\quad -1.60235 \Theta _3 \Theta _4-1.24227 \Theta _4{}^2+1\Bigr) ,\nonumber\\  
m^2_{R} \equiv m^2_{ac} &= m_{3/2}^2\Bigl(-0.93815 \Theta _1{}^2-0.630132 \Theta _1 \Theta _2+1.49955 \Theta _1 \Theta _3+1.27302 \Theta _1 \Theta _4 \nonumber \\  &\quad  -2.84613 \Theta _2{}^2+1.49955 \Theta _2 \Theta _3+1.28512 \Theta _2 \Theta _4-0.482776 \Theta _3{}^2 \nonumber \\  &\quad -1.68827 \Theta _3 \Theta _4-0.472685 \Theta _4{}^2+1\Bigr). 
\end{align}
All soft terms are subject to the constraint \eqref{constraint}.
\FloatBarrier

\subsection{Model 15-dual}\label{sec:model-15.5}
\begin{figure}[htb]
\centering
\includegraphics[width=\textwidth]{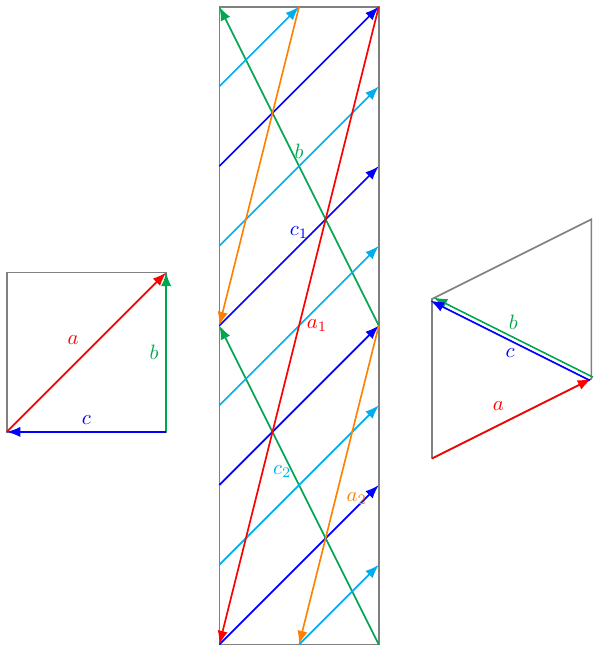}
\caption{Brane configuration for the three two-tori in Model~\hyperref[model15.5]{15-dual} where the third two-torus is tilted. Fermion mass hierarchies result from the intersections on the second two-torus.}  \label{Fig.15.5}
\end{figure}

In Model~\hyperref[model15.5]{15-dual} the three-point Yukawa couplings arise from the triplet intersections from the branes $a$, $b$ and $c$ on the second two-torus with 6 pairs of Higgs from the $\mathcal{N}=2$ sector.
Yukawa matrices for the Model~\hyperref[model15.5]{15-dual} are of rank 3 and the three intersections required to form the disk diagrams for the Yukawa couplings all occur on the second torus as shown in figure~\ref{Fig.15.5}.  
\begin{table}[htb]\footnotesize\centering
\renewcommand{\arraystretch}{1.8}
$\begin{array}{|c|c|c|c|}\hline
 \text{\hyperref[model15.5]{15-dual}} & \theta ^1 & \theta ^2 & \theta ^3 \\\hline
 a & \tan ^{-1}\left(\frac{5}{\sqrt{2}}\right) & \frac{\pi }{2} & 0 \\\hline
 b & \tan ^{-1}\left(2 \sqrt{2}\right) & -\tan ^{-1}\left(\sqrt{2}\right) & \tan ^{-1}\left(\frac{1}{\sqrt{2}}\right) \\\hline
 c & \tan ^{-1}\left(\frac{1}{\sqrt{2}}\right) & -\tan ^{-1}\left(\frac{1}{\sqrt{2}}\right) & -\tan ^{-1}\left(\frac{1}{\sqrt{2}}\right) \\\hline
\end{array}$
\caption{The angles with respect to the orientifold plane made by the cycle wrapped by stack of D6-branes on each of the three two-tori in Model~\hyperref[model15.5]{15-dual}.}
\label{Angles15.5}
\end{table}

The complex structure moduli $U^i$ \eqref{U-moduli} are given as,
\begin{align}\label{eqn:U-moduli15.5}
\{U^1,U^2,U^3\} &=\left\{\frac{5 i}{\sqrt{2}},2 i \sqrt{2},\frac{2}{3} \left(1+i \sqrt{2}\right)\right\},
\end{align}
and the corresponding $u$-moduli and $s$-modulus in supergravity basis from \eqref{eq:moduli} are,
\begin{align}\label{s-u-moduli15.5}
\{u^1,u^2,u^3\} & = \left\{\frac{\sqrt[4]{2} e^{-\phi _4}}{\sqrt{5} \pi },\frac{\sqrt{5} e^{-\phi _4}}{2\ 2^{3/4} \pi },\frac{\sqrt{5} e^{-\phi _4}}{2^{3/4} \pi }\right\}, \nonumber\\
s & = \frac{e^{-\phi _4}}{2\ 2^{3/4} \sqrt{5} \pi } .
\end{align}
Using \eqref{kingauagefun} and the values from the table~\ref{model15.5}, the gauge kinetic function becomes,
\begin{align}\label{fx15.5}
\{f_a,f_b,f_c\} & = \left\{\frac{27 e^{-\phi _4}}{16\ 2^{3/4} \sqrt{5} \pi },\frac{3 \sqrt{5} e^{-\phi _4}}{8\ 2^{3/4} \pi },\frac{3 e^{-\phi _4}}{4\ 2^{3/4} \sqrt{5} \pi }\right\},
\end{align}
To calculate the gaugino masses $\{M_Y,M_b,M_a\}$ for the respective gauge groups $\U(1)_Y$, $\SU(2)_L$, and $\SU(3)_C$, we first compute $\{M_a,M_b,M_c\}$ using \eqref{gaugino-masses} as,
\begin{align}\label{Gauginosabc15.5}
M_a &= \frac{m_{3/2} (4 \Theta _1+5 \Theta _2+20 \Theta _3+2 \Theta _4)}{9 \sqrt{3}},\nonumber \\
M_b &= \frac{m_{3/2} (\Theta _2+2 \Theta _3)}{\sqrt{3}},\nonumber \\
M_c &= \frac{m_{3/2} (\Theta _1-2 \Theta _4)}{\sqrt{3}}.
\end{align} 

Next, to compute the trilinear coupling and the sleptons mass-squared we require the angles, the differences of angles and their first and second order derivatives with respect to the moduli. In table~\ref{Angles15.5} we show the angles \eqref{eq:angle} made by the cycles wrapped by each stack of D6-branes with respect to the orientifold plane on each two-torus.
The differences of the angles, $\theta_{xy}^{i}= \theta_{y}^{i} -\theta_{x}^{i}$ are,
\begin{align}\label{angle-diff15.5}\arraycolsep=0pt
\hskip -1em \left[
\begin{array}{ccc}
 \{0.,0.,0.\} & \{0.275643,0.0969093,-0.0893668\} & \{-0.153561,0.526113,-0.0893668\} \\
 \{-0.275643,-0.0969093,0.0893668\} & \{0.,0.,0.\} & \{-0.429204,0.429204,0.\} \\
 \{0.153561,-0.526113,0.0893668\} & \{0.429204,-0.429204,0.\} & \{0.,0.,0.\} \\
\end{array}
\right]
\end{align}

To account for the negative angle differences we employ the sign function $\sigma_{xy}^{i}$, which is $-1$ only for negative angle difference and $+1$ otherwise,
\begin{align}\label{sigmaK15.5}
\sigma_{xy}^{i} & = \left(
\begin{array}{ccc}
 \{1,1,1\} & \{1,1,-1\} & \{-1,1,-1\} \\
 \{-1,-1,1\} & \{1,1,1\} & \{-1,1,1\} \\
 \{1,-1,1\} & \{1,-1,1\} & \{1,1,1\} \\
\end{array}
\right),
\end{align}
and the function $\eta_{xy}$ is evaluated by taking the product on the torus index $i$ as,
\begin{equation}\label{eta15.5}
\eta_{xy} = \left(
\begin{array}{ccc}
 1 & -1 & 1 \\
 1 & 1 & -1 \\
 -1 & -1 & 1 \\
\end{array}
\right).
\end{equation}
Using the values of $\sigma_{xy}^{i}$ and $\eta_{xy}$ in \eqref{Psi} we can compute the four cases of functions $\Psi(\theta_{xy})$ defined in \eqref{eqn:Psi1} and \eqref{eqn:Psi2}. Similarly we calculate the derivative $\Psi'(\theta^j_{xy}) = \frac{d\Psi(\theta^j_{xy})}{d \theta^j_{xy}}$ using equations \eqref{DPsi}, \eqref{derivative-angles1}, \eqref{derivative-angles2} and the properties of digamma function $\psi^{(0)}(z)$ \eqref{eq:property} while neglecting the contribution of the $t$-moduli. 

Utilizing above results while ignoring the CP-violating phases $\gamma_m$, the gaugino masses; the trilinear coupling \eqref{tri-coupling}; and the squared-masses of squarks and sleptons \eqref{slepton-mass} are obtained as,
\begin{align}\label{GauginosYba123-model15.5}  
M_{\tilde B} &\equiv M_Y = m_{3/2}\Bigl(\frac{2 \Theta _1+\Theta _2+4 \Theta _3-2 \Theta _4}{3 \sqrt{3}}\Bigr), \nonumber\\
M_{\tilde W} &\equiv M_b = m_{3/2}\Bigl(\frac{\Theta _2+2 \Theta _3}{\sqrt{3}}\Bigr),\nonumber \\
M_{\tilde g} &\equiv M_a = m_{3/2}\Bigl(\frac{4 \Theta _1+5 \Theta _2+20 \Theta _3+2 \Theta _4}{9 \sqrt{3}}\Bigr),\nonumber \\
A_0 \equiv A_{abc} &= m_{3/2}\Bigl(-0.0762806 \Theta _1-1.65577 \Theta _2+0.222917 \Theta _3-0.222917 \Theta _4\Bigr),\nonumber\\  
m^2_{L} \equiv m^2_{ab} &= m_{3/2}^2\Bigl(-2.50015 \Theta _1{}^2-1.28868 \Theta _1 \Theta _2+0.56326 \Theta _1 \Theta _3+1.78275 \Theta _1 \Theta _4 \nonumber \\   &\quad -0.272045 \Theta _2{}^2+1.18291 \Theta _2 \Theta _3+0.336724 \Theta _2 \Theta _4+0.10754 \Theta _3{}^2 \nonumber \\  &\quad -2.1859 \Theta _3 \Theta _4-0.202498 \Theta _4{}^2+1\Bigr) ,\nonumber\\  
m^2_{R} \equiv m^2_{ac} &= m_{3/2}^2\Bigl(-0.229829 \Theta _1{}^2+1.01289 \Theta _1 \Theta _2-0.749437 \Theta _1 \Theta _3-1.33965 \Theta _1 \Theta _4 \nonumber \\  &\quad  +0.277823 \Theta _2{}^2-0.150917 \Theta _2 \Theta _3-0.779826 \Theta _2 \Theta _4-1.85574 \Theta _3{}^2 \nonumber \\  &\quad +1.9101 \Theta _3 \Theta _4-0.7735 \Theta _4{}^2+1\Bigr). 
\end{align}
All soft terms are subject to the constraint \eqref{constraint}.
\FloatBarrier

\subsection{Model 16}\label{sec:model-16}
\begin{figure}[htb]
\centering
\includegraphics[width=\textwidth]{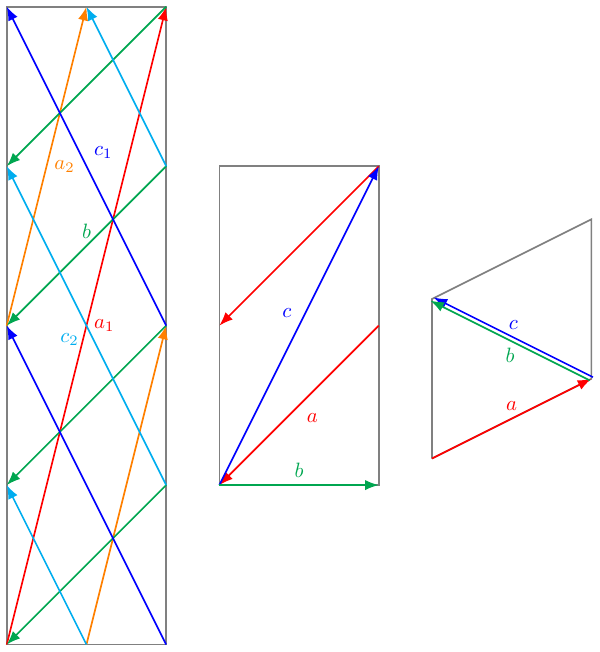}
\caption{Brane configuration for the three two-tori in Model~\hyperref[model16]{16} where the third two-torus is tilted. Fermion mass hierarchies result from the intersections on the first two-torus.}  \label{Fig.16}
\end{figure}

In Model~\hyperref[model16]{16} the three-point Yukawa couplings arise from the triplet intersections from the branes $a$, $b$ and $c$ on the first two-torus with 6 pairs of Higgs from the $\mathcal{N}=2$ sector.
Yukawa matrices for the Model~\hyperref[model16]{16} are of rank 3 and the three intersections required to form the disk diagrams for the Yukawa couplings all occur on the first torus as shown in figure~\ref{Fig.16}.  
\begin{table}[htb]\footnotesize\centering
\renewcommand{\arraystretch}{1.8}
$\begin{array}{|c|c|c|c|}\hline
 \text{\hyperref[model16]{16}} & \theta ^1 & \theta ^2 & \theta ^3 \\\hline
 a & \tan ^{-1}\left(\sqrt{\frac{13}{2}}\right) & \tan ^{-1}\left(\frac{\sqrt{\frac{13}{2}}}{4}\right) & -\tan ^{-1}\left(\frac{\sqrt{\frac{13}{2}}}{2}\right) \\\hline
 b & \tan ^{-1}\left(\sqrt{26}\right) & 0 & \tan ^{-1}\left(2 \sqrt{26}\right) \\\hline
 c & \tan ^{-1}\left(\frac{\sqrt{\frac{13}{2}}}{4}\right) & -\tan ^{-1}\left(\frac{\sqrt{\frac{13}{2}}}{4}\right) & -\tan ^{-1}\left(\frac{\sqrt{\frac{13}{2}}}{4}\right) \\\hline
\end{array}$
\caption{The angles with respect to the orientifold plane made by the cycle wrapped by stack of D6-branes on each of the three two-tori in Model~\hyperref[model16]{16}.}
\label{Angles16}
\end{table}

The complex structure moduli $U^i$ \eqref{U-moduli} are given as,
\begin{align}\label{eqn:U-moduli16}
\{U^1,U^2,U^3\} &=\left\{i \sqrt{\frac{13}{2}},2 i \sqrt{26},\frac{2}{45} \left(13+4 i \sqrt{26}\right)\right\},
\end{align}
and the corresponding $u$-moduli and $s$-modulus in supergravity basis from \eqref{eq:moduli} are,
\begin{align}\label{s-u-moduli16}
\{u^1,u^2,u^3\} & = \left\{\frac{\sqrt[4]{13} e^{-\phi _4}}{2^{3/4} \pi },\frac{\sqrt[4]{13} e^{-\phi _4}}{4\ 2^{3/4} \pi },\frac{\sqrt[4]{26} e^{-\phi _4}}{\pi }\right\}, \nonumber\\
s & = \frac{e^{-\phi _4}}{26^{3/4} \pi } .
\end{align}
Using \eqref{kingauagefun} and the values from the table~\ref{model16}, the gauge kinetic function becomes,
\begin{align}\label{fx16}
\{f_a,f_b,f_c\} & = \left\{\frac{135 e^{-\phi _4}}{16\ 26^{3/4} \pi },\frac{45 e^{-\phi _4}}{32\ 26^{3/4} \pi },\frac{315 e^{-\phi _4}}{32\ 26^{3/4} \pi }\right\},
\end{align}
To calculate the gaugino masses $\{M_Y,M_b,M_a\}$ for the respective gauge groups $\U(1)_Y$, $\SU(2)_L$, and $\SU(3)_C$, we first compute $\{M_a,M_b,M_c\}$ using \eqref{gaugino-masses} as,
\begin{align}\label{Gauginosabc16}
M_a &= \frac{m_{3/2} (26 \Theta _1+13 \Theta _2+8 (13 \Theta _3+\Theta _4))}{45 \sqrt{3}},\nonumber \\
M_b &= \frac{m_{3/2} (13 \Theta _2-32 \Theta _4)}{15 \sqrt{3}},\nonumber \\
M_c &= \frac{m_{3/2} (104 \Theta _1-13 \Theta _2+208 \Theta _3-16 \Theta _4)}{105 \sqrt{3}}.
\end{align} 

Next, to compute the trilinear coupling and the sleptons mass-squared we require the angles, the differences of angles and their first and second order derivatives with respect to the moduli. In table~\ref{Angles16} we show the angles \eqref{eq:angle} made by the cycles wrapped by each stack of D6-branes with respect to the orientifold plane on each two-torus.
The differences of the angles, $\theta_{xy}^{i}= \theta_{y}^{i} -\theta_{x}^{i}$ are,
\begin{align}\label{angle-diff16}\arraycolsep=0pt
\hskip -1em \left[
\begin{array}{ccc}
 \{0.,0.,0.\} & \{0.228854,-0.235545,0.0066917\} & \{0.0389882,0.237505,0.0066917\} \\
 \{-0.228854,0.235545,-0.0066917\} & \{0.,0.,0.\} & \{-0.189865,0.473051,0.\} \\
 \{-0.0389882,-0.237505,-0.0066917\} & \{0.189865,-0.473051,0.\} & \{0.,0.,0.\} \\
\end{array}
\right]
\end{align}

To account for the negative angle differences we employ the sign function $\sigma_{xy}^{i}$, which is $-1$ only for negative angle difference and $+1$ otherwise,
\begin{align}\label{sigmaK16}
\sigma_{xy}^{i} & = \left(
\begin{array}{ccc}
 \{1,1,1\} & \{1,-1,1\} & \{1,1,1\} \\
 \{-1,1,-1\} & \{1,1,1\} & \{-1,1,1\} \\
 \{-1,-1,-1\} & \{1,-1,1\} & \{1,1,1\} \\
\end{array}
\right),
\end{align}
and the function $\eta_{xy}$ is evaluated by taking the product on the torus index $i$ as,
\begin{equation}\label{eta16}
\eta_{xy} = \left(
\begin{array}{ccc}
 1 & -1 & 1 \\
 1 & 1 & -1 \\
 -1 & -1 & 1 \\
\end{array}
\right).
\end{equation}
Using the values of $\sigma_{xy}^{i}$ and $\eta_{xy}$ in \eqref{Psi} we can compute the four cases of functions $\Psi(\theta_{xy})$ defined in \eqref{eqn:Psi1} and \eqref{eqn:Psi2}. Similarly we calculate the derivative $\Psi'(\theta^j_{xy}) = \frac{d\Psi(\theta^j_{xy})}{d \theta^j_{xy}}$ using equations \eqref{DPsi}, \eqref{derivative-angles1}, \eqref{derivative-angles2} and the properties of digamma function $\psi^{(0)}(z)$ \eqref{eq:property} while neglecting the contribution of the $t$-moduli. 

Utilizing above results while ignoring the CP-violating phases $\gamma_m$, the gaugino masses; the trilinear coupling \eqref{tri-coupling}; and the squared-masses of squarks and sleptons \eqref{slepton-mass} are obtained as,
\begin{align}\label{GauginosYba123-model16}  
M_{\tilde B} &\equiv M_Y = m_{3/2}\Bigl(\frac{416 \Theta _1+13 \Theta _2+1040 \Theta _3-16 \Theta _4}{495 \sqrt{3}}\Bigr), \nonumber\\
M_{\tilde W} &\equiv M_b = m_{3/2}\Bigl(\frac{13 \Theta _2-32 \Theta _4}{15 \sqrt{3}}\Bigr),\nonumber \\
M_{\tilde g} &\equiv M_a = m_{3/2}\Bigl(\frac{26 \Theta _1+13 \Theta _2+8 (13 \Theta _3+\Theta _4)}{45 \sqrt{3}}\Bigr),\nonumber \\
A_0 \equiv A_{abc} &= m_{3/2}\Bigl(-0.674714 \Theta _1-1.05734 \Theta _2-0.296892 \Theta _3+0.296892 \Theta _4\Bigr),\nonumber\\  
m^2_{L} \equiv m^2_{ab} &= m_{3/2}^2\Bigl(0.357659 \Theta _1{}^2+0.87063 \Theta _1 \Theta _2-1.16965 \Theta _1 \Theta _3-0.902287 \Theta _1 \Theta _4 \nonumber \\   &\quad -0.152099 \Theta _2{}^2-0.139037 \Theta _2 \Theta _3-1.26422 \Theta _2 \Theta _4-0.463359 \Theta _3{}^2 \nonumber \\  &\quad +1.81488 \Theta _3 \Theta _4-2.03894 \Theta _4{}^2+1\Bigr) ,\nonumber\\  
m^2_{R} \equiv m^2_{ac} &= m_{3/2}^2\Bigl(-2.30496 \Theta _1{}^2-0.638749 \Theta _1 \Theta _2+0.0264302 \Theta _1 \Theta _3+1.87359 \Theta _1 \Theta _4 \nonumber \\  &\quad  -1.13248 \Theta _2{}^2+0.518024 \Theta _2 \Theta _3+0.292916 \Theta _2 \Theta _4-0.63835 \Theta _3{}^2 \nonumber \\  &\quad -1.58299 \Theta _3 \Theta _4-0.0912327 \Theta _4{}^2+1\Bigr). 
\end{align}
All soft terms are subject to the constraint \eqref{constraint}.
\FloatBarrier

\subsection{Model 16-dual}\label{sec:model-16.5}
\begin{figure}[htb]
\centering
\includegraphics[width=\textwidth]{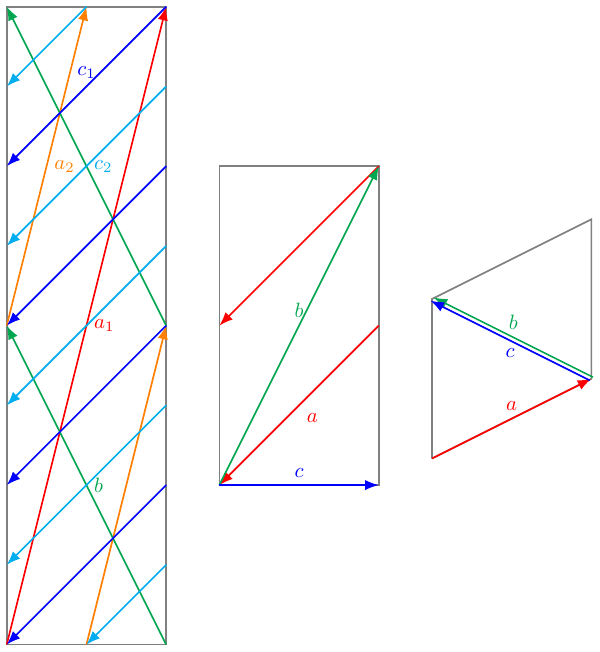}
\caption{Brane configuration for the three two-tori in Model~\hyperref[model16.5]{16-dual} where the third two-torus is tilted. Fermion mass hierarchies result from the intersections on the first two-torus.}  \label{Fig.16.5}
\end{figure}

In Model~\hyperref[model16.5]{16-dual} the three-point Yukawa couplings arise from the triplet intersections from the branes $a$, $b$ and $c$ on the first two-torus with 6 pairs of Higgs from the $\mathcal{N}=2$ sector.
Yukawa matrices for the Model~\hyperref[model16.5]{16-dual} are of rank 3 and the three intersections required to form the disk diagrams for the Yukawa couplings all occur on the first torus as shown in figure~\ref{Fig.16.5}.  
\begin{table}[htb]\footnotesize\centering
\renewcommand{\arraystretch}{1.8}
$\begin{array}{|c|c|c|c|}\hline
 \text{\hyperref[model16.5]{16-dual}} & \theta ^1 & \theta ^2 & \theta ^3 \\\hline
 a & \tan ^{-1}\left(\sqrt{\frac{13}{2}}\right) & -\tan ^{-1}\left(\frac{\sqrt{\frac{13}{2}}}{2}\right) & \tan ^{-1}\left(\frac{\sqrt{\frac{13}{2}}}{4}\right) \\\hline
 b & \tan ^{-1}\left(\sqrt{26}\right) & \tan ^{-1}\left(2 \sqrt{26}\right) & 0 \\\hline
 c & \tan ^{-1}\left(\frac{\sqrt{\frac{13}{2}}}{4}\right) & -\tan ^{-1}\left(\frac{\sqrt{\frac{13}{2}}}{4}\right) & -\tan ^{-1}\left(\frac{\sqrt{\frac{13}{2}}}{4}\right) \\\hline
\end{array}$
\caption{The angles with respect to the orientifold plane made by the cycle wrapped by stack of D6-branes on each of the three two-tori in Model~\hyperref[model16.5]{16-dual}.}
\label{Angles16.5}
\end{table}

The complex structure moduli $U^i$ \eqref{U-moduli} are given as,
\begin{align}\label{eqn:U-moduli16.5}
\{U^1,U^2,U^3\} &=\left\{i \sqrt{\frac{13}{2}},2 i \sqrt{26},\frac{2}{45} \left(13+4 i \sqrt{26}\right)\right\},
\end{align}
and the corresponding $u$-moduli and $s$-modulus in supergravity basis from \eqref{eq:moduli} are,
\begin{align}\label{s-u-moduli16.5}
\{u^1,u^2,u^3\} & = \left\{\frac{\sqrt[4]{13} e^{-\phi _4}}{2^{3/4} \pi },\frac{\sqrt[4]{13} e^{-\phi _4}}{4\ 2^{3/4} \pi },\frac{\sqrt[4]{26} e^{-\phi _4}}{\pi }\right\}, \nonumber\\
s & = \frac{e^{-\phi _4}}{26^{3/4} \pi } .
\end{align}
Using \eqref{kingauagefun} and the values from the table~\ref{model16.5}, the gauge kinetic function becomes,
\begin{align}\label{fx16.5}
\{f_a,f_b,f_c\} & = \left\{\frac{135 e^{-\phi _4}}{16\ 26^{3/4} \pi },\frac{315 e^{-\phi _4}}{32\ 26^{3/4} \pi },\frac{45 e^{-\phi _4}}{32\ 26^{3/4} \pi }\right\},
\end{align}
To calculate the gaugino masses $\{M_Y,M_b,M_a\}$ for the respective gauge groups $\U(1)_Y$, $\SU(2)_L$, and $\SU(3)_C$, we first compute $\{M_a,M_b,M_c\}$ using \eqref{gaugino-masses} as,
\begin{align}\label{Gauginosabc16.5}
M_a &= \frac{m_{3/2} (26 \Theta _1+13 \Theta _2+8 (13 \Theta _3+\Theta _4))}{45 \sqrt{3}},\nonumber \\
M_b &= \frac{m_{3/2} (104 \Theta _1-13 \Theta _2+208 \Theta _3-16 \Theta _4)}{105 \sqrt{3}},\nonumber \\
M_c &= \frac{m_{3/2} (13 \Theta _2-32 \Theta _4)}{15 \sqrt{3}}.
\end{align} 

Next, to compute the trilinear coupling and the sleptons mass-squared we require the angles, the differences of angles and their first and second order derivatives with respect to the moduli. In table~\ref{Angles16.5} we show the angles \eqref{eq:angle} made by the cycles wrapped by each stack of D6-branes with respect to the orientifold plane on each two-torus.
The differences of the angles, $\theta_{xy}^{i}= \theta_{y}^{i} -\theta_{x}^{i}$ are,
\begin{align}\label{angle-diff16.5}\arraycolsep=0pt
\hskip -1em \left[
\begin{array}{ccc}
 \{0.,0.,0.\} & \{0.0389882,0.237505,0.0066917\} & \{0.228854,-0.235545,0.0066917\} \\
 \{-0.0389882,-0.237505,-0.0066917\} & \{0.,0.,0.\} & \{0.189865,-0.473051,0.\} \\
 \{-0.228854,0.235545,-0.0066917\} & \{-0.189865,0.473051,0.\} & \{0.,0.,0.\} \\
\end{array}
\right]
\end{align}

To account for the negative angle differences we employ the sign function $\sigma_{xy}^{i}$, which is $-1$ only for negative angle difference and $+1$ otherwise,
\begin{align}\label{sigmaK16.5}
\sigma_{xy}^{i} & = \left(
\begin{array}{ccc}
 \{1,1,1\} & \{1,1,1\} & \{1,-1,1\} \\
 \{-1,-1,-1\} & \{1,1,1\} & \{1,-1,1\} \\
 \{-1,1,-1\} & \{-1,1,1\} & \{1,1,1\} \\
\end{array}
\right),
\end{align}
and the function $\eta_{xy}$ is evaluated by taking the product on the torus index $i$ as,
\begin{equation}\label{eta16.5}
\eta_{xy} = \left(
\begin{array}{ccc}
 1 & 1 & -1 \\
 -1 & 1 & -1 \\
 1 & -1 & 1 \\
\end{array}
\right).
\end{equation}
Using the values of $\sigma_{xy}^{i}$ and $\eta_{xy}$ in \eqref{Psi} we can compute the four cases of functions $\Psi(\theta_{xy})$ defined in \eqref{eqn:Psi1} and \eqref{eqn:Psi2}. Similarly we calculate the derivative $\Psi'(\theta^j_{xy}) = \frac{d\Psi(\theta^j_{xy})}{d \theta^j_{xy}}$ using equations \eqref{DPsi}, \eqref{derivative-angles1}, \eqref{derivative-angles2} and the properties of digamma function $\psi^{(0)}(z)$ \eqref{eq:property} while neglecting the contribution of the $t$-moduli. 

Utilizing above results while ignoring the CP-violating phases $\gamma_m$, the gaugino masses; the trilinear coupling \eqref{tri-coupling}; and the squared-masses of squarks and sleptons \eqref{slepton-mass} are obtained as,
\begin{align}\label{GauginosYba123-model16.5}  
M_{\tilde B} &\equiv M_Y = m_{3/2}\Bigl(\frac{104 \Theta _1+91 \Theta _2+416 \Theta _3-64 \Theta _4}{225 \sqrt{3}}\Bigr), \nonumber\\
M_{\tilde W} &\equiv M_b = m_{3/2}\Bigl(\frac{104 \Theta _1-13 \Theta _2+208 \Theta _3-16 \Theta _4}{105 \sqrt{3}}\Bigr),\nonumber \\
M_{\tilde g} &\equiv M_a = m_{3/2}\Bigl(\frac{26 \Theta _1+13 \Theta _2+8 (13 \Theta _3+\Theta _4)}{45 \sqrt{3}}\Bigr),\nonumber \\
A_0 \equiv A_{abc} &= m_{3/2}\Bigl(-0.674714 \Theta _1-1.05734 \Theta _2-0.296892 \Theta _3+0.296892 \Theta _4\Bigr),\nonumber\\  
m^2_{L} \equiv m^2_{ab} &= m_{3/2}^2\Bigl(-2.30496 \Theta _1{}^2-0.638749 \Theta _1 \Theta _2+0.0264302 \Theta _1 \Theta _3+1.87359 \Theta _1 \Theta _4 \nonumber \\   &\quad -1.13248 \Theta _2{}^2+0.518024 \Theta _2 \Theta _3+0.292916 \Theta _2 \Theta _4-0.63835 \Theta _3{}^2 \nonumber \\  &\quad -1.58299 \Theta _3 \Theta _4-0.0912327 \Theta _4{}^2+1\Bigr) ,\nonumber\\  
m^2_{R} \equiv m^2_{ac} &= m_{3/2}^2\Bigl(0.357659 \Theta _1{}^2+0.87063 \Theta _1 \Theta _2-1.16965 \Theta _1 \Theta _3-0.902287 \Theta _1 \Theta _4 \nonumber \\  &\quad  -0.152099 \Theta _2{}^2-0.139037 \Theta _2 \Theta _3-1.26422 \Theta _2 \Theta _4-0.463359 \Theta _3{}^2 \nonumber \\  &\quad +1.81488 \Theta _3 \Theta _4-2.03894 \Theta _4{}^2+1\Bigr). 
\end{align}
All soft terms are subject to the constraint \eqref{constraint}.
\FloatBarrier

\subsection{Model 17}\label{sec:model-17}
\begin{figure}[htb]
\centering
\includegraphics[width=\textwidth]{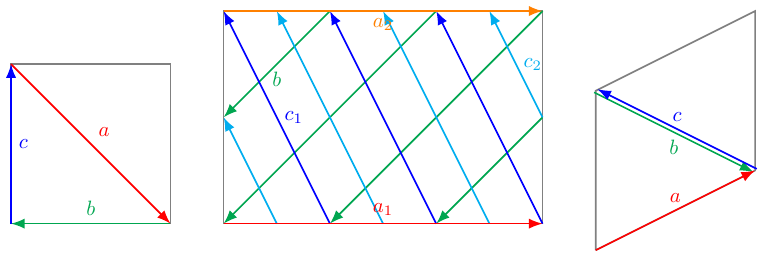}
\caption{Brane configuration for the three two-tori in Model~\hyperref[model17]{17} where the third two-torus is tilted. Fermion mass hierarchies result from the intersections on the second two-torus.}  \label{Fig.17}
\end{figure}

In Model~\hyperref[model17]{17} the three-point Yukawa couplings arise from the triplet intersections from the branes $a$, $b$ and $c$ on the second two-torus with 9 pairs of Higgs from the $\mathcal{N}=2$ sector.
Yukawa matrices for the Model~\hyperref[model17]{17} are of rank 3 and the three intersections required to form the disk diagrams for the Yukawa couplings all occur on the second torus as shown in figure~\ref{Fig.17}.  
\begin{table}[htb]\footnotesize\centering
\renewcommand{\arraystretch}{1.8}
$\begin{array}{|c|c|c|c|}\hline
 \text{\hyperref[model17]{17}} & \theta ^1 & \theta ^2 & \theta ^3 \\\hline
 a & -\tan ^{-1}\left(\frac{1}{\sqrt{2}}\right) & 0 & \frac{\pi }{2} \\\hline
 b & 0 & \tan ^{-1}\left(\frac{1}{\sqrt{2}}\right) & -\tan ^{-1}\left(\sqrt{2}\right) \\\hline
 c & \tan ^{-1}\left(\frac{1}{\sqrt{2}}\right) & -\tan ^{-1}\left(\frac{1}{\sqrt{2}}\right) & -\tan ^{-1}\left(\frac{1}{\sqrt{2}}\right) \\\hline
\end{array}$
\caption{The angles with respect to the orientifold plane made by the cycle wrapped by stack of D6-branes on each of the three two-tori in Model~\hyperref[model17]{17}.}
\label{Angles17}
\end{table}

The complex structure moduli $U^i$ \eqref{U-moduli} are given as,
\begin{align}\label{eqn:U-moduli17}
\{U^1,U^2,U^3\} &=\left\{\frac{i}{\sqrt{2}},\frac{i \sqrt{2}}{3},\frac{2}{3} \left(1+i \sqrt{2}\right)\right\},
\end{align}
and the corresponding $u$-moduli and $s$-modulus in supergravity basis from \eqref{eq:moduli} are,
\begin{align}\label{s-u-moduli17}
\{u^1,u^2,u^3\} & = \left\{\frac{e^{-\phi _4}}{\sqrt[4]{2} \sqrt{3} \pi },\frac{\sqrt{3} e^{-\phi _4}}{2 \sqrt[4]{2} \pi },\frac{e^{-\phi _4}}{2 \sqrt[4]{2} \sqrt{3} \pi }\right\}, \nonumber\\
s & = \frac{\sqrt{3} e^{-\phi _4}}{2 \sqrt[4]{2} \pi } .
\end{align}
Using \eqref{kingauagefun} and the values from the table~\ref{model17}, the gauge kinetic function becomes,
\begin{align}\label{fx17}
\{f_a,f_b,f_c\} & = \left\{\frac{3 \sqrt{3} e^{-\phi _4}}{16 \sqrt[4]{2} \pi },\frac{3 \sqrt{3} e^{-\phi _4}}{8 \sqrt[4]{2} \pi },\frac{3 \sqrt{3} e^{-\phi _4}}{16 \sqrt[4]{2} \pi }\right\},
\end{align}
To calculate the gaugino masses $\{M_Y,M_b,M_a\}$ for the respective gauge groups $\U(1)_Y$, $\SU(2)_L$, and $\SU(3)_C$, we first compute $\{M_a,M_b,M_c\}$ using \eqref{gaugino-masses} as,
\begin{align}\label{Gauginosabc17}
M_a &= \frac{m_{3/2} (\Theta _2-2 \Theta _4)}{\sqrt{3}},\nonumber \\
M_b &= \frac{m_{3/2} (\Theta _1-2 \Theta _4)}{\sqrt{3}},\nonumber \\
M_c &= \frac{m_{3/2} (\Theta _2+2 \Theta _3)}{\sqrt{3}}.
\end{align} 

Next, to compute the trilinear coupling and the sleptons mass-squared we require the angles, the differences of angles and their first and second order derivatives with respect to the moduli. In table~\ref{Angles17} we show the angles \eqref{eq:angle} made by the cycles wrapped by each stack of D6-branes with respect to the orientifold plane on each two-torus.
The differences of the angles, $\theta_{xy}^{i}= \theta_{y}^{i} -\theta_{x}^{i}$ are,
\begin{align}\label{angle-diff17}\arraycolsep=0pt
\hskip -1em \left[
\begin{array}{ccc}
 \{0.,0.,0.\} & \{-0.242928,0.473887,-0.230959\} & \{0.186276,0.186276,-0.0893668\} \\
 \{0.242928,-0.473887,0.230959\} & \{0.,0.,0.\} & \{0.429204,-0.287611,0.141593\} \\
 \{-0.186276,-0.186276,0.0893668\} & \{-0.429204,0.287611,-0.141593\} & \{0.,0.,0.\} \\
\end{array}
\right]
\end{align}

To account for the negative angle differences we employ the sign function $\sigma_{xy}^{i}$, which is $-1$ only for negative angle difference and $+1$ otherwise,
\begin{align}\label{sigmaK17}
\sigma_{xy}^{i} & = \left(
\begin{array}{ccc}
 \{1,1,1\} & \{-1,1,-1\} & \{1,1,-1\} \\
 \{1,-1,1\} & \{1,1,1\} & \{1,-1,1\} \\
 \{-1,-1,1\} & \{-1,1,-1\} & \{1,1,1\} \\
\end{array}
\right),
\end{align}
and the function $\eta_{xy}$ is evaluated by taking the product on the torus index $i$ as,
\begin{equation}\label{eta17}
\eta_{xy} = \left(
\begin{array}{ccc}
 1 & 1 & -1 \\
 -1 & 1 & -1 \\
 1 & 1 & 1 \\
\end{array}
\right).
\end{equation}
Using the values of $\sigma_{xy}^{i}$ and $\eta_{xy}$ in \eqref{Psi} we can compute the four cases of functions $\Psi(\theta_{xy})$ defined in \eqref{eqn:Psi1} and \eqref{eqn:Psi2}. Similarly we calculate the derivative $\Psi'(\theta^j_{xy}) = \frac{d\Psi(\theta^j_{xy})}{d \theta^j_{xy}}$ using equations \eqref{DPsi}, \eqref{derivative-angles1}, \eqref{derivative-angles2} and the properties of digamma function $\psi^{(0)}(z)$ \eqref{eq:property} while neglecting the contribution of the $t$-moduli. 

Utilizing above results while ignoring the CP-violating phases $\gamma_m$, the gaugino masses; the trilinear coupling \eqref{tri-coupling}; and the squared-masses of squarks and sleptons \eqref{slepton-mass} are obtained as,
\begin{align}\label{GauginosYba123-model17}  
M_{\tilde B} &\equiv M_Y = m_{3/2}\Bigl(\frac{5 \Theta _2+6 \Theta _3-4 \Theta _4}{5 \sqrt{3}}\Bigr), \nonumber\\
M_{\tilde W} &\equiv M_b = m_{3/2}\Bigl(\frac{\Theta _1-2 \Theta _4}{\sqrt{3}}\Bigr),\nonumber \\
M_{\tilde g} &\equiv M_a = m_{3/2}\Bigl(\frac{\Theta _2-2 \Theta _4}{\sqrt{3}}\Bigr),\nonumber \\
A_0 \equiv A_{abc} &= m_{3/2}\Bigl(-0.128593 \Theta _1-1.64721 \Theta _2-0.107781 \Theta _3+0.151538 \Theta _4\Bigr),\nonumber\\  
m^2_{L} \equiv m^2_{ab} &= m_{3/2}^2\Bigl(0.36917 \Theta _1{}^2+0.669259 \Theta _1 \Theta _2-0.840813 \Theta _1 \Theta _3-1.82443 \Theta _1 \Theta _4 \nonumber \\   &\quad +0.375979 \Theta _2{}^2-0.820669 \Theta _2 \Theta _3-0.733709 \Theta _2 \Theta _4-0.929866 \Theta _3{}^2 \nonumber \\  &\quad +1.7274 \Theta _3 \Theta _4-1.1538 \Theta _4{}^2+1\Bigr) ,\nonumber\\  
m^2_{R} \equiv m^2_{ac} &= m_{3/2}^2\Bigl(-2.2278 \Theta _1{}^2-2.22881 \Theta _1 \Theta _2+0.0357856 \Theta _1 \Theta _3+1.61629 \Theta _1 \Theta _4 \nonumber \\  &\quad  +0.395725 \Theta _2{}^2+0.464088 \Theta _2 \Theta _3+0.503187 \Theta _2 \Theta _4+0.535193 \Theta _3{}^2 \nonumber \\  &\quad -3.12602 \Theta _3 \Theta _4+0.309973 \Theta _4{}^2+1\Bigr). 
\end{align}
All soft terms are subject to the constraint \eqref{constraint}.
\FloatBarrier

\subsection{Model 17-dual}\label{sec:model-17.5}
\begin{figure}[htb]
\centering
\includegraphics[width=\textwidth]{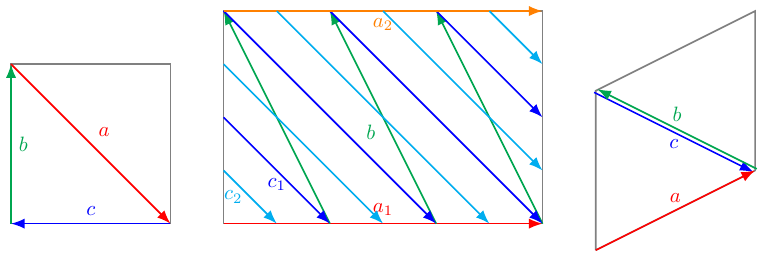}
\caption{Brane configuration for the three two-tori in Model~\hyperref[model17.5]{17-dual} where the third two-torus is tilted. Fermion mass hierarchies result from the intersections on the second two-torus.}  \label{Fig.17.5}
\end{figure}

In Model~\hyperref[model17.5]{17-dual} the three-point Yukawa couplings arise from the triplet intersections from the branes $a$, $b$ and $c$ on the second two-torus with 9 pairs of Higgs from the $\mathcal{N}=2$ sector.
Yukawa matrices for the Model~\hyperref[model17.5]{17-dual} are of rank 3 and the three intersections required to form the disk diagrams for the Yukawa couplings all occur on the second torus as shown in figure~\ref{Fig.17.5}.  
\begin{table}[htb]\footnotesize\centering
\renewcommand{\arraystretch}{1.8}
$\begin{array}{|c|c|c|c|}\hline
 \text{\hyperref[model17.5]{17-dual}} & \theta ^1 & \theta ^2 & \theta ^3 \\\hline
 a & -\tan ^{-1}\left(\frac{1}{\sqrt{2}}\right) & \frac{\pi }{2} & 0 \\\hline
 b & 0 & -\tan ^{-1}\left(\sqrt{2}\right) & \tan ^{-1}\left(\frac{1}{\sqrt{2}}\right) \\\hline
 c & \tan ^{-1}\left(\frac{1}{\sqrt{2}}\right) & -\tan ^{-1}\left(\frac{1}{\sqrt{2}}\right) & -\tan ^{-1}\left(\frac{1}{\sqrt{2}}\right) \\\hline
\end{array}$
\caption{The angles with respect to the orientifold plane made by the cycle wrapped by stack of D6-branes on each of the three two-tori in Model~\hyperref[model17.5]{17-dual}.}
\label{Angles17.5}
\end{table}

The complex structure moduli $U^i$ \eqref{U-moduli} are given as,
\begin{align}\label{eqn:U-moduli17.5}
\{U^1,U^2,U^3\} &=\left\{\frac{i}{\sqrt{2}},\frac{i \sqrt{2}}{3},\frac{2}{3} \left(1+i \sqrt{2}\right)\right\},
\end{align}
and the corresponding $u$-moduli and $s$-modulus in supergravity basis from \eqref{eq:moduli} are,
\begin{align}\label{s-u-moduli17.5}
\{u^1,u^2,u^3\} & = \left\{\frac{e^{-\phi _4}}{\sqrt[4]{2} \sqrt{3} \pi },\frac{\sqrt{3} e^{-\phi _4}}{2 \sqrt[4]{2} \pi },\frac{e^{-\phi _4}}{2 \sqrt[4]{2} \sqrt{3} \pi }\right\}, \nonumber\\
s & = \frac{\sqrt{3} e^{-\phi _4}}{2 \sqrt[4]{2} \pi } .
\end{align}
Using \eqref{kingauagefun} and the values from the table~\ref{model17.5}, the gauge kinetic function becomes,
\begin{align}\label{fx17.5}
\{f_a,f_b,f_c\} & = \left\{\frac{3 \sqrt{3} e^{-\phi _4}}{16 \sqrt[4]{2} \pi },\frac{3 \sqrt{3} e^{-\phi _4}}{16 \sqrt[4]{2} \pi },\frac{3 \sqrt{3} e^{-\phi _4}}{8 \sqrt[4]{2} \pi }\right\},
\end{align}
To calculate the gaugino masses $\{M_Y,M_b,M_a\}$ for the respective gauge groups $\U(1)_Y$, $\SU(2)_L$, and $\SU(3)_C$, we first compute $\{M_a,M_b,M_c\}$ using \eqref{gaugino-masses} as,
\begin{align}\label{Gauginosabc17.5}
M_a &= \frac{m_{3/2} (\Theta _2-2 \Theta _4)}{\sqrt{3}},\nonumber \\
M_b &= \frac{m_{3/2} (\Theta _2+2 \Theta _3)}{\sqrt{3}},\nonumber \\
M_c &= \frac{m_{3/2} (\Theta _1-2 \Theta _4)}{\sqrt{3}}.
\end{align} 

Next, to compute the trilinear coupling and the sleptons mass-squared we require the angles, the differences of angles and their first and second order derivatives with respect to the moduli. In table~\ref{Angles17.5} we show the angles \eqref{eq:angle} made by the cycles wrapped by each stack of D6-branes with respect to the orientifold plane on each two-torus.
The differences of the angles, $\theta_{xy}^{i}= \theta_{y}^{i} -\theta_{x}^{i}$ are,
\begin{align}\label{angle-diff17.5}\arraycolsep=0pt
\hskip -1em \left[
\begin{array}{ccc}
 \{0.,0.,0.\} & \{0.186276,0.186276,-0.0893668\} & \{-0.242928,0.473887,-0.230959\} \\
 \{-0.186276,-0.186276,0.0893668\} & \{0.,0.,0.\} & \{-0.429204,0.287611,-0.141593\} \\
 \{0.242928,-0.473887,0.230959\} & \{0.429204,-0.287611,0.141593\} & \{0.,0.,0.\} \\
\end{array}
\right]
\end{align}

To account for the negative angle differences we employ the sign function $\sigma_{xy}^{i}$, which is $-1$ only for negative angle difference and $+1$ otherwise,
\begin{align}\label{sigmaK17.5}
\sigma_{xy}^{i} & = \left(
\begin{array}{ccc}
 \{1,1,1\} & \{1,1,-1\} & \{-1,1,-1\} \\
 \{-1,-1,1\} & \{1,1,1\} & \{-1,1,-1\} \\
 \{1,-1,1\} & \{1,-1,1\} & \{1,1,1\} \\
\end{array}
\right),
\end{align}
and the function $\eta_{xy}$ is evaluated by taking the product on the torus index $i$ as,
\begin{equation}\label{eta17.5}
\eta_{xy} = \left(
\begin{array}{ccc}
 1 & -1 & 1 \\
 1 & 1 & 1 \\
 -1 & -1 & 1 \\
\end{array}
\right).
\end{equation}
Using the values of $\sigma_{xy}^{i}$ and $\eta_{xy}$ in \eqref{Psi} we can compute the four cases of functions $\Psi(\theta_{xy})$ defined in \eqref{eqn:Psi1} and \eqref{eqn:Psi2}. Similarly we calculate the derivative $\Psi'(\theta^j_{xy}) = \frac{d\Psi(\theta^j_{xy})}{d \theta^j_{xy}}$ using equations \eqref{DPsi}, \eqref{derivative-angles1}, \eqref{derivative-angles2} and the properties of digamma function $\psi^{(0)}(z)$ \eqref{eq:property} while neglecting the contribution of the $t$-moduli. 

Utilizing above results while ignoring the CP-violating phases $\gamma_m$, the gaugino masses; the trilinear coupling \eqref{tri-coupling}; and the squared-masses of squarks and sleptons \eqref{slepton-mass} are obtained as,
\begin{align}\label{GauginosYba123-model17.5}  
M_{\tilde B} &\equiv M_Y = m_{3/2}\Bigl(\frac{3 \Theta _1+\Theta _2-8 \Theta _4}{4 \sqrt{3}}\Bigr), \nonumber\\
M_{\tilde W} &\equiv M_b = m_{3/2}\Bigl(\frac{\Theta _2+2 \Theta _3}{\sqrt{3}}\Bigr),\nonumber \\
M_{\tilde g} &\equiv M_a = m_{3/2}\Bigl(\frac{\Theta _2-2 \Theta _4}{\sqrt{3}}\Bigr),\nonumber \\
A_0 \equiv A_{abc} &= m_{3/2}\Bigl(-0.128593 \Theta _1-1.64721 \Theta _2-0.107781 \Theta _3+0.151538 \Theta _4\Bigr),\nonumber\\  
m^2_{L} \equiv m^2_{ab} &= m_{3/2}^2\Bigl(-2.2278 \Theta _1{}^2-2.22881 \Theta _1 \Theta _2+0.0357856 \Theta _1 \Theta _3+1.61629 \Theta _1 \Theta _4 \nonumber \\   &\quad +0.395725 \Theta _2{}^2+0.464088 \Theta _2 \Theta _3+0.503187 \Theta _2 \Theta _4+0.535193 \Theta _3{}^2 \nonumber \\  &\quad -3.12602 \Theta _3 \Theta _4+0.309973 \Theta _4{}^2+1\Bigr) ,\nonumber\\  
m^2_{R} \equiv m^2_{ac} &= m_{3/2}^2\Bigl(0.36917 \Theta _1{}^2+0.669259 \Theta _1 \Theta _2-0.840813 \Theta _1 \Theta _3-1.82443 \Theta _1 \Theta _4 \nonumber \\  &\quad  +0.375979 \Theta _2{}^2-0.820669 \Theta _2 \Theta _3-0.733709 \Theta _2 \Theta _4-0.929866 \Theta _3{}^2 \nonumber \\  &\quad +1.7274 \Theta _3 \Theta _4-1.1538 \Theta _4{}^2+1\Bigr). 
\end{align}
All soft terms are subject to the constraint \eqref{constraint}.
\FloatBarrier

\subsection{Model 18}\label{sec:model-18}
\begin{figure}[htb]
\centering
\includegraphics[width=\textwidth]{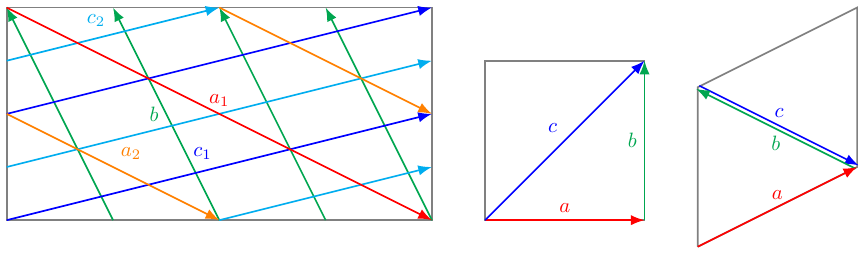}
\caption{Brane configuration for the three two-tori in Model~\hyperref[model18]{18} where the third two-torus is tilted. Fermion mass hierarchies result from the intersections on the first two-torus.}  \label{Fig.18}
\end{figure}

In Model~\hyperref[model18]{18} the three-point Yukawa couplings arise from the triplet intersections from the branes $a$, $b$ and $c$ on the first two-torus with 9 pairs of Higgs from the $\mathcal{N}=2$ sector.
Yukawa matrices for the Model~\hyperref[model18]{18} are of rank 3 and the three intersections required to form the disk diagrams for the Yukawa couplings all occur on the first torus as shown in figure~\ref{Fig.18}.  
\begin{table}[htb]\footnotesize\centering
\renewcommand{\arraystretch}{1.8}
$\begin{array}{|c|c|c|c|}\hline
 \text{\hyperref[model18]{18}} & \theta ^1 & \theta ^2 & \theta ^3 \\\hline
 a & -\tan ^{-1}\left(\frac{1}{2}\right) & -\tan ^{-1}(2) & \tan ^{-1}\left(\frac{1}{4}\right) \\\hline
 b & 0 & \frac{\pi }{2} & \tan ^{-1}\left(\frac{2}{9}\right) \\\hline
 c & \tan ^{-1}\left(\frac{1}{2}\right) & -\tan ^{-1}\left(\frac{1}{2}\right) & -\tan ^{-1}\left(\frac{1}{2}\right) \\\hline
\end{array}$
\caption{The angles with respect to the orientifold plane made by the cycle wrapped by stack of D6-branes on each of the three two-tori in Model~\hyperref[model18]{18}.}
\label{Angles18}
\end{table}

The complex structure moduli $U^i$ \eqref{U-moduli} are given as,
\begin{align}\label{eqn:U-moduli18}
\{U^1,U^2,U^3\} &=\left\{\frac{i}{2},\frac{2 i}{9},\frac{2}{5}+\frac{4 i}{5}\right\},
\end{align}
and the corresponding $u$-moduli and $s$-modulus in supergravity basis from \eqref{eq:moduli} are,
\begin{align}\label{s-u-moduli18}
\{u^1,u^2,u^3\} & = \left\{\frac{e^{-\phi _4}}{3 \pi },\frac{3 e^{-\phi _4}}{4 \pi },\frac{e^{-\phi _4}}{6 \pi }\right\}, \nonumber\\
s & = \frac{3 e^{-\phi _4}}{2 \pi } .
\end{align}
Using \eqref{kingauagefun} and the values from the table~\ref{model18}, the gauge kinetic function becomes,
\begin{align}\label{fx18}
\{f_a,f_b,f_c\} & = \left\{\frac{15 e^{-\phi _4}}{32 \pi },\frac{5 e^{-\phi _4}}{24 \pi },\frac{85 e^{-\phi _4}}{96 \pi }\right\},
\end{align}
To calculate the gaugino masses $\{M_Y,M_b,M_a\}$ for the respective gauge groups $\U(1)_Y$, $\SU(2)_L$, and $\SU(3)_C$, we first compute $\{M_a,M_b,M_c\}$ using \eqref{gaugino-masses} as,
\begin{align}\label{Gauginosabc18}
M_a &= \frac{1}{5} \sqrt{3} m_{3/2} (\Theta _2-4 \Theta _4),\nonumber \\
M_b &= \frac{1}{5} \sqrt{3} m_{3/2} (\Theta _1+4 \Theta _3),\nonumber \\
M_c &= \frac{1}{85} \sqrt{3} m_{3/2} (8 \Theta _1+9 \Theta _2-4 (\Theta _3+18 \Theta _4)).
\end{align} 

Next, to compute the trilinear coupling and the sleptons mass-squared we require the angles, the differences of angles and their first and second order derivatives with respect to the moduli. In table~\ref{Angles18} we show the angles \eqref{eq:angle} made by the cycles wrapped by each stack of D6-branes with respect to the orientifold plane on each two-torus.
The differences of the angles, $\theta_{xy}^{i}= \theta_{y}^{i} -\theta_{x}^{i}$ are,
\begin{align}\label{angle-diff18}\arraycolsep=0pt
\hskip -1em \left[
\begin{array}{ccc}
 \{0.,0.,0.\} & \{-0.501908,0.570796,0.214297\} & \{-0.291374,0.218669,0.0727048\} \\
 \{0.501908,-0.570796,-0.214297\} & \{0.,0.,0.\} & \{0.210535,-0.352127,-0.141593\} \\
 \{0.291374,-0.218669,-0.0727048\} & \{-0.210535,0.352127,0.141593\} & \{0.,0.,0.\} \\
\end{array}
\right]
\end{align}

To account for the negative angle differences we employ the sign function $\sigma_{xy}^{i}$, which is $-1$ only for negative angle difference and $+1$ otherwise,
\begin{align}\label{sigmaK18}
\sigma_{xy}^{i} & = \left(
\begin{array}{ccc}
 \{1,1,1\} & \{-1,1,1\} & \{-1,1,1\} \\
 \{1,-1,-1\} & \{1,1,1\} & \{1,-1,-1\} \\
 \{1,-1,-1\} & \{-1,1,1\} & \{1,1,1\} \\
\end{array}
\right),
\end{align}
and the function $\eta_{xy}$ is evaluated by taking the product on the torus index $i$ as,
\begin{equation}\label{eta18}
\eta_{xy} = \left(
\begin{array}{ccc}
 1 & -1 & -1 \\
 1 & 1 & 1 \\
 1 & -1 & 1 \\
\end{array}
\right).
\end{equation}
Using the values of $\sigma_{xy}^{i}$ and $\eta_{xy}$ in \eqref{Psi} we can compute the four cases of functions $\Psi(\theta_{xy})$ defined in \eqref{eqn:Psi1} and \eqref{eqn:Psi2}. Similarly we calculate the derivative $\Psi'(\theta^j_{xy}) = \frac{d\Psi(\theta^j_{xy})}{d \theta^j_{xy}}$ using equations \eqref{DPsi}, \eqref{derivative-angles1}, \eqref{derivative-angles2} and the properties of digamma function $\psi^{(0)}(z)$ \eqref{eq:property} while neglecting the contribution of the $t$-moduli. 

Utilizing above results while ignoring the CP-violating phases $\gamma_m$, the gaugino masses; the trilinear coupling \eqref{tri-coupling}; and the squared-masses of squarks and sleptons \eqref{slepton-mass} are obtained as,
\begin{align}\label{GauginosYba123-model18}  
M_{\tilde B} &\equiv M_Y = m_{3/2}\Bigl(\frac{1}{115} \sqrt{3} (8 \Theta _1+15 \Theta _2-4 (\Theta _3+24 \Theta _4))\Bigr), \nonumber\\
M_{\tilde W} &\equiv M_b = m_{3/2}\Bigl(\frac{1}{5} \sqrt{3} (\Theta _1+4 \Theta _3)\Bigr),\nonumber \\
M_{\tilde g} &\equiv M_a = m_{3/2}\Bigl(\frac{1}{5} \sqrt{3} (\Theta _2-4 \Theta _4)\Bigr),\nonumber \\
A_0 \equiv A_{abc} &= m_{3/2}\Bigl(-1.82708 \Theta _1-1.50614 \Theta _2+1.42587 \Theta _3+0.175299 \Theta _4\Bigr),\nonumber\\  
m^2_{L} \equiv m^2_{ab} &= m_{3/2}^2\Bigl(0.730534 \Theta _1{}^2+0.131797 \Theta _1 \Theta _2-1.01748 \Theta _1 \Theta _3-1.17062 \Theta _1 \Theta _4 \nonumber \\   &\quad +0.144867 \Theta _2{}^2-1.16345 \Theta _2 \Theta _3-0.752402 \Theta _2 \Theta _4-0.846416 \Theta _3{}^2 \nonumber \\  &\quad +0.832553 \Theta _3 \Theta _4-0.848056 \Theta _4{}^2+1\Bigr) ,\nonumber\\  
m^2_{R} \equiv m^2_{ac} &= m_{3/2}^2\Bigl(0.72924 \Theta _1{}^2+0.0000979797 \Theta _1 \Theta _2-0.60321 \Theta _1 \Theta _3-2.56141 \Theta _1 \Theta _4 \nonumber \\  &\quad  +0.759033 \Theta _2{}^2-2.39659 \Theta _2 \Theta _3-0.388307 \Theta _2 \Theta _4-1.82808 \Theta _3{}^2 \nonumber \\  &\quad +1.44981 \Theta _3 \Theta _4+0.339613 \Theta _4{}^2+1\Bigr). 
\end{align}
All soft terms are subject to the constraint \eqref{constraint}.
\FloatBarrier

\subsection{Model 18-dual}\label{sec:model-18.5}
\begin{figure}[htb]
\centering
\includegraphics[width=\textwidth]{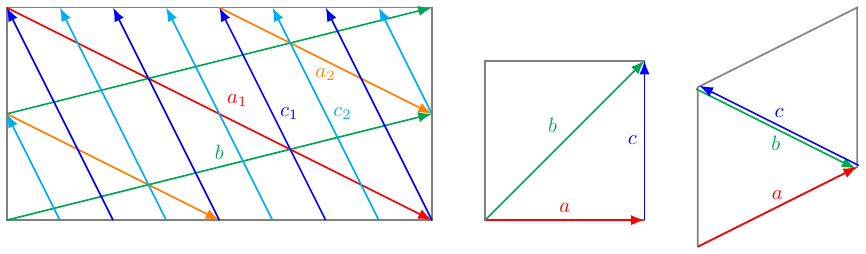}
\caption{Brane configuration for the three two-tori in Model~\hyperref[model18.5]{18-dual} where the third two-torus is tilted. Fermion mass hierarchies result from the intersections on the first two-torus.}  \label{Fig.18.5}
\end{figure}

In Model~\hyperref[model18.5]{18-dual} the three-point Yukawa couplings arise from the triplet intersections from the branes $a$, $b$ and $c$ on the first two-torus with 9 pairs of Higgs from the $\mathcal{N}=2$ sector.
Yukawa matrices for the Model~\hyperref[model18.5]{18-dual} are of rank 3 and the three intersections required to form the disk diagrams for the Yukawa couplings all occur on the first torus as shown in figure~\ref{Fig.18.5}.  
\begin{table}[htb]\footnotesize\centering
\renewcommand{\arraystretch}{1.8}
$\begin{array}{|c|c|c|c|}\hline
 \text{\hyperref[model18.5]{18-dual}} & \theta ^1 & \theta ^2 & \theta ^3 \\\hline
 a & -\tan ^{-1}\left(\frac{1}{2}\right) & \tan ^{-1}\left(\frac{1}{4}\right) & -\tan ^{-1}(2) \\\hline
 b & 0 & \tan ^{-1}\left(\frac{2}{9}\right) & \frac{\pi }{2} \\\hline
 c & \tan ^{-1}\left(\frac{1}{2}\right) & -\tan ^{-1}\left(\frac{1}{2}\right) & -\tan ^{-1}\left(\frac{1}{2}\right) \\\hline
\end{array}$
\caption{The angles with respect to the orientifold plane made by the cycle wrapped by stack of D6-branes on each of the three two-tori in Model~\hyperref[model18.5]{18-dual}.}
\label{Angles18.5}
\end{table}

The complex structure moduli $U^i$ \eqref{U-moduli} are given as,
\begin{align}\label{eqn:U-moduli18.5}
\{U^1,U^2,U^3\} &=\left\{\frac{i}{2},\frac{2 i}{9},\frac{2}{5}+\frac{4 i}{5}\right\},
\end{align}
and the corresponding $u$-moduli and $s$-modulus in supergravity basis from \eqref{eq:moduli} are,
\begin{align}\label{s-u-moduli18.5}
\{u^1,u^2,u^3\} & = \left\{\frac{e^{-\phi _4}}{3 \pi },\frac{3 e^{-\phi _4}}{4 \pi },\frac{e^{-\phi _4}}{6 \pi }\right\}, \nonumber\\
s & = \frac{3 e^{-\phi _4}}{2 \pi } .
\end{align}
Using \eqref{kingauagefun} and the values from the table~\ref{model18.5}, the gauge kinetic function becomes,
\begin{align}\label{fx18.5}
\{f_a,f_b,f_c\} & = \left\{\frac{15 e^{-\phi _4}}{32 \pi },\frac{85 e^{-\phi _4}}{96 \pi },\frac{5 e^{-\phi _4}}{24 \pi }\right\},
\end{align}
To calculate the gaugino masses $\{M_Y,M_b,M_a\}$ for the respective gauge groups $\U(1)_Y$, $\SU(2)_L$, and $\SU(3)_C$, we first compute $\{M_a,M_b,M_c\}$ using \eqref{gaugino-masses} as,
\begin{align}\label{Gauginosabc18.5}
M_a &= \frac{1}{5} \sqrt{3} m_{3/2} (\Theta _2-4 \Theta _4),\nonumber \\
M_b &= \frac{1}{85} \sqrt{3} m_{3/2} (8 \Theta _1+9 \Theta _2-4 (\Theta _3+18 \Theta _4)),\nonumber \\
M_c &= \frac{1}{5} \sqrt{3} m_{3/2} (\Theta _1+4 \Theta _3).
\end{align} 

Next, to compute the trilinear coupling and the sleptons mass-squared we require the angles, the differences of angles and their first and second order derivatives with respect to the moduli. In table~\ref{Angles18.5} we show the angles \eqref{eq:angle} made by the cycles wrapped by each stack of D6-branes with respect to the orientifold plane on each two-torus.
The differences of the angles, $\theta_{xy}^{i}= \theta_{y}^{i} -\theta_{x}^{i}$ are,
\begin{align}\label{angle-diff18.5}\arraycolsep=0pt
\hskip -1em \left[
\begin{array}{ccc}
 \{0.,0.,0.\} & \{-0.291374,0.218669,0.0727048\} & \{-0.501908,0.570796,0.214297\} \\
 \{0.291374,-0.218669,-0.0727048\} & \{0.,0.,0.\} & \{-0.210535,0.352127,0.141593\} \\
 \{0.501908,-0.570796,-0.214297\} & \{0.210535,-0.352127,-0.141593\} & \{0.,0.,0.\} \\
\end{array}
\right]
\end{align}

To account for the negative angle differences we employ the sign function $\sigma_{xy}^{i}$, which is $-1$ only for negative angle difference and $+1$ otherwise,
\begin{align}\label{sigmaK18.5}
\sigma_{xy}^{i} & = \left(
\begin{array}{ccc}
 \{1,1,1\} & \{-1,1,1\} & \{-1,1,1\} \\
 \{1,-1,-1\} & \{1,1,1\} & \{-1,1,1\} \\
 \{1,-1,-1\} & \{1,-1,-1\} & \{1,1,1\} \\
\end{array}
\right),
\end{align}
and the function $\eta_{xy}$ is evaluated by taking the product on the torus index $i$ as,
\begin{equation}\label{eta18.5}
\eta_{xy} = \left(
\begin{array}{ccc}
 1 & -1 & -1 \\
 1 & 1 & -1 \\
 1 & 1 & 1 \\
\end{array}
\right).
\end{equation}
Using the values of $\sigma_{xy}^{i}$ and $\eta_{xy}$ in \eqref{Psi} we can compute the four cases of functions $\Psi(\theta_{xy})$ defined in \eqref{eqn:Psi1} and \eqref{eqn:Psi2}. Similarly we calculate the derivative $\Psi'(\theta^j_{xy}) = \frac{d\Psi(\theta^j_{xy})}{d \theta^j_{xy}}$ using equations \eqref{DPsi}, \eqref{derivative-angles1}, \eqref{derivative-angles2} and the properties of digamma function $\psi^{(0)}(z)$ \eqref{eq:property} while neglecting the contribution of the $t$-moduli. 

Utilizing above results while ignoring the CP-violating phases $\gamma_m$, the gaugino masses; the trilinear coupling \eqref{tri-coupling}; and the squared-masses of squarks and sleptons \eqref{slepton-mass} are obtained as,
\begin{align}\label{GauginosYba123-model18.5}  
M_{\tilde B} &\equiv M_Y = m_{3/2}\Bigl(\frac{1}{25} \sqrt{3} (2 \Theta _1+3 \Theta _2+8 \Theta _3-12 \Theta _4)\Bigr), \nonumber\\
M_{\tilde W} &\equiv M_b = m_{3/2}\Bigl(\frac{1}{85} \sqrt{3} (8 \Theta _1+9 \Theta _2-4 (\Theta _3+18 \Theta _4))\Bigr),\nonumber \\
M_{\tilde g} &\equiv M_a = m_{3/2}\Bigl(\frac{1}{5} \sqrt{3} (\Theta _2-4 \Theta _4)\Bigr),\nonumber \\
A_0 \equiv A_{abc} &= m_{3/2}\Bigl(-1.82708 \Theta _1-1.50614 \Theta _2+1.42587 \Theta _3+0.175299 \Theta _4\Bigr),\nonumber\\  
m^2_{L} \equiv m^2_{ab} &= m_{3/2}^2\Bigl(0.72924 \Theta _1{}^2+0.0000979797 \Theta _1 \Theta _2-0.60321 \Theta _1 \Theta _3-2.56141 \Theta _1 \Theta _4 \nonumber \\   &\quad +0.759033 \Theta _2{}^2-2.39659 \Theta _2 \Theta _3-0.388307 \Theta _2 \Theta _4-1.82808 \Theta _3{}^2 \nonumber \\  &\quad +1.44981 \Theta _3 \Theta _4+0.339613 \Theta _4{}^2+1\Bigr) ,\nonumber\\  
m^2_{R} \equiv m^2_{ac} &= m_{3/2}^2\Bigl(0.730534 \Theta _1{}^2+0.131797 \Theta _1 \Theta _2-1.01748 \Theta _1 \Theta _3-1.17062 \Theta _1 \Theta _4 \nonumber \\  &\quad  +0.144867 \Theta _2{}^2-1.16345 \Theta _2 \Theta _3-0.752402 \Theta _2 \Theta _4-0.846416 \Theta _3{}^2 \nonumber \\  &\quad +0.832553 \Theta _3 \Theta _4-0.848056 \Theta _4{}^2+1\Bigr). 
\end{align}
All soft terms are subject to the constraint \eqref{constraint}.
\FloatBarrier

\subsection{Model 19}\label{sec:model-19}
\begin{figure}[htb]
\centering
\includegraphics[width=\textwidth]{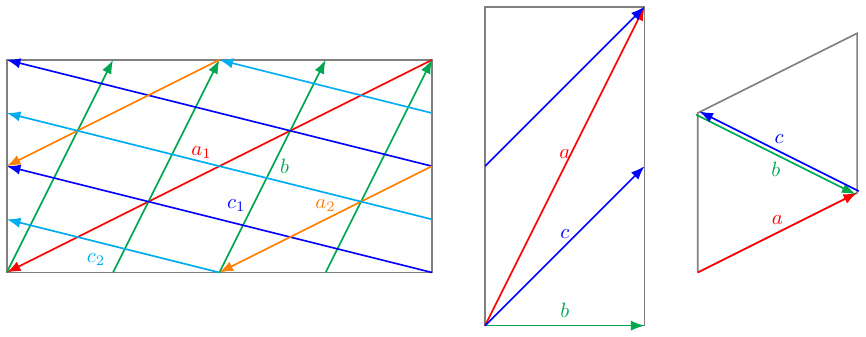}
\caption{Brane configuration for the three two-tori in Model~\hyperref[model19]{19} where the third two-torus is tilted. Fermion mass hierarchies result from the intersections on the first two-torus.}  \label{Fig.19}
\end{figure}

In Model~\hyperref[model19]{19} the three-point Yukawa couplings arise from the triplet intersections from the branes $a$, $b$ and $c$ on the first two-torus with 9 pairs of Higgs from the $\mathcal{N}=2$ sector.
Yukawa matrices for the Model~\hyperref[model19]{19} are of rank 3 and the three intersections required to form the disk diagrams for the Yukawa couplings all occur on the first torus as shown in figure~\ref{Fig.19}.  
\begin{table}[htb]\footnotesize\centering
\renewcommand{\arraystretch}{1.8}
$\begin{array}{|c|c|c|c|}\hline
 \text{\hyperref[model19]{19}} & \theta ^1 & \theta ^2 & \theta ^3 \\\hline
 a & \tan ^{-1}\left(\sqrt{\frac{7}{23}}\right) & \tan ^{-1}\left(4 \sqrt{\frac{7}{23}}\right) & -\tan ^{-1}\left(\frac{\sqrt{\frac{7}{23}}}{2}\right) \\\hline
 b & \tan ^{-1}\left(\sqrt{161}\right) & 0 & \tan ^{-1}\left(\frac{\sqrt{161}}{2}\right) \\\hline
 c & \tan ^{-1}\left(4 \sqrt{\frac{7}{23}}\right) & -\tan ^{-1}\left(4 \sqrt{\frac{7}{23}}\right) & -\tan ^{-1}\left(4 \sqrt{\frac{7}{23}}\right) \\\hline
\end{array}$
\caption{The angles with respect to the orientifold plane made by the cycle wrapped by stack of D6-branes on each of the three two-tori in Model~\hyperref[model19]{19}.}
\label{Angles19}
\end{table}

The complex structure moduli $U^i$ \eqref{U-moduli} are given as,
\begin{align}\label{eqn:U-moduli19}
\{U^1,U^2,U^3\} &=\left\{i \sqrt{\frac{7}{23}},i \sqrt{161},\frac{8}{135} \left(28+i \sqrt{161}\right)\right\},
\end{align}
and the corresponding $u$-moduli and $s$-modulus in supergravity basis from \eqref{eq:moduli} are,
\begin{align}\label{s-u-moduli19}
\{u^1,u^2,u^3\} & = \left\{\frac{\sqrt{2} \sqrt[4]{161} e^{-\phi _4}}{\pi },\frac{\sqrt{2} \sqrt[4]{7} e^{-\phi _4}}{23^{3/4} \pi },\frac{\sqrt[4]{161} e^{-\phi _4}}{4 \sqrt{2} \pi }\right\}, \nonumber\\
s & = \frac{\sqrt[4]{23} e^{-\phi _4}}{4 \sqrt{2} 7^{3/4} \pi } .
\end{align}
Using \eqref{kingauagefun} and the values from the table~\ref{model19}, the gauge kinetic function becomes,
\begin{align}\label{fx19}
\{f_a,f_b,f_c\} & = \left\{\frac{405 e^{-\phi _4}}{8 \sqrt{2} 161^{3/4} \pi },\frac{135 e^{-\phi _4}}{16 \sqrt{2} 161^{3/4} \pi },\frac{1485 e^{-\phi _4}}{16 \sqrt{2} 161^{3/4} \pi }\right\},
\end{align}
To calculate the gaugino masses $\{M_Y,M_b,M_a\}$ for the respective gauge groups $\U(1)_Y$, $\SU(2)_L$, and $\SU(3)_C$, we first compute $\{M_a,M_b,M_c\}$ using \eqref{gaugino-masses} as,
\begin{align}\label{Gauginosabc19}
M_a &= \frac{m_{3/2} (644 \Theta _1+28 \Theta _2+23 (7 \Theta _3+\Theta _4))}{270 \sqrt{3}},\nonumber \\
M_b &= \frac{m_{3/2} (112 \Theta _2-23 \Theta _4)}{45 \sqrt{3}},\nonumber \\
M_c &= \frac{m_{3/2} (1288 \Theta _1-56 \Theta _2+161 \Theta _3-92 \Theta _4)}{495 \sqrt{3}}.
\end{align} 

Next, to compute the trilinear coupling and the sleptons mass-squared we require the angles, the differences of angles and their first and second order derivatives with respect to the moduli. In table~\ref{Angles19} we show the angles \eqref{eq:angle} made by the cycles wrapped by each stack of D6-branes with respect to the orientifold plane on each two-torus.
The differences of the angles, $\theta_{xy}^{i}= \theta_{y}^{i} -\theta_{x}^{i}$ are,
\begin{align}\label{angle-diff19}\arraycolsep=0pt
\hskip -1em \left[
\begin{array}{ccc}
 \{0.,0.,0.\} & \{-0.217223,-0.492148,0.70937\} & \{0.50991,-0.0776874,0.850963\} \\
 \{0.217223,0.492148,-0.70937\} & \{0.,0.,0.\} & \{0.727132,0.41446,0.141593\} \\
 \{-0.50991,0.0776874,-0.850963\} & \{-0.727132,-0.41446,-0.141593\} & \{0.,0.,0.\} \\
\end{array}
\right]
\end{align}

To account for the negative angle differences we employ the sign function $\sigma_{xy}^{i}$, which is $-1$ only for negative angle difference and $+1$ otherwise,
\begin{align}\label{sigmaK19}
\sigma_{xy}^{i} & = \left(
\begin{array}{ccc}
 \{1,1,1\} & \{-1,-1,1\} & \{1,-1,1\} \\
 \{1,1,-1\} & \{1,1,1\} & \{1,1,1\} \\
 \{-1,1,-1\} & \{-1,-1,-1\} & \{1,1,1\} \\
\end{array}
\right),
\end{align}
and the function $\eta_{xy}$ is evaluated by taking the product on the torus index $i$ as,
\begin{equation}\label{eta19}
\eta_{xy} = \left(
\begin{array}{ccc}
 1 & 1 & -1 \\
 -1 & 1 & 1 \\
 1 & -1 & 1 \\
\end{array}
\right).
\end{equation}
Using the values of $\sigma_{xy}^{i}$ and $\eta_{xy}$ in \eqref{Psi} we can compute the four cases of functions $\Psi(\theta_{xy})$ defined in \eqref{eqn:Psi1} and \eqref{eqn:Psi2}. Similarly we calculate the derivative $\Psi'(\theta^j_{xy}) = \frac{d\Psi(\theta^j_{xy})}{d \theta^j_{xy}}$ using equations \eqref{DPsi}, \eqref{derivative-angles1}, \eqref{derivative-angles2} and the properties of digamma function $\psi^{(0)}(z)$ \eqref{eq:property} while neglecting the contribution of the $t$-moduli. 

Utilizing above results while ignoring the CP-violating phases $\gamma_m$, the gaugino masses; the trilinear coupling \eqref{tri-coupling}; and the squared-masses of squarks and sleptons \eqref{slepton-mass} are obtained as,
\begin{align}\label{GauginosYba123-model19}  
M_{\tilde B} &\equiv M_Y = m_{3/2}\Bigl(\frac{5152 \Theta _1-112 \Theta _2+805 \Theta _3-230 \Theta _4}{2025 \sqrt{3}}\Bigr), \nonumber\\
M_{\tilde W} &\equiv M_b = m_{3/2}\Bigl(\frac{112 \Theta _2-23 \Theta _4}{45 \sqrt{3}}\Bigr),\nonumber \\
M_{\tilde g} &\equiv M_a = m_{3/2}\Bigl(\frac{644 \Theta _1+28 \Theta _2+23 (7 \Theta _3+\Theta _4)}{270 \sqrt{3}}\Bigr),\nonumber \\
A_0 \equiv A_{abc} &= m_{3/2}\Bigl(-0.133497 \Theta _1-1.55973 \Theta _2+0.108067 \Theta _3-0.146888 \Theta _4\Bigr),\nonumber\\  
m^2_{L} \equiv m^2_{ab} &= m_{3/2}^2\Bigl(-1.92219 \Theta _1{}^2-0.606279 \Theta _1 \Theta _2+0.894634 \Theta _1 \Theta _3-0.207461 \Theta _1 \Theta _4 \nonumber \\   &\quad -1.85446 \Theta _2{}^2+0.894634 \Theta _2 \Theta _3+0.862019 \Theta _2 \Theta _4-0.50419 \Theta _3{}^2 \nonumber \\  &\quad +0.239164 \Theta _3 \Theta _4-0.506606 \Theta _4{}^2+1\Bigr) ,\nonumber\\  
m^2_{R} \equiv m^2_{ac} &= m_{3/2}^2\Bigl(-1.63243 \Theta _1{}^2+1.90023 \Theta _1 \Theta _2-0.696574 \Theta _1 \Theta _3+1.49474 \Theta _1 \Theta _4 \nonumber \\  &\quad  +0.770153 \Theta _2{}^2+1.49862 \Theta _2 \Theta _3-1.898 \Theta _2 \Theta _4-1.96883 \Theta _3{}^2 \nonumber \\  &\quad +1.08122 \Theta _3 \Theta _4-1.52482 \Theta _4{}^2+1\Bigr). 
\end{align}
All soft terms are subject to the constraint \eqref{constraint}.
\FloatBarrier

\subsection{Model 19-dual}\label{sec:model-19.5}
\begin{figure}[htb]
\centering
\includegraphics[width=\textwidth]{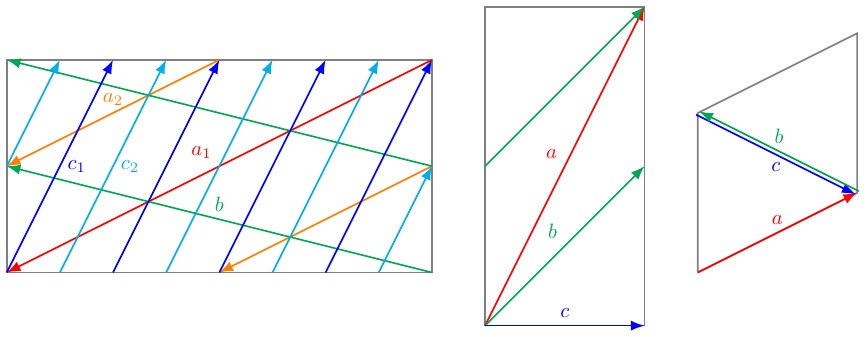}
\caption{Brane configuration for the three two-tori in Model~\hyperref[model19.5]{19-dual} where the third two-torus is tilted. Fermion mass hierarchies result from the intersections on the first two-torus.}  \label{Fig.19.5}
\end{figure}

In Model~\hyperref[model19.5]{19-dual} the three-point Yukawa couplings arise from the triplet intersections from the branes $a$, $b$ and $c$ on the first two-torus with 9 pairs of Higgs from the $\mathcal{N}=2$ sector.
Yukawa matrices for the Model~\hyperref[model19.5]{19-dual} are of rank 3 and the three intersections required to form the disk diagrams for the Yukawa couplings all occur on the first torus as shown in figure~\ref{Fig.19.5}.  
\begin{table}[htb]\footnotesize\centering
\renewcommand{\arraystretch}{1.8}
$\begin{array}{|c|c|c|c|}\hline
 \text{\hyperref[model19.5]{19-dual}} & \theta ^1 & \theta ^2 & \theta ^3 \\\hline
 a & \tan ^{-1}\left(\sqrt{\frac{7}{23}}\right) & -\tan ^{-1}\left(\frac{\sqrt{\frac{7}{23}}}{2}\right) & \tan ^{-1}\left(4 \sqrt{\frac{7}{23}}\right) \\\hline
 b & \tan ^{-1}\left(\sqrt{161}\right) & \tan ^{-1}\left(\frac{\sqrt{161}}{2}\right) & 0 \\\hline
 c & \tan ^{-1}\left(4 \sqrt{\frac{7}{23}}\right) & -\tan ^{-1}\left(4 \sqrt{\frac{7}{23}}\right) & -\tan ^{-1}\left(4 \sqrt{\frac{7}{23}}\right) \\\hline
\end{array}$
\caption{The angles with respect to the orientifold plane made by the cycle wrapped by stack of D6-branes on each of the three two-tori in Model~\hyperref[model19.5]{19-dual}.}
\label{Angles19.5}
\end{table}

The complex structure moduli $U^i$ \eqref{U-moduli} are given as,
\begin{align}\label{eqn:U-moduli19.5}
\{U^1,U^2,U^3\} &=\left\{i \sqrt{\frac{7}{23}},i \sqrt{161},\frac{8}{135} \left(28+i \sqrt{161}\right)\right\},
\end{align}
and the corresponding $u$-moduli and $s$-modulus in supergravity basis from \eqref{eq:moduli} are,
\begin{align}\label{s-u-moduli19.5}
\{u^1,u^2,u^3\} & = \left\{\frac{\sqrt{2} \sqrt[4]{161} e^{-\phi _4}}{\pi },\frac{\sqrt{2} \sqrt[4]{7} e^{-\phi _4}}{23^{3/4} \pi },\frac{\sqrt[4]{161} e^{-\phi _4}}{4 \sqrt{2} \pi }\right\}, \nonumber\\
s & = \frac{\sqrt[4]{23} e^{-\phi _4}}{4 \sqrt{2} 7^{3/4} \pi } .
\end{align}
Using \eqref{kingauagefun} and the values from the table~\ref{model19.5}, the gauge kinetic function becomes,
\begin{align}\label{fx19.5}
\{f_a,f_b,f_c\} & = \left\{\frac{405 e^{-\phi _4}}{8 \sqrt{2} 161^{3/4} \pi },\frac{1485 e^{-\phi _4}}{16 \sqrt{2} 161^{3/4} \pi },\frac{135 e^{-\phi _4}}{16 \sqrt{2} 161^{3/4} \pi }\right\},
\end{align}
To calculate the gaugino masses $\{M_Y,M_b,M_a\}$ for the respective gauge groups $\U(1)_Y$, $\SU(2)_L$, and $\SU(3)_C$, we first compute $\{M_a,M_b,M_c\}$ using \eqref{gaugino-masses} as,
\begin{align}\label{Gauginosabc19.5}
M_a &= \frac{m_{3/2} (644 \Theta _1+28 \Theta _2+23 (7 \Theta _3+\Theta _4))}{270 \sqrt{3}},\nonumber \\
M_b &= \frac{m_{3/2} (1288 \Theta _1-56 \Theta _2+161 \Theta _3-92 \Theta _4)}{495 \sqrt{3}},\nonumber \\
M_c &= \frac{m_{3/2} (112 \Theta _2-23 \Theta _4)}{45 \sqrt{3}}.
\end{align} 

Next, to compute the trilinear coupling and the sleptons mass-squared we require the angles, the differences of angles and their first and second order derivatives with respect to the moduli. In table~\ref{Angles19.5} we show the angles \eqref{eq:angle} made by the cycles wrapped by each stack of D6-branes with respect to the orientifold plane on each two-torus.
The differences of the angles, $\theta_{xy}^{i}= \theta_{y}^{i} -\theta_{x}^{i}$ are,
\begin{align}\label{angle-diff19.5}\arraycolsep=0pt
\hskip -1em \left[
\begin{array}{ccc}
 \{0.,0.,0.\} & \{0.50991,-0.0776874,0.850963\} & \{-0.217223,-0.492148,0.70937\} \\
 \{-0.50991,0.0776874,-0.850963\} & \{0.,0.,0.\} & \{-0.727132,-0.41446,-0.141593\} \\
 \{0.217223,0.492148,-0.70937\} & \{0.727132,0.41446,0.141593\} & \{0.,0.,0.\} \\
\end{array}
\right]
\end{align}

To account for the negative angle differences we employ the sign function $\sigma_{xy}^{i}$, which is $-1$ only for negative angle difference and $+1$ otherwise,
\begin{align}\label{sigmaK19.5}
\sigma_{xy}^{i} & = \left(
\begin{array}{ccc}
 \{1,1,1\} & \{1,-1,1\} & \{-1,-1,1\} \\
 \{-1,1,-1\} & \{1,1,1\} & \{-1,-1,-1\} \\
 \{1,1,-1\} & \{1,1,1\} & \{1,1,1\} \\
\end{array}
\right),
\end{align}
and the function $\eta_{xy}$ is evaluated by taking the product on the torus index $i$ as,
\begin{equation}\label{eta19.5}
\eta_{xy} = \left(
\begin{array}{ccc}
 1 & -1 & 1 \\
 1 & 1 & -1 \\
 -1 & 1 & 1 \\
\end{array}
\right).
\end{equation}
Using the values of $\sigma_{xy}^{i}$ and $\eta_{xy}$ in \eqref{Psi} we can compute the four cases of functions $\Psi(\theta_{xy})$ defined in \eqref{eqn:Psi1} and \eqref{eqn:Psi2}. Similarly we calculate the derivative $\Psi'(\theta^j_{xy}) = \frac{d\Psi(\theta^j_{xy})}{d \theta^j_{xy}}$ using equations \eqref{DPsi}, \eqref{derivative-angles1}, \eqref{derivative-angles2} and the properties of digamma function $\psi^{(0)}(z)$ \eqref{eq:property} while neglecting the contribution of the $t$-moduli. 

Utilizing above results while ignoring the CP-violating phases $\gamma_m$, the gaugino masses; the trilinear coupling \eqref{tri-coupling}; and the squared-masses of squarks and sleptons \eqref{slepton-mass} are obtained as,
\begin{align}\label{GauginosYba123-model19.5}  
M_{\tilde B} &\equiv M_Y = m_{3/2}\Bigl(\frac{1288 \Theta _1+392 \Theta _2+322 \Theta _3-23 \Theta _4}{675 \sqrt{3}}\Bigr), \nonumber\\
M_{\tilde W} &\equiv M_b = m_{3/2}\Bigl(\frac{1288 \Theta _1-56 \Theta _2+161 \Theta _3-92 \Theta _4}{495 \sqrt{3}}\Bigr),\nonumber \\
M_{\tilde g} &\equiv M_a = m_{3/2}\Bigl(\frac{644 \Theta _1+28 \Theta _2+23 (7 \Theta _3+\Theta _4)}{270 \sqrt{3}}\Bigr),\nonumber \\
A_0 \equiv A_{abc} &= m_{3/2}\Bigl(-0.133497 \Theta _1-1.55973 \Theta _2+0.108067 \Theta _3-0.146888 \Theta _4\Bigr),\nonumber\\  
m^2_{L} \equiv m^2_{ab} &= m_{3/2}^2\Bigl(-1.63243 \Theta _1{}^2+1.90023 \Theta _1 \Theta _2-0.696574 \Theta _1 \Theta _3+1.49474 \Theta _1 \Theta _4 \nonumber \\   &\quad +0.770153 \Theta _2{}^2+1.49862 \Theta _2 \Theta _3-1.898 \Theta _2 \Theta _4-1.96883 \Theta _3{}^2 \nonumber \\  &\quad +1.08122 \Theta _3 \Theta _4-1.52482 \Theta _4{}^2+1\Bigr) ,\nonumber\\  
m^2_{R} \equiv m^2_{ac} &= m_{3/2}^2\Bigl(-1.92219 \Theta _1{}^2-0.606279 \Theta _1 \Theta _2+0.894634 \Theta _1 \Theta _3-0.207461 \Theta _1 \Theta _4 \nonumber \\  &\quad  -1.85446 \Theta _2{}^2+0.894634 \Theta _2 \Theta _3+0.862019 \Theta _2 \Theta _4-0.50419 \Theta _3{}^2 \nonumber \\  &\quad +0.239164 \Theta _3 \Theta _4-0.506606 \Theta _4{}^2+1\Bigr). 
\end{align}
All soft terms are subject to the constraint \eqref{constraint}.
\FloatBarrier

\subsection{Model 20}\label{sec:model-20}
\begin{figure}[htb]
\centering
\includegraphics[width=\textwidth]{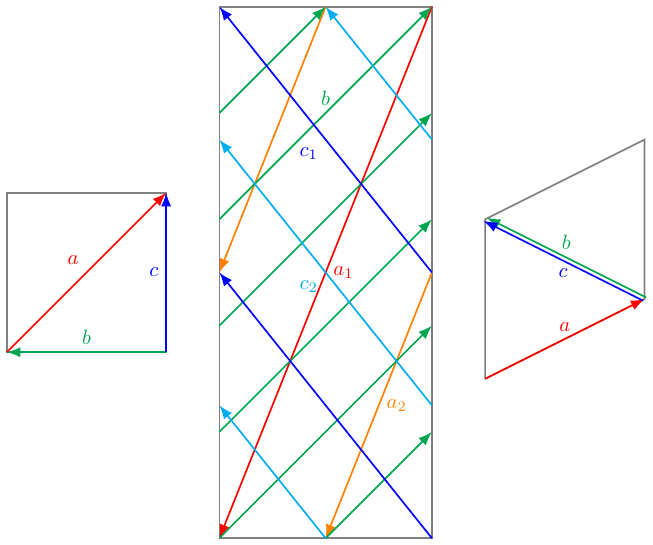}
\caption{Brane configuration for the three two-tori in Model~\hyperref[model20]{20} where the third two-torus is tilted. Fermion mass hierarchies result from the intersections on the second two-torus.}  \label{Fig.20}
\end{figure}

In Model~\hyperref[model20]{20} the three-point Yukawa couplings arise from the triplet intersections from the branes $a$, $b$ and $c$ on the second two-torus with 9 pairs of Higgs from the $\mathcal{N}=2$ sector.
Yukawa matrices for the Model~\hyperref[model20]{20} are of rank 3 and the three intersections required to form the disk diagrams for the Yukawa couplings all occur on the second torus as shown in figure~\ref{Fig.20}.  
\begin{table}[htb]\footnotesize\centering
\renewcommand{\arraystretch}{1.8}
$\begin{array}{|c|c|c|c|}\hline
 \text{\hyperref[model20]{20}} & \theta ^1 & \theta ^2 & \theta ^3 \\\hline
 a & \tan ^{-1}\left(\frac{7}{\sqrt{5}}\right) & 0 & \frac{\pi }{2} \\\hline
 b & \tan ^{-1}\left(\sqrt{5}\right) & \tan ^{-1}\left(\frac{2}{\sqrt{5}}\right) & -\tan ^{-1}\left(\frac{\sqrt{5}}{2}\right) \\\hline
 c & \tan ^{-1}\left(\frac{2}{\sqrt{5}}\right) & -\tan ^{-1}\left(\frac{2}{\sqrt{5}}\right) & -\tan ^{-1}\left(\frac{2}{\sqrt{5}}\right) \\\hline
\end{array}$
\caption{The angles with respect to the orientifold plane made by the cycle wrapped by stack of D6-branes on each of the three two-tori in Model~\hyperref[model20]{20}.}
\label{Angles20}
\end{table}

The complex structure moduli $U^i$ \eqref{U-moduli} are given as,
\begin{align}\label{eqn:U-moduli20}
\{U^1,U^2,U^3\} &=\left\{\frac{7 i}{\sqrt{5}},i \sqrt{5},\frac{4}{9} \left(2+i \sqrt{5}\right)\right\},
\end{align}
and the corresponding $u$-moduli and $s$-modulus in supergravity basis from \eqref{eq:moduli} are,
\begin{align}\label{s-u-moduli20}
\{u^1,u^2,u^3\} & = \left\{\frac{\sqrt[4]{5} e^{-\phi _4}}{\sqrt{7} \pi },\frac{\sqrt{7} e^{-\phi _4}}{5^{3/4} \pi },\frac{\sqrt[4]{5} \sqrt{7} e^{-\phi _4}}{4 \pi }\right\}, \nonumber\\
s & = \frac{\sqrt[4]{5} e^{-\phi _4}}{4 \sqrt{7} \pi } .
\end{align}
Using \eqref{kingauagefun} and the values from the table~\ref{model20}, the gauge kinetic function becomes,
\begin{align}\label{fx20}
\{f_a,f_b,f_c\} & = \left\{\frac{27 e^{-\phi _4}}{8\ 5^{3/4} \sqrt{7} \pi },\frac{9 \sqrt[4]{5} e^{-\phi _4}}{16 \sqrt{7} \pi },\frac{9 \sqrt{7} e^{-\phi _4}}{16\ 5^{3/4} \pi }\right\},
\end{align}
To calculate the gaugino masses $\{M_Y,M_b,M_a\}$ for the respective gauge groups $\U(1)_Y$, $\SU(2)_L$, and $\SU(3)_C$, we first compute $\{M_a,M_b,M_c\}$ using \eqref{gaugino-masses} as,
\begin{align}\label{Gauginosabc20}
M_a &= \frac{m_{3/2} (10 \Theta _1+14 \Theta _2+5 (7 \Theta _3+\Theta _4))}{18 \sqrt{3}},\nonumber \\
M_b &= \frac{m_{3/2} (4 \Theta _1-5 \Theta _4)}{3 \sqrt{3}},\nonumber \\
M_c &= \frac{m_{3/2} (4 \Theta _2+5 \Theta _3)}{3 \sqrt{3}}.
\end{align} 

Next, to compute the trilinear coupling and the sleptons mass-squared we require the angles, the differences of angles and their first and second order derivatives with respect to the moduli. In table~\ref{Angles20} we show the angles \eqref{eq:angle} made by the cycles wrapped by each stack of D6-branes with respect to the orientifold plane on each two-torus.
The differences of the angles, $\theta_{xy}^{i}= \theta_{y}^{i} -\theta_{x}^{i}$ are,
\begin{align}\label{angle-diff20}\arraycolsep=0pt
\hskip -1em \left[
\begin{array}{ccc}
 \{0.,0.,0.\} & \{-0.12001,0.721058,-0.317863\} & \{0.309193,0.291855,-0.317863\} \\
 \{0.12001,-0.721058,0.317863\} & \{0.,0.,0.\} & \{0.429204,-0.429204,0.\} \\
 \{-0.309193,-0.291855,0.317863\} & \{-0.429204,0.429204,0.\} & \{0.,0.,0.\} \\
\end{array}
\right]
\end{align}

To account for the negative angle differences we employ the sign function $\sigma_{xy}^{i}$, which is $-1$ only for negative angle difference and $+1$ otherwise,
\begin{align}\label{sigmaK20}
\sigma_{xy}^{i} & = \left(
\begin{array}{ccc}
 \{1,1,1\} & \{-1,1,-1\} & \{1,1,-1\} \\
 \{1,-1,1\} & \{1,1,1\} & \{1,-1,1\} \\
 \{-1,-1,1\} & \{-1,1,1\} & \{1,1,1\} \\
\end{array}
\right),
\end{align}
and the function $\eta_{xy}$ is evaluated by taking the product on the torus index $i$ as,
\begin{equation}\label{eta20}
\eta_{xy} = \left(
\begin{array}{ccc}
 1 & 1 & -1 \\
 -1 & 1 & -1 \\
 1 & -1 & 1 \\
\end{array}
\right).
\end{equation}
Using the values of $\sigma_{xy}^{i}$ and $\eta_{xy}$ in \eqref{Psi} we can compute the four cases of functions $\Psi(\theta_{xy})$ defined in \eqref{eqn:Psi1} and \eqref{eqn:Psi2}. Similarly we calculate the derivative $\Psi'(\theta^j_{xy}) = \frac{d\Psi(\theta^j_{xy})}{d \theta^j_{xy}}$ using equations \eqref{DPsi}, \eqref{derivative-angles1}, \eqref{derivative-angles2} and the properties of digamma function $\psi^{(0)}(z)$ \eqref{eq:property} while neglecting the contribution of the $t$-moduli. 

Utilizing above results while ignoring the CP-violating phases $\gamma_m$, the gaugino masses; the trilinear coupling \eqref{tri-coupling}; and the squared-masses of squarks and sleptons \eqref{slepton-mass} are obtained as,
\begin{align}\label{GauginosYba123-model20}  
M_{\tilde B} &\equiv M_Y = m_{3/2}\Bigl(\frac{20 \Theta _1+112 \Theta _2+175 \Theta _3+10 \Theta _4}{99 \sqrt{3}}\Bigr), \nonumber\\
M_{\tilde W} &\equiv M_b = m_{3/2}\Bigl(\frac{4 \Theta _1-5 \Theta _4}{3 \sqrt{3}}\Bigr),\nonumber \\
M_{\tilde g} &\equiv M_a = m_{3/2}\Bigl(\frac{10 \Theta _1+14 \Theta _2+5 (7 \Theta _3+\Theta _4)}{18 \sqrt{3}}\Bigr),\nonumber \\
A_0 \equiv A_{abc} &= m_{3/2}\Bigl(-0.152188 \Theta _1-1.57986 \Theta _2+0.282383 \Theta _3-0.282383 \Theta _4\Bigr),\nonumber\\  
m^2_{L} \equiv m^2_{ab} &= m_{3/2}^2\Bigl(-0.399003 \Theta _1{}^2+1.25156 \Theta _1 \Theta _2-0.666067 \Theta _1 \Theta _3-0.958492 \Theta _1 \Theta _4 \nonumber \\   &\quad +0.106172 \Theta _2{}^2+0.163812 \Theta _2 \Theta _3-0.73081 \Theta _2 \Theta _4-2.07593 \Theta _3{}^2 \nonumber \\  &\quad +1.71861 \Theta _3 \Theta _4-0.689841 \Theta _4{}^2+1\Bigr) ,\nonumber\\  
m^2_{R} \equiv m^2_{ac} &= m_{3/2}^2\Bigl(-2.35521 \Theta _1{}^2-1.21594 \Theta _1 \Theta _2+0.483768 \Theta _1 \Theta _3+1.51111 \Theta _1 \Theta _4 \nonumber \\  &\quad  -0.387583 \Theta _2{}^2+1.12751 \Theta _2 \Theta _3+0.178197 \Theta _2 \Theta _4+0.0690201 \Theta _3{}^2 \nonumber \\  &\quad -1.68298 \Theta _3 \Theta _4-0.338871 \Theta _4{}^2+1\Bigr). 
\end{align}
All soft terms are subject to the constraint \eqref{constraint}.
\FloatBarrier

\subsection{Model 20-dual}\label{sec:model-20.5}
\begin{figure}[htb]
\centering
\includegraphics[width=\textwidth]{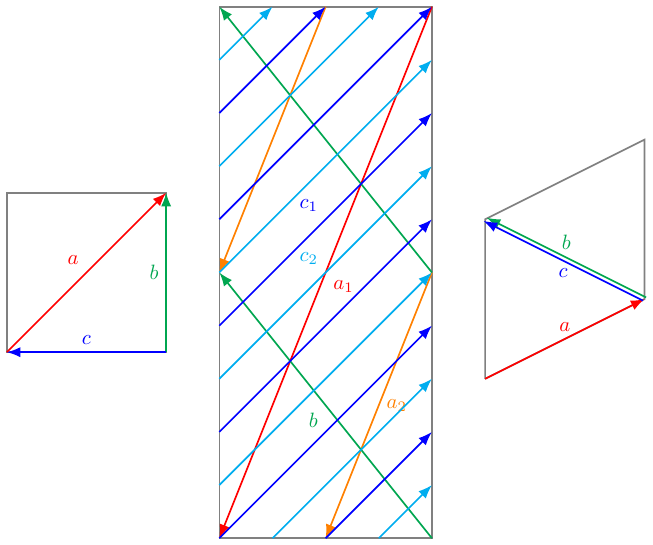}
\caption{Brane configuration for the three two-tori in Model~\hyperref[model20.5]{20-dual} where the third two-torus is tilted. Fermion mass hierarchies result from the intersections on the second two-torus.}  \label{Fig.20.5}
\end{figure}

In Model~\hyperref[model20.5]{20-dual} the three-point Yukawa couplings arise from the triplet intersections from the branes $a$, $b$ and $c$ on the second two-torus with 9 pairs of Higgs from the $\mathcal{N}=2$ sector.
Yukawa matrices for the Model~\hyperref[model20.5]{20-dual} are of rank 3 and the three intersections required to form the disk diagrams for the Yukawa couplings all occur on the second torus as shown in figure~\ref{Fig.20.5}.  
\begin{table}[htb]\footnotesize\centering
\renewcommand{\arraystretch}{1.8}
$\begin{array}{|c|c|c|c|}\hline
 \text{\hyperref[model20.5]{20-dual}} & \theta ^1 & \theta ^2 & \theta ^3 \\\hline
 a & \tan ^{-1}\left(\frac{7}{\sqrt{5}}\right) & \frac{\pi }{2} & 0 \\\hline
 b & \tan ^{-1}\left(\sqrt{5}\right) & -\tan ^{-1}\left(\frac{\sqrt{5}}{2}\right) & \tan ^{-1}\left(\frac{2}{\sqrt{5}}\right) \\\hline
 c & \tan ^{-1}\left(\frac{2}{\sqrt{5}}\right) & -\tan ^{-1}\left(\frac{2}{\sqrt{5}}\right) & -\tan ^{-1}\left(\frac{2}{\sqrt{5}}\right) \\\hline
\end{array}$
\caption{The angles with respect to the orientifold plane made by the cycle wrapped by stack of D6-branes on each of the three two-tori in Model~\hyperref[model20.5]{20-dual}.}
\label{Angles20.5}
\end{table}

The complex structure moduli $U^i$ \eqref{U-moduli} are given as,
\begin{align}\label{eqn:U-moduli20.5}
\{U^1,U^2,U^3\} &=\left\{\frac{7 i}{\sqrt{5}},i \sqrt{5},\frac{4}{9} \left(2+i \sqrt{5}\right)\right\},
\end{align}
and the corresponding $u$-moduli and $s$-modulus in supergravity basis from \eqref{eq:moduli} are,
\begin{align}\label{s-u-moduli20.5}
\{u^1,u^2,u^3\} & = \left\{\frac{\sqrt[4]{5} e^{-\phi _4}}{\sqrt{7} \pi },\frac{\sqrt{7} e^{-\phi _4}}{5^{3/4} \pi },\frac{\sqrt[4]{5} \sqrt{7} e^{-\phi _4}}{4 \pi }\right\}, \nonumber\\
s & = \frac{\sqrt[4]{5} e^{-\phi _4}}{4 \sqrt{7} \pi } .
\end{align}
Using \eqref{kingauagefun} and the values from the table~\ref{model20.5}, the gauge kinetic function becomes,
\begin{align}\label{fx20.5}
\{f_a,f_b,f_c\} & = \left\{\frac{27 e^{-\phi _4}}{8\ 5^{3/4} \sqrt{7} \pi },\frac{9 \sqrt{7} e^{-\phi _4}}{16\ 5^{3/4} \pi },\frac{9 \sqrt[4]{5} e^{-\phi _4}}{16 \sqrt{7} \pi }\right\},
\end{align}
To calculate the gaugino masses $\{M_Y,M_b,M_a\}$ for the respective gauge groups $\U(1)_Y$, $\SU(2)_L$, and $\SU(3)_C$, we first compute $\{M_a,M_b,M_c\}$ using \eqref{gaugino-masses} as,
\begin{align}\label{Gauginosabc20.5}
M_a &= \frac{m_{3/2} (10 \Theta _1+14 \Theta _2+5 (7 \Theta _3+\Theta _4))}{18 \sqrt{3}},\nonumber \\
M_b &= \frac{m_{3/2} (4 \Theta _2+5 \Theta _3)}{3 \sqrt{3}},\nonumber \\
M_c &= \frac{m_{3/2} (4 \Theta _1-5 \Theta _4)}{3 \sqrt{3}}.
\end{align} 

Next, to compute the trilinear coupling and the sleptons mass-squared we require the angles, the differences of angles and their first and second order derivatives with respect to the moduli. In table~\ref{Angles20.5} we show the angles \eqref{eq:angle} made by the cycles wrapped by each stack of D6-branes with respect to the orientifold plane on each two-torus.
The differences of the angles, $\theta_{xy}^{i}= \theta_{y}^{i} -\theta_{x}^{i}$ are,
\begin{align}\label{angle-diff20.5}\arraycolsep=0pt
\hskip -1em \left[
\begin{array}{ccc}
 \{0.,0.,0.\} & \{0.309193,0.291855,-0.317863\} & \{-0.12001,0.721058,-0.317863\} \\
 \{-0.309193,-0.291855,0.317863\} & \{0.,0.,0.\} & \{-0.429204,0.429204,0.\} \\
 \{0.12001,-0.721058,0.317863\} & \{0.429204,-0.429204,0.\} & \{0.,0.,0.\} \\
\end{array}
\right]
\end{align}

To account for the negative angle differences we employ the sign function $\sigma_{xy}^{i}$, which is $-1$ only for negative angle difference and $+1$ otherwise,
\begin{align}\label{sigmaK20.5}
\sigma_{xy}^{i} & = \left(
\begin{array}{ccc}
 \{1,1,1\} & \{1,1,-1\} & \{-1,1,-1\} \\
 \{-1,-1,1\} & \{1,1,1\} & \{-1,1,1\} \\
 \{1,-1,1\} & \{1,-1,1\} & \{1,1,1\} \\
\end{array}
\right),
\end{align}
and the function $\eta_{xy}$ is evaluated by taking the product on the torus index $i$ as,
\begin{equation}\label{eta20.5}
\eta_{xy} = \left(
\begin{array}{ccc}
 1 & -1 & 1 \\
 1 & 1 & -1 \\
 -1 & -1 & 1 \\
\end{array}
\right).
\end{equation}
Using the values of $\sigma_{xy}^{i}$ and $\eta_{xy}$ in \eqref{Psi} we can compute the four cases of functions $\Psi(\theta_{xy})$ defined in \eqref{eqn:Psi1} and \eqref{eqn:Psi2}. Similarly we calculate the derivative $\Psi'(\theta^j_{xy}) = \frac{d\Psi(\theta^j_{xy})}{d \theta^j_{xy}}$ using equations \eqref{DPsi}, \eqref{derivative-angles1}, \eqref{derivative-angles2} and the properties of digamma function $\psi^{(0)}(z)$ \eqref{eq:property} while neglecting the contribution of the $t$-moduli. 

Utilizing above results while ignoring the CP-violating phases $\gamma_m$, the gaugino masses; the trilinear coupling \eqref{tri-coupling}; and the squared-masses of squarks and sleptons \eqref{slepton-mass} are obtained as,
\begin{align}\label{GauginosYba123-model20.5}  
M_{\tilde B} &\equiv M_Y = m_{3/2}\Bigl(\frac{80 \Theta _1+28 \Theta _2+70 \Theta _3-65 \Theta _4}{81 \sqrt{3}}\Bigr), \nonumber\\
M_{\tilde W} &\equiv M_b = m_{3/2}\Bigl(\frac{4 \Theta _2+5 \Theta _3}{3 \sqrt{3}}\Bigr),\nonumber \\
M_{\tilde g} &\equiv M_a = m_{3/2}\Bigl(\frac{10 \Theta _1+14 \Theta _2+5 (7 \Theta _3+\Theta _4)}{18 \sqrt{3}}\Bigr),\nonumber \\
A_0 \equiv A_{abc} &= m_{3/2}\Bigl(-0.152188 \Theta _1-1.57986 \Theta _2+0.282383 \Theta _3-0.282383 \Theta _4\Bigr),\nonumber\\  
m^2_{L} \equiv m^2_{ab} &= m_{3/2}^2\Bigl(-2.35521 \Theta _1{}^2-1.21594 \Theta _1 \Theta _2+0.483768 \Theta _1 \Theta _3+1.51111 \Theta _1 \Theta _4 \nonumber \\   &\quad -0.387583 \Theta _2{}^2+1.12751 \Theta _2 \Theta _3+0.178197 \Theta _2 \Theta _4+0.0690201 \Theta _3{}^2 \nonumber \\  &\quad -1.68298 \Theta _3 \Theta _4-0.338871 \Theta _4{}^2+1\Bigr) ,\nonumber\\  
m^2_{R} \equiv m^2_{ac} &= m_{3/2}^2\Bigl(-0.399003 \Theta _1{}^2+1.25156 \Theta _1 \Theta _2-0.666067 \Theta _1 \Theta _3-0.958492 \Theta _1 \Theta _4 \nonumber \\  &\quad  +0.106172 \Theta _2{}^2+0.163812 \Theta _2 \Theta _3-0.73081 \Theta _2 \Theta _4-2.07593 \Theta _3{}^2 \nonumber \\  &\quad +1.71861 \Theta _3 \Theta _4-0.689841 \Theta _4{}^2+1\Bigr). 
\end{align}
All soft terms are subject to the constraint \eqref{constraint}.
\FloatBarrier

\subsection{Model 21}\label{sec:model-21}
\begin{figure}[htb]
\centering
\includegraphics[width=\textwidth]{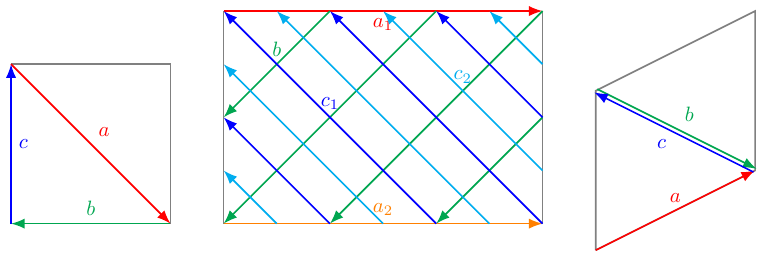}
\caption{Brane configuration for the three two-tori in Model~\hyperref[model21]{21} where the third two-torus is tilted. Fermion mass hierarchies result from the intersections on the second two-torus.}  \label{Fig.21}
\end{figure}

In Model~\hyperref[model21]{21} the three-point Yukawa couplings arise from the triplet intersections from the branes $a$, $b$ and $c$ on the second two-torus with 12 pairs of Higgs from the $\mathcal{N}=2$ sector.
Yukawa matrices for the Model~\hyperref[model21]{21} are of rank 3 and the three intersections required to form the disk diagrams for the Yukawa couplings all occur on the second torus as shown in figure~\ref{Fig.21}.  
\begin{table}[htb]\footnotesize\centering
\renewcommand{\arraystretch}{1.8}
$\begin{array}{|c|c|c|c|}\hline
 \text{\hyperref[model21]{21}} & \theta ^1 & \theta ^2 & \theta ^3 \\\hline
 a & \frac{3 \pi }{4} & 0 & \frac{\pi }{2} \\\hline
 b & 0 & \frac{\pi }{4} & \frac{3 \pi }{4} \\\hline
 c & \frac{\pi }{4} & \frac{3 \pi }{4} & \frac{3 \pi }{4} \\\hline
\end{array}$
\caption{The angles with respect to the orientifold plane made by the cycle wrapped by stack of D6-branes on each of the three two-tori in Model~\hyperref[model21]{21}.}
\label{Angles21}
\end{table}

The complex structure moduli $U^i$ \eqref{U-moduli} are given as,
\begin{align}\label{eqn:U-moduli21}
\{U^1,U^2,U^3\} &=\left\{i,\frac{2 i}{3},1+i\right\},
\end{align}
and the corresponding $u$-moduli and $s$-modulus in supergravity basis from \eqref{eq:moduli} are,
\begin{align}\label{s-u-moduli21}
\{u^1,u^2,u^3\} & = \left\{\frac{e^{-\phi _4}}{\sqrt{3} \pi },\frac{\sqrt{3} e^{-\phi _4}}{2 \pi },\frac{e^{-\phi _4}}{2 \sqrt{3} \pi }\right\}, \nonumber\\
s & = \frac{\sqrt{3} e^{-\phi _4}}{4 \pi } .
\end{align}
Using \eqref{kingauagefun} and the values from the table~\ref{model21}, the gauge kinetic function becomes,
\begin{align}\label{fx21}
\{f_a,f_b,f_c\} & = \left\{\frac{\sqrt{3} e^{-\phi _4}}{8 \pi },\frac{\sqrt{3} e^{-\phi _4}}{4 \pi },\frac{\sqrt{3} e^{-\phi _4}}{4 \pi }\right\},
\end{align}
To calculate the gaugino masses $\{M_Y,M_b,M_a\}$ for the respective gauge groups $\U(1)_Y$, $\SU(2)_L$, and $\SU(3)_C$, we first compute $\{M_a,M_b,M_c\}$ using \eqref{gaugino-masses} as,
\begin{align}\label{Gauginosabc21}
M_a &= \frac{1}{2} \sqrt{3} m_{3/2} (\Theta _2-\Theta _4),\nonumber \\
M_b &= \frac{1}{2} \sqrt{3} m_{3/2} (\Theta _1-\Theta _4),\nonumber \\
M_c &= \frac{1}{2} \sqrt{3} m_{3/2} (\Theta _2+\Theta _3).
\end{align} 

Next, to compute the trilinear coupling and the sleptons mass-squared we require the angles, the differences of angles and their first and second order derivatives with respect to the moduli. In table~\ref{Angles21} we show the angles \eqref{eq:angle} made by the cycles wrapped by each stack of D6-branes with respect to the orientifold plane on each two-torus.
The differences of the angles, $\theta_{xy}^{i}= \theta_{y}^{i} -\theta_{x}^{i}$ are,
\begin{align}\label{angle-diff21}\arraycolsep=0pt
\hskip -1em \left[
\begin{array}{ccc}
 \{0.,0.,0.\} & \{-0.0730092,0.643806,-0.570796\} & \{0.356194,0.356194,-0.429204\} \\
 \{0.0730092,-0.643806,0.570796\} & \{0.,0.,0.\} & \{0.429204,-0.287611,0.141593\} \\
 \{-0.356194,-0.356194,0.429204\} & \{-0.429204,0.287611,-0.141593\} & \{0.,0.,0.\} \\
\end{array}
\right]
\end{align}

To account for the negative angle differences we employ the sign function $\sigma_{xy}^{i}$, which is $-1$ only for negative angle difference and $+1$ otherwise,
\begin{align}\label{sigmaK21}
\sigma_{xy}^{i} & = \left(
\begin{array}{ccc}
 \{1,1,1\} & \{-1,1,-1\} & \{1,1,-1\} \\
 \{1,-1,1\} & \{1,1,1\} & \{1,-1,1\} \\
 \{-1,-1,1\} & \{-1,1,-1\} & \{1,1,1\} \\
\end{array}
\right),
\end{align}
and the function $\eta_{xy}$ is evaluated by taking the product on the torus index $i$ as,
\begin{equation}\label{eta21}
\eta_{xy} = \left(
\begin{array}{ccc}
 1 & 1 & -1 \\
 -1 & 1 & -1 \\
 1 & 1 & 1 \\
\end{array}
\right).
\end{equation}
Using the values of $\sigma_{xy}^{i}$ and $\eta_{xy}$ in \eqref{Psi} we can compute the four cases of functions $\Psi(\theta_{xy})$ defined in \eqref{eqn:Psi1} and \eqref{eqn:Psi2}. Similarly we calculate the derivative $\Psi'(\theta^j_{xy}) = \frac{d\Psi(\theta^j_{xy})}{d \theta^j_{xy}}$ using equations \eqref{DPsi}, \eqref{derivative-angles1}, \eqref{derivative-angles2} and the properties of digamma function $\psi^{(0)}(z)$ \eqref{eq:property} while neglecting the contribution of the $t$-moduli. 

Utilizing above results while ignoring the CP-violating phases $\gamma_m$, the gaugino masses; the trilinear coupling \eqref{tri-coupling}; and the squared-masses of squarks and sleptons \eqref{slepton-mass} are obtained as,
\begin{align}\label{GauginosYba123-model21}  
M_{\tilde B} &\equiv M_Y = m_{3/2}\Bigl(\frac{1}{8} \sqrt{3} (4 \Theta _2+3 \Theta _3-\Theta _4)\Bigr), \nonumber\\
M_{\tilde W} &\equiv M_b = m_{3/2}\Bigl(\frac{1}{2} \sqrt{3} (\Theta _1-\Theta _4)\Bigr),\nonumber \\
M_{\tilde g} &\equiv M_a = m_{3/2}\Bigl(\frac{1}{2} \sqrt{3} (\Theta _2-\Theta _4)\Bigr),\nonumber \\
A_0 \equiv A_{abc} &= m_{3/2}\Bigl(-0.444661 \Theta _1-1.36219 \Theta _2+0.260871 \Theta _3-0.186075 \Theta _4\Bigr),\nonumber\\  
m^2_{L} \equiv m^2_{ab} &= m_{3/2}^2\Bigl(-0.0689562 \Theta _1{}^2+1.52807 \Theta _1 \Theta _2-0.25782 \Theta _1 \Theta _3-0.582907 \Theta _1 \Theta _4 \nonumber \\   &\quad -0.127855 \Theta _2{}^2+0.31395 \Theta _2 \Theta _3-0.623439 \Theta _2 \Theta _4-2.04994 \Theta _3{}^2 \nonumber \\  &\quad +1.23441 \Theta _3 \Theta _4-0.809377 \Theta _4{}^2+1\Bigr) ,\nonumber\\  
m^2_{R} \equiv m^2_{ac} &= m_{3/2}^2\Bigl(-2.28519 \Theta _1{}^2-1.00169 \Theta _1 \Theta _2+0.506454 \Theta _1 \Theta _3+1.26634 \Theta _1 \Theta _4 \nonumber \\  &\quad  -0.637087 \Theta _2{}^2+1.13601 \Theta _2 \Theta _3+0.140834 \Theta _2 \Theta _4-0.0262168 \Theta _3{}^2 \nonumber \\  &\quad -1.1866 \Theta _3 \Theta _4-0.492648 \Theta _4{}^2+1\Bigr). 
\end{align}
All soft terms are subject to the constraint \eqref{constraint}.
\FloatBarrier

\subsection{Model 22}\label{sec:model-22}
\begin{figure}[htb]
\centering
\includegraphics[width=\textwidth]{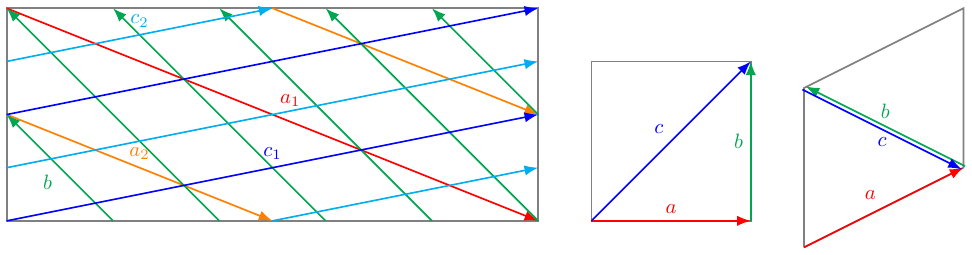}
\caption{Brane configuration for the three two-tori in Model~\hyperref[model22]{22} where the third two-torus is tilted. Fermion mass hierarchies result from the intersections on the first two-torus.}  \label{Fig.22}
\end{figure}

In Model~\hyperref[model22]{22} the three-point Yukawa couplings arise from the triplet intersections from the branes $a$, $b$ and $c$ on the first two-torus with 12 pairs of Higgs from the $\mathcal{N}=2$ sector.
Yukawa matrices for the Model~\hyperref[model22]{22} are of rank 3 and the three intersections required to form the disk diagrams for the Yukawa couplings all occur on the first torus as shown in figure~\ref{Fig.22}.  
\begin{table}[htb]\footnotesize\centering
\renewcommand{\arraystretch}{1.8}
$\begin{array}{|c|c|c|c|}\hline
 \text{\hyperref[model22]{22}} & \theta ^1 & \theta ^2 & \theta ^3 \\\hline
 a & -\tan ^{-1}\left(\sqrt{\frac{2}{5}}\right) & -\tan ^{-1}\left(\sqrt{\frac{5}{2}}\right) & \tan ^{-1}\left(\frac{1}{\sqrt{10}}\right) \\\hline
 b & 0 & \frac{\pi }{2} & \tan ^{-1}\left(\frac{\sqrt{\frac{5}{2}}}{6}\right) \\\hline
 c & \tan ^{-1}\left(\sqrt{\frac{2}{5}}\right) & -\tan ^{-1}\left(\sqrt{\frac{2}{5}}\right) & -\tan ^{-1}\left(\sqrt{\frac{2}{5}}\right) \\\hline
\end{array}$
\caption{The angles with respect to the orientifold plane made by the cycle wrapped by stack of D6-branes on each of the three two-tori in Model~\hyperref[model22]{22}.}
\label{Angles22}
\end{table}

The complex structure moduli $U^i$ \eqref{U-moduli} are given as,
\begin{align}\label{eqn:U-moduli22}
\{U^1,U^2,U^3\} &=\left\{i \sqrt{\frac{2}{5}},\frac{1}{6} i \sqrt{\frac{5}{2}},\frac{2}{7} \left(2+i \sqrt{10}\right)\right\},
\end{align}
and the corresponding $u$-moduli and $s$-modulus in supergravity basis from \eqref{eq:moduli} are,
\begin{align}\label{s-u-moduli22}
\{u^1,u^2,u^3\} & = \left\{\frac{\sqrt[4]{\frac{5}{2}} e^{-\phi _4}}{2 \sqrt{3} \pi },\frac{2^{3/4} \sqrt{3} e^{-\phi _4}}{5^{3/4} \pi },\frac{\sqrt[4]{\frac{5}{2}} e^{-\phi _4}}{4 \sqrt{3} \pi }\right\}, \nonumber\\
s & = \frac{\sqrt[4]{\frac{5}{2}} \sqrt{3} e^{-\phi _4}}{2 \pi } .
\end{align}
Using \eqref{kingauagefun} and the values from the table~\ref{model22}, the gauge kinetic function becomes,
\begin{align}\label{fx22}
\{f_a,f_b,f_c\} & = \left\{\frac{7 \sqrt{3} e^{-\phi _4}}{8 \sqrt[4]{2} 5^{3/4} \pi },\frac{7 \sqrt[4]{\frac{5}{2}} e^{-\phi _4}}{16 \sqrt{3} \pi },\frac{77 e^{-\phi _4}}{16 \sqrt[4]{2} \sqrt{3} 5^{3/4} \pi }\right\},
\end{align}
To calculate the gaugino masses $\{M_Y,M_b,M_a\}$ for the respective gauge groups $\U(1)_Y$, $\SU(2)_L$, and $\SU(3)_C$, we first compute $\{M_a,M_b,M_c\}$ using \eqref{gaugino-masses} as,
\begin{align}\label{Gauginosabc22}
M_a &= \frac{1}{7} \sqrt{3} m_{3/2} (2 \Theta _2-5 \Theta _4),\nonumber \\
M_b &= \frac{1}{7} \sqrt{3} m_{3/2} (2 \Theta _1+5 \Theta _3),\nonumber \\
M_c &= \frac{1}{77} \sqrt{3} m_{3/2} (10 \Theta _1+12 \Theta _2-5 (\Theta _3+12 \Theta _4)).
\end{align} 

Next, to compute the trilinear coupling and the sleptons mass-squared we require the angles, the differences of angles and their first and second order derivatives with respect to the moduli. In table~\ref{Angles22} we show the angles \eqref{eq:angle} made by the cycles wrapped by each stack of D6-branes with respect to the orientifold plane on each two-torus.
The differences of the angles, $\theta_{xy}^{i}= \theta_{y}^{i} -\theta_{x}^{i}$ are,
\begin{align}\label{angle-diff22}\arraycolsep=0pt
\hskip -1em \left[
\begin{array}{ccc}
 \{0.,0.,0.\} & \{-0.301318,0.570796,0.0137074\} & \{-0.12978,0.257665,-0.127885\} \\
 \{0.301318,-0.570796,-0.0137074\} & \{0.,0.,0.\} & \{0.171538,-0.313131,-0.141593\} \\
 \{0.12978,-0.257665,0.127885\} & \{-0.171538,0.313131,0.141593\} & \{0.,0.,0.\} \\
\end{array}
\right]
\end{align}

To account for the negative angle differences we employ the sign function $\sigma_{xy}^{i}$, which is $-1$ only for negative angle difference and $+1$ otherwise,
\begin{align}\label{sigmaK22}
\sigma_{xy}^{i} & = \left(
\begin{array}{ccc}
 \{1,1,1\} & \{-1,1,1\} & \{-1,1,-1\} \\
 \{1,-1,-1\} & \{1,1,1\} & \{1,-1,-1\} \\
 \{1,-1,1\} & \{-1,1,1\} & \{1,1,1\} \\
\end{array}
\right),
\end{align}
and the function $\eta_{xy}$ is evaluated by taking the product on the torus index $i$ as,
\begin{equation}\label{eta22}
\eta_{xy} = \left(
\begin{array}{ccc}
 1 & -1 & 1 \\
 1 & 1 & 1 \\
 -1 & -1 & 1 \\
\end{array}
\right).
\end{equation}
Using the values of $\sigma_{xy}^{i}$ and $\eta_{xy}$ in \eqref{Psi} we can compute the four cases of functions $\Psi(\theta_{xy})$ defined in \eqref{eqn:Psi1} and \eqref{eqn:Psi2}. Similarly we calculate the derivative $\Psi'(\theta^j_{xy}) = \frac{d\Psi(\theta^j_{xy})}{d \theta^j_{xy}}$ using equations \eqref{DPsi}, \eqref{derivative-angles1}, \eqref{derivative-angles2} and the properties of digamma function $\psi^{(0)}(z)$ \eqref{eq:property} while neglecting the contribution of the $t$-moduli. 

Utilizing above results while ignoring the CP-violating phases $\gamma_m$, the gaugino masses; the trilinear coupling \eqref{tri-coupling}; and the squared-masses of squarks and sleptons \eqref{slepton-mass} are obtained as,
\begin{align}\label{GauginosYba123-model22}  
M_{\tilde B} &\equiv M_Y = m_{3/2}\Bigl(\frac{2 \Theta _1+4 \Theta _2-\Theta _3-16 \Theta _4}{7 \sqrt{3}}\Bigr), \nonumber\\
M_{\tilde W} &\equiv M_b = m_{3/2}\Bigl(\frac{1}{7} \sqrt{3} (2 \Theta _1+5 \Theta _3)\Bigr),\nonumber \\
M_{\tilde g} &\equiv M_a = m_{3/2}\Bigl(\frac{1}{7} \sqrt{3} (2 \Theta _2-5 \Theta _4)\Bigr),\nonumber \\
A_0 \equiv A_{abc} &= m_{3/2}\Bigl(-1.83963 \Theta _1-1.52115 \Theta _2+0.246437 \Theta _3+1.38229 \Theta _4\Bigr),\nonumber\\  
m^2_{L} \equiv m^2_{ab} &= m_{3/2}^2\Bigl(1.01495 \Theta _1{}^2-0.356479 \Theta _1 \Theta _2-1.50575 \Theta _1 \Theta _3-0.924945 \Theta _1 \Theta _4 \nonumber \\   &\quad +0.41741 \Theta _2{}^2-1.65173 \Theta _2 \Theta _3-1.24068 \Theta _2 \Theta _4-0.642555 \Theta _3{}^2 \nonumber \\  &\quad +0.586882 \Theta _3 \Theta _4-0.632323 \Theta _4{}^2+1\Bigr) ,\nonumber\\  
m^2_{R} \equiv m^2_{ac} &= m_{3/2}^2\Bigl(0.668817 \Theta _1{}^2+0.0838741 \Theta _1 \Theta _2-2.13002 \Theta _1 \Theta _3-0.767052 \Theta _1 \Theta _4 \nonumber \\  &\quad  +0.714726 \Theta _2{}^2-0.702234 \Theta _2 \Theta _3-2.08352 \Theta _2 \Theta _4+0.213147 \Theta _3{}^2 \nonumber \\  &\quad +1.43445 \Theta _3 \Theta _4-1.76444 \Theta _4{}^2+1\Bigr). 
\end{align}
All soft terms are subject to the constraint \eqref{constraint}.
\FloatBarrier

\subsection{Model 22-dual}\label{sec:model-22.5}
\begin{figure}[htb]
\centering
\includegraphics[width=\textwidth]{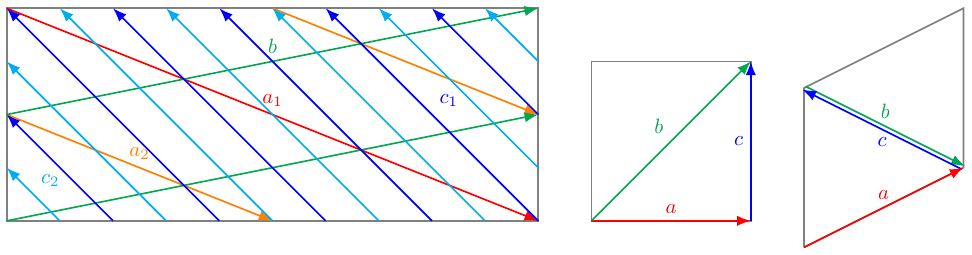}
\caption{Brane configuration for the three two-tori in Model~\hyperref[model22.5]{22-dual} where the third two-torus is tilted. Fermion mass hierarchies result from the intersections on the first two-torus.}  \label{Fig.22.5}
\end{figure}

In Model~\hyperref[model22.5]{22-dual} the three-point Yukawa couplings arise from the triplet intersections from the branes $a$, $b$ and $c$ on the first two-torus with 12 pairs of Higgs from the $\mathcal{N}=2$ sector.
Yukawa matrices for the Model~\hyperref[model22.5]{22-dual} are of rank 3 and the three intersections required to form the disk diagrams for the Yukawa couplings all occur on the first torus as shown in figure~\ref{Fig.22.5}.  
\begin{table}[htb]\footnotesize\centering
\renewcommand{\arraystretch}{1.8}
$\begin{array}{|c|c|c|c|}\hline
 \text{\hyperref[model22.5]{22-dual}} & \theta ^1 & \theta ^2 & \theta ^3 \\\hline
 a & -\tan ^{-1}\left(\sqrt{\frac{2}{5}}\right) & \tan ^{-1}\left(\frac{1}{\sqrt{10}}\right) & -\tan ^{-1}\left(\sqrt{\frac{5}{2}}\right) \\\hline
 b & 0 & \tan ^{-1}\left(\frac{\sqrt{\frac{5}{2}}}{6}\right) & \frac{\pi }{2} \\\hline
 c & \tan ^{-1}\left(\sqrt{\frac{2}{5}}\right) & -\tan ^{-1}\left(\sqrt{\frac{2}{5}}\right) & -\tan ^{-1}\left(\sqrt{\frac{2}{5}}\right) \\\hline
\end{array}$
\caption{The angles with respect to the orientifold plane made by the cycle wrapped by stack of D6-branes on each of the three two-tori in Model~\hyperref[model22.5]{22-dual}.}
\label{Angles22.5}
\end{table}

The complex structure moduli $U^i$ \eqref{U-moduli} are given as,
\begin{align}\label{eqn:U-moduli22.5}
\{U^1,U^2,U^3\} &=\left\{i \sqrt{\frac{2}{5}},\frac{1}{6} i \sqrt{\frac{5}{2}},\frac{2}{7} \left(2+i \sqrt{10}\right)\right\},
\end{align}
and the corresponding $u$-moduli and $s$-modulus in supergravity basis from \eqref{eq:moduli} are,
\begin{align}\label{s-u-moduli22.5}
\{u^1,u^2,u^3\} & = \left\{\frac{\sqrt[4]{\frac{5}{2}} e^{-\phi _4}}{2 \sqrt{3} \pi },\frac{2^{3/4} \sqrt{3} e^{-\phi _4}}{5^{3/4} \pi },\frac{\sqrt[4]{\frac{5}{2}} e^{-\phi _4}}{4 \sqrt{3} \pi }\right\}, \nonumber\\
s & = \frac{\sqrt[4]{\frac{5}{2}} \sqrt{3} e^{-\phi _4}}{2 \pi } .
\end{align}
Using \eqref{kingauagefun} and the values from the table~\ref{model22.5}, the gauge kinetic function becomes,
\begin{align}\label{fx22.5}
\{f_a,f_b,f_c\} & = \left\{\frac{7 \sqrt{3} e^{-\phi _4}}{8 \sqrt[4]{2} 5^{3/4} \pi },\frac{77 e^{-\phi _4}}{16 \sqrt[4]{2} \sqrt{3} 5^{3/4} \pi },\frac{7 \sqrt[4]{\frac{5}{2}} e^{-\phi _4}}{16 \sqrt{3} \pi }\right\},
\end{align}
To calculate the gaugino masses $\{M_Y,M_b,M_a\}$ for the respective gauge groups $\U(1)_Y$, $\SU(2)_L$, and $\SU(3)_C$, we first compute $\{M_a,M_b,M_c\}$ using \eqref{gaugino-masses} as,
\begin{align}\label{Gauginosabc22.5}
M_a &= \frac{1}{7} \sqrt{3} m_{3/2} (2 \Theta _2-5 \Theta _4),\nonumber \\
M_b &= \frac{1}{77} \sqrt{3} m_{3/2} (10 \Theta _1+12 \Theta _2-5 (\Theta _3+12 \Theta _4)),\nonumber \\
M_c &= \frac{1}{7} \sqrt{3} m_{3/2} (2 \Theta _1+5 \Theta _3).
\end{align} 

Next, to compute the trilinear coupling and the sleptons mass-squared we require the angles, the differences of angles and their first and second order derivatives with respect to the moduli. In table~\ref{Angles22.5} we show the angles \eqref{eq:angle} made by the cycles wrapped by each stack of D6-branes with respect to the orientifold plane on each two-torus.
The differences of the angles, $\theta_{xy}^{i}= \theta_{y}^{i} -\theta_{x}^{i}$ are,
\begin{align}\label{angle-diff22.5}\arraycolsep=0pt
\hskip -1em \left[
\begin{array}{ccc}
 \{0.,0.,0.\} & \{-0.12978,0.257665,-0.127885\} & \{-0.301318,0.570796,0.0137074\} \\
 \{0.12978,-0.257665,0.127885\} & \{0.,0.,0.\} & \{-0.171538,0.313131,0.141593\} \\
 \{0.301318,-0.570796,-0.0137074\} & \{0.171538,-0.313131,-0.141593\} & \{0.,0.,0.\} \\
\end{array}
\right]
\end{align}

To account for the negative angle differences we employ the sign function $\sigma_{xy}^{i}$, which is $-1$ only for negative angle difference and $+1$ otherwise,
\begin{align}\label{sigmaK22.5}
\sigma_{xy}^{i} & = \left(
\begin{array}{ccc}
 \{1,1,1\} & \{-1,1,-1\} & \{-1,1,1\} \\
 \{1,-1,1\} & \{1,1,1\} & \{-1,1,1\} \\
 \{1,-1,-1\} & \{1,-1,-1\} & \{1,1,1\} \\
\end{array}
\right),
\end{align}
and the function $\eta_{xy}$ is evaluated by taking the product on the torus index $i$ as,
\begin{equation}\label{eta22.5}
\eta_{xy} = \left(
\begin{array}{ccc}
 1 & 1 & -1 \\
 -1 & 1 & -1 \\
 1 & 1 & 1 \\
\end{array}
\right).
\end{equation}
Using the values of $\sigma_{xy}^{i}$ and $\eta_{xy}$ in \eqref{Psi} we can compute the four cases of functions $\Psi(\theta_{xy})$ defined in \eqref{eqn:Psi1} and \eqref{eqn:Psi2}. Similarly we calculate the derivative $\Psi'(\theta^j_{xy}) = \frac{d\Psi(\theta^j_{xy})}{d \theta^j_{xy}}$ using equations \eqref{DPsi}, \eqref{derivative-angles1}, \eqref{derivative-angles2} and the properties of digamma function $\psi^{(0)}(z)$ \eqref{eq:property} while neglecting the contribution of the $t$-moduli. 

Utilizing above results while ignoring the CP-violating phases $\gamma_m$, the gaugino masses; the trilinear coupling \eqref{tri-coupling}; and the squared-masses of squarks and sleptons \eqref{slepton-mass} are obtained as,
\begin{align}\label{GauginosYba123-model22.5}  
M_{\tilde B} &\equiv M_Y = m_{3/2}\Bigl(\frac{10 \Theta _1+8 \Theta _2+25 \Theta _3-20 \Theta _4}{21 \sqrt{3}}\Bigr), \nonumber\\
M_{\tilde W} &\equiv M_b = m_{3/2}\Bigl(\frac{1}{77} \sqrt{3} (10 \Theta _1+12 \Theta _2-5 (\Theta _3+12 \Theta _4))\Bigr),\nonumber \\
M_{\tilde g} &\equiv M_a = m_{3/2}\Bigl(\frac{1}{7} \sqrt{3} (2 \Theta _2-5 \Theta _4)\Bigr),\nonumber \\
A_0 \equiv A_{abc} &= m_{3/2}\Bigl(-1.83963 \Theta _1-1.52115 \Theta _2+0.246437 \Theta _3+1.38229 \Theta _4\Bigr),\nonumber\\  
m^2_{L} \equiv m^2_{ab} &= m_{3/2}^2\Bigl(0.668817 \Theta _1{}^2+0.0838741 \Theta _1 \Theta _2-2.13002 \Theta _1 \Theta _3-0.767052 \Theta _1 \Theta _4 \nonumber \\   &\quad +0.714726 \Theta _2{}^2-0.702234 \Theta _2 \Theta _3-2.08352 \Theta _2 \Theta _4+0.213147 \Theta _3{}^2 \nonumber \\  &\quad +1.43445 \Theta _3 \Theta _4-1.76444 \Theta _4{}^2+1\Bigr) ,\nonumber\\  
m^2_{R} \equiv m^2_{ac} &= m_{3/2}^2\Bigl(1.01495 \Theta _1{}^2-0.356479 \Theta _1 \Theta _2-1.50575 \Theta _1 \Theta _3-0.924945 \Theta _1 \Theta _4 \nonumber \\  &\quad  +0.41741 \Theta _2{}^2-1.65173 \Theta _2 \Theta _3-1.24068 \Theta _2 \Theta _4-0.642555 \Theta _3{}^2 \nonumber \\  &\quad +0.586882 \Theta _3 \Theta _4-0.632323 \Theta _4{}^2+1\Bigr). 
\end{align}
All soft terms are subject to the constraint \eqref{constraint}.
\FloatBarrier

\section{Discussion and conclusion}\label{sec:conclusion}

We have studied the supersymmetry breaking soft terms for all the viable models in the complete landscape of three-family supersymmetric Pati-Salam models arising from intersecting D6-branes on a $\T^6/(\Z_2\times \Z_2)$ orientifold in type IIA string theory, comprising 33 independent models with distinct string-scale gauge coupling relations. It is found that only 17 models contain viable Yukawa textures. We have focused on the $u$-moduli dominated case together with the $s$-moduli turned on, where the soft terms remain independent of the Yukawa couplings and the Wilson lines. The results for the trilinear coupling, gaugino-masses, squared-mass parameters of squarks, sleptons and Higgs depend on the brane wrapping numbers as well as supersymmetry breaking parameters which include the gravitino-mass parameter $m_{3/2}$ and the goldstino angles $\Theta_{m}$, $m=1,2,3,4$. 

In particular for the case of dual models under the exchange of two SU(2) sectors, the Yukawa couplings remain unchanged as discussed in \cite{Sabir:2024cgt}, however, the corresponding soft term parameters only match for the trilinear coupling and the mass of the gluino. This can be explained by the internal geometry of the compact space where the Yukawa interactions depend only on the triangular area of the worldsheet instantons while the soft term parameters have an additional dependence on the orientation-angles of D6-branes in the three two-tori. Thus, the dual models besides having different gauge coupling relations also possess distinct masses of gauginos and squared-masses of squarks and sleptons. While it was already clear from the analysis of the Yukawa coupling \cite{Sabir:2024cgt} that the best performing models in the entire landscape are the models \hyperref[model22]{22} and \hyperref[model22.5]{22-dual}, the different result for susy breaking soft terms combined with experimental data for superpartners can in principle distinguish between the two models.

Our analysis reveals that the susy breaking soft terms are in general highly model dependent as evident from the results of section~\ref{sec:soft_terms}. Nonetheless, we have found a special limit under which the Higgs and the gaugino-masses in all of the models become degenerate. This limit corresponds to setting three goldstino angles to be equal to $1/2$ and taking dilaton angle to be $-1/2$, i.e. $\Theta_1 = \Theta_2 = \Theta_3 = -\Theta_s = 1/2$, provided that all CP-violating phases $\gamma_i$ are set to zero. Note that the special limit satisfies the crucial constraint \eqref{constraint} on the values of $\Theta_i$. This results in the following universal masses for the gauginos (viz. bino, wino and gluino) and the Higgs,
\begin{gather}\label{eq:Universal}
m_{1/2} \equiv M_{\tilde{B}} = M_{\tilde{W}} = M_{\tilde{g}} = \frac{\sqrt{3}}{2} m_{3/2},\nonumber\\ 
m_H = \frac{m_{3/2}}{2} = \frac{m_{1/2}}{\sqrt{3}}.
\end{gather}
Unlike the gaugino masses, the trilinear coupling $A_0$ and the scalar squared-mass parameters of squarks and the sleptons, $m^2_{L}$ and $m^2_{R}$, do not exhibit such universality and in general depend on the specific model parameters. 

Given that the precise scale of supersymmetry remains uncertain, we defer further empirical investigations into supersymmetric partner spectra and the resultant neutralino relic density to future studies. It is noteworthy that string theory offers insights into the intricate phenomenology of superpartners through the wrapping numbers and the angles of intersecting D6-branes in the internal compact dimensions. However, additional experimental input is essential to fully leverage this mathematical framework.

\FloatBarrier

\acknowledgments{TL is supported in part by the National Key Research and Development Program of China Grant No. 2020YFC2201504, by the Projects No. 11875062, No. 11947302, No. 12047503, and No. 12275333 supported by the National Natural Science Foundation of China, by the Key Research Program of the Chinese Academy of Sciences, Grant No. XDPB15, by the Scientific Instrument Developing Project of the Chinese Academy of Sciences, Grant No. YJKYYQ20190049, and by the International Partnership Program of Chinese Academy of Sciences for Grand Challenges, Grant No. 112311KYSB20210012. Z.-W. Wang is supported in part by the hundred talented program at University of Electronic Science and Technology of China. AM is supported by the Guangdong Basic and Applied Basic Research Foundation (Grant No. 2021B1515130007), Shenzhen Natural Science Fund (the Stable Support Plan Program 20220810130956001).}

\appendix
\section{Independent supersymmetric Pati-Salam models}\label{appA}

We tabulate the $33$ independent models with distinct allowed gauge coupling relations in the three-family $\mathcal{N}=1$ supersymmetric Pati-Salam landscape arising from intersecting D6-branes on a type IIA $\T^6/(\Z_2\times \Z_2)$ orientifold \cite{He:2021gug, Sabir:2024cgt}. 

%%%%%%%%%%%%%%%%%%%%%%%%%%%%%%%%% Model 1 %%%%%%%%%%%%%%%%%%%%%%%%%%%%%%%%%%%%%%%%%%%%%%%%%%%%%%%

\begin{table}[ht]\footnotesize\centering 
$\begin{array}{|c|c|c|c|c|c|c|c|c|c|c|}
\hline\multicolumn{2}{|c|}{\text{Model~\hyperref[spec1]{1}}} & \multicolumn{9}{c|}{ {\U(4)_C\times \U(2)_L \times \U(2)_R\times \USp(4)_2\times \USp(4)_4} }\\
\hline\rm{stack} & N & (n^1,l^1)\times(n^2,l^2)\times(n^3,l^3) & n_{\yng(2)}& n_{\yng(1,1)_{}} & b & b' & c & c' & 2 &  4\\
\hline
a & 8 & (1, -1)\times (1, 0)\times (1, 1) & 0 & 0  & 3 & 0 & -3 & 0 & -1 &  1\\
b & 4 &  (0, 1)\times (-1, 3)\times (-1, 1) & 2 & -2  & \text{-} & \text{-} & 0 & -4 & 1 &  0\\
c & 4 &  (-2, -1)\times (-1, -1)\times (1, -1) & 2 & 6  & \text{-} & \text{-} & \text{-} & \text{-} & 1 &  -2\\
\hline
 &   &  & \multicolumn{8}{c|}{ x_A = \frac{1}{3}x_B = x_C = \frac{1}{3}x_D }\\
2 & 4 &  (1, 0)\times (0, -1)\times (0, 2) & \multicolumn{8}{c|}{ ~\beta^g_2=-2,~\beta^g_4=-2, }\\
 &   &   & \multicolumn{8}{c|}{ \chi_1=1 , ~ \chi_2=\frac{1}{3} , ~ \chi_3=2 }\\
4 & 4  &   (0, -1)\times (0, 1)\times (2, 0) & \multicolumn{8}{c|}{}\\
\hline  
\end{array}$
\caption{D6-brane configurations and intersection numbers of Model~\hyperref[spec1]{1}, and its MSSM gauge coupling relation is $g_a^2=g_b^2=\frac{5}{3}g_c^2=\frac{25}{19}\frac{5 g_Y^2}{3}=4 \sqrt{\frac{2}{3}} \, \pi \,  e^{\phi _4}$.}
\label{model1}
\end{table}

%%%%%%%%%%%%%%%%%%%%%%%%%%%%%%%%% Model 1-dual %%%%%%%%%%%%%%%%%%%%%%%%%%%%%%%%%%%%%%%%%%%%%%%%%%%%%%%

\begin{table}[ht]\footnotesize\centering 
$\begin{array}{|c|c|c|c|c|c|c|c|c|c|c|}
\hline\multicolumn{2}{|c|}{\text{Model~\hyperref[spec1.5]{1-dual}}} & \multicolumn{9}{c|}{ {\U(4)_C\times \U(2)_L \times \U(2)_R\times \USp(4)_2\times \USp(4)_4} }\\
\hline\rm{stack} & N & (n^1,l^1)\times(n^2,l^2)\times(n^3,l^3) & n_{\yng(2)}& n_{\yng(1,1)_{}} & b & b' & c & c' & 2 &  4\\
\hline
a & 8 & (1, -1)\times (1, 0)\times (1, 1) & 0 & 0  & -3 & 0 & 3 & 0 & -1 &  1\\
b & 4 &  (-2, -1)\times (-1, -1)\times (1, -1) & 2 & 6  & \text{-} & \text{-} & 0 & -4 & 1 &  -2\\
c & 4 &  (0, 1)\times (-1, 3)\times (-1, 1) & 2 & -2  & \text{-} & \text{-} & \text{-} & \text{-} & 1 &  0\\
\hline
 &   &  & \multicolumn{8}{c|}{ x_A = \frac{1}{3}x_B = x_C = \frac{1}{3}x_D }\\
2 & 4 &  (1, 0)\times (0, -1)\times (0, 2) & \multicolumn{8}{c|}{ ~\beta^g_2=-2,~\beta^g_4=-2, }\\
 &   &   & \multicolumn{8}{c|}{ \chi_1=1 , ~ \chi_2=\frac{1}{3} , ~ \chi_3=2 }\\
4 & 4  &   (0, -1)\times (0, 1)\times (2, 0) & \multicolumn{8}{c|}{}\\
\hline  
\end{array}$
\caption{D6-brane configurations and intersection numbers of Model~\hyperref[spec1.5]{1-dual}, and its MSSM gauge coupling relation is $g_a^2=\frac{5}{3}g_b^2=g_c^2=\frac{5 g_Y^2}{3}=4 \sqrt{\frac{2}{3}} \, \pi \,  e^{\phi _4}$.}
\label{model1.5}
\end{table}

%%%%%%%%%%%%%%%%%%%%%%%%%%%%%%%%% Model 2 %%%%%%%%%%%%%%%%%%%%%%%%%%%%%%%%%%%%%%%%%%%%%%%%%%%%%%%

\begin{table}[ht]\footnotesize\centering 
$\begin{array}{|c|c|c|c|c|c|c|c|c|c|c|c|c|}
\hline\multicolumn{2}{|c|}{\text{Model~\hyperref[spec2]{2}}} & \multicolumn{11}{c|}{ {\U(4)_C\times \U(2)_L \times \U(2)_R\times \USp(2)_1\times \USp(2)_2\times \USp(2)_3\times \USp(2)_4} }\\
\hline\rm{stack} & N & (n^1,l^1)\times(n^2,l^2)\times(n^3,l^3) & n_{\yng(2)}& n_{\yng(1,1)_{}} & b & b' & c & c' & 1 &  2 &  3 &  4\\
\hline
a & 8 & (1, -1)\times (1, 0)\times (1, 1) & 0 & 0  & -2 & -1 & 3 & 0 & 0 &  -1 &  0 &  1\\
b & 4 &  (-1, 0)\times (-1, -1)\times (1, -3) & 2 & -2  & \text{-} & \text{-} & -4 & -4 & 0 &  0 &  1 &  -3\\
c & 4 &  (0, 1)\times (-1, 3)\times (-1, 1) & 2 & -2  & \text{-} & \text{-} & \text{-} & \text{-} & -3 &  1 &  0 &  0\\
\hline
1 & 2  &  (1, 0)\times (1, 0)\times (2, 0) & \multicolumn{10}{c|}{ x_A = \frac{1}{3}x_B = \frac{1}{9}x_C = \frac{1}{3}x_D }\\
2 & 2 &  (1, 0)\times (0, -1)\times (0, 2) & \multicolumn{10}{c|}{ \beta^g_1=-3,~\beta^g_2=-3,~\beta^g_3=-5,~\beta^g_4=-1, }\\
3 & 2  &  (0, -1)\times (1, 0)\times (0, 2)  & \multicolumn{10}{c|}{ \chi_1=\frac{1}{3} , ~ \chi_2=1 , ~ \chi_3=\frac{2}{3} }\\
4 & 2  &   (0, -1)\times (0, 1)\times (2, 0) & \multicolumn{10}{c|}{}\\
\hline  
\end{array}$
\caption{D6-brane configurations and intersection numbers of Model~\hyperref[spec2]{2}, and its MSSM gauge coupling relation is $g_a^2=\frac{9}{5}g_b^2=g_c^2=\frac{5 g_Y^2}{3}=\frac{12}{5} \sqrt{2} \, \pi \,  e^{\phi _4}$.}
\label{model2}
\end{table}

%%%%%%%%%%%%%%%%%%%%%%%%%%%%%%%%% Model 3 %%%%%%%%%%%%%%%%%%%%%%%%%%%%%%%%%%%%%%%%%%%%%%%%%%%%%%%

\begin{table}[ht]\footnotesize\centering 
$\begin{array}{|c|c|c|c|c|c|c|c|c|c|}
\hline\multicolumn{2}{|c|}{\text{Model~\hyperref[spec3]{3}}} & \multicolumn{8}{c|}{ {\U(4)_C\times \U(2)_L \times \U(2)_R\times \USp(4)_4} }\\
\hline\rm{stack} & N & (n^1,l^1)\times(n^2,l^2)\times(n^3,l^3) & n_{\yng(2)}& n_{\yng(1,1)_{}} & b & b' & c & c' & 4\\
\hline
a & 8 & (-2, 1)\times (-1, 0)\times (1, 1) & -1 & 1  & -3 & 0 & 3 & 0 & 2\\
b & 4 &  (1, 0)\times (1, 3)\times (1, -1) & -2 & 2  & \text{-} & \text{-} & 0 & 4 & -1\\
c & 4 &  (-1, 2)\times (-1, 1)\times (-1, 1) & 2 & 6  & \text{-} & \text{-} & \text{-} & \text{-} & 1\\
\hline
 &   &  & \multicolumn{7}{c|}{ x_A = \frac{4}{3}x_B = 8x_C = \frac{8}{3}x_D }\\
  &   &   & \multicolumn{7}{c|}{ ~\beta^g_4=0, }\\
 &   &   & \multicolumn{7}{c|}{ \chi_1=4 , ~ \chi_2=\frac{2}{3} , ~ \chi_3=4 }\\
4 & 4  &   (0, -1)\times (0, 1)\times (2, 0) & \multicolumn{7}{c|}{}\\
\hline  
\end{array}$
\caption{D6-brane configurations and intersection numbers of Model~\hyperref[spec3]{3}, and its MSSM gauge coupling relation is $g_a^2=\frac{1}{2}g_b^2=\frac{13}{6}g_c^2=\frac{65}{44}\frac{5 g_Y^2}{3}=\frac{16}{5} \sqrt{\frac{2}{3}} \, \pi \,  e^{\phi _4}$.}
\label{model3}
\end{table}

%%%%%%%%%%%%%%%%%%%%%%%%%%%%%%%%% Model 3-dual %%%%%%%%%%%%%%%%%%%%%%%%%%%%%%%%%%%%%%%%%%%%%%%%%%%%%%%

\begin{table}[ht]\footnotesize\centering 
$\begin{array}{|c|c|c|c|c|c|c|c|c|c|}
\hline\multicolumn{2}{|c|}{\text{Model~\hyperref[spec3.5]{3-dual}}} & \multicolumn{8}{c|}{ {\U(4)_C\times \U(2)_L \times \U(2)_R\times \USp(4)_4} }\\
\hline\rm{stack} & N & (n^1,l^1)\times(n^2,l^2)\times(n^3,l^3) & n_{\yng(2)}& n_{\yng(1,1)_{}} & b & b' & c & c' & 4\\
\hline
a & 8 & (-2, 1)\times (-1, 0)\times (1, 1) & -1 & 1  & 3 & 0 & -3 & 0 & 2\\
b & 4 &  (-1, 2)\times (-1, 1)\times (-1, 1) & 2 & 6  & \text{-} & \text{-} & 0 & 4 & 1\\
c & 4 &  (1, 0)\times (1, 3)\times (1, -1) & -2 & 2  & \text{-} & \text{-} & \text{-} & \text{-} & -1\\
\hline
 &   &  & \multicolumn{7}{c|}{ x_A = \frac{4}{3}x_B = 8x_C = \frac{8}{3}x_D }\\
  &   &   & \multicolumn{7}{c|}{ ~\beta^g_4=0, }\\
 &   &   & \multicolumn{7}{c|}{ \chi_1=4 , ~ \chi_2=\frac{2}{3} , ~ \chi_3=4 }\\
4 & 4  &   (0, -1)\times (0, 1)\times (2, 0) & \multicolumn{7}{c|}{}\\
\hline  
\end{array}$
\caption{D6-brane configurations and intersection numbers of Model~\hyperref[spec3.5]{3-dual}, and its MSSM gauge coupling relation is $g_a^2=\frac{13}{6}g_b^2=\frac{1}{2}g_c^2=\frac{5}{8}\frac{5 g_Y^2}{3}=\frac{16}{5} \sqrt{\frac{2}{3}} \, \pi \,  e^{\phi _4}$.}
\label{model3.5}
\end{table}

%%%%%%%%%%%%%%%%%%%%%%%%%%%%%%%%% Model 4 %%%%%%%%%%%%%%%%%%%%%%%%%%%%%%%%%%%%%%%%%%%%%%%%%%%%%%%

\begin{table}[ht]\footnotesize\centering 
$\begin{array}{|c|c|c|c|c|c|c|c|c|c|}
\hline\multicolumn{2}{|c|}{\text{Model~\hyperref[spec4]{4}}} & \multicolumn{8}{c|}{ {\U(4)_C\times \U(2)_L \times \U(2)_R\times \USp(4)_4} }\\
\hline\rm{stack} & N & (n^1,l^1)\times(n^2,l^2)\times(n^3,l^3) & n_{\yng(2)}& n_{\yng(1,1)_{}} & b & b' & c & c' & 4\\
\hline
a & 8 & (2, -1)\times (1, 0)\times (1, 1) & -1 & 1  & -2 & -1 & 3 & 0 & 2\\
b & 4 &  (1, 0)\times (1, 1)\times (1, -3) & 2 & -2  & \text{-} & \text{-} & -4 & 0 & -3\\
c & 4 &  (-1, 2)\times (-1, 1)\times (-1, 1) & 2 & 6  & \text{-} & \text{-} & \text{-} & \text{-} & 1\\
\hline
 &   &  & \multicolumn{7}{c|}{ x_A = 2x_B = \frac{4}{3}x_C = 4x_D }\\
  &   &   & \multicolumn{7}{c|}{ ~\beta^g_4=2, }\\
 &   &   & \multicolumn{7}{c|}{ \chi_1=2 \sqrt{\frac{2}{3}} , ~ \chi_2=\sqrt{6} , ~ \chi_3=2 \sqrt{\frac{2}{3}} }\\
4 & 4  &   (0, -1)\times (0, 1)\times (2, 0) & \multicolumn{7}{c|}{}\\
\hline  
\end{array}$
\caption{D6-brane configurations and intersection numbers of Model~\hyperref[spec4]{4}, and its MSSM gauge coupling relation is $g_a^2=\frac{21}{10}g_b^2=\frac{7}{2}g_c^2=\frac{7}{4}\frac{5 g_Y^2}{3}=\frac{8}{5} 6^{3/4} \, \pi \,  e^{\phi _4}$.}
\label{model4}
\end{table}

%%%%%%%%%%%%%%%%%%%%%%%%%%%%%%%%% Model 5 %%%%%%%%%%%%%%%%%%%%%%%%%%%%%%%%%%%%%%%%%%%%%%%%%%%%%%%

\begin{table}[ht]\footnotesize\centering 
$\begin{array}{|c|c|c|c|c|c|c|c|c|c|c|c|}
\hline\multicolumn{2}{|c|}{\text{Model~\hyperref[spec5]{5}}} & \multicolumn{10}{c|}{ {\U(4)_C\times \U(2)_L \times \U(2)_R\times \USp(2)_1\times \USp(2)_2\times \USp(2)_4} }\\
\hline\rm{stack} & N & (n^1,l^1)\times(n^2,l^2)\times(n^3,l^3) & n_{\yng(2)}& n_{\yng(1,1)_{}} & b & b' & c & c' & 1 &  2 &  4\\
\hline
a & 8 & (1, -1)\times (1, 0)\times (1, 1) & 0 & 0  & -2 & -1 & 3 & 0 & 0 &  -1 &  1\\
b & 4 &  (-1, 0)\times (-1, -1)\times (1, -3) & 2 & -2  & \text{-} & \text{-} & -5 & -2 & 0 &  0 &  -3\\
c & 4 &  (0, 1)\times (-2, 3)\times (-1, 1) & 1 & -1  & \text{-} & \text{-} & \text{-} & \text{-} & -3 &  2 &  0\\
\hline
1 & 2  &  (1, 0)\times (1, 0)\times (2, 0) & \multicolumn{9}{c|}{ x_A = \frac{2}{3}x_B = \frac{2}{9}x_C = \frac{2}{3}x_D }\\
2 & 2 &  (1, 0)\times (0, -1)\times (0, 2) & \multicolumn{9}{c|}{ \beta^g_1=-3,~\beta^g_2=-2,~\beta^g_4=-1, }\\
 &   &   & \multicolumn{9}{c|}{ \chi_1=\frac{\sqrt{2}}{3} , ~ \chi_2=\sqrt{2} , ~ \chi_3=\frac{2 \sqrt{2}}{3} }\\
4 & 2  &   (0, -1)\times (0, 1)\times (2, 0) & \multicolumn{9}{c|}{}\\
\hline  
\end{array}$
\caption{D6-brane configurations and intersection numbers of Model~\hyperref[spec5]{5}, and its MSSM gauge coupling relation is $g_a^2=\frac{27}{11}g_b^2=2g_c^2=\frac{10}{7}\frac{5 g_Y^2}{3}=\frac{48}{11} \sqrt[4]{2} \, \pi \,  e^{\phi _4}$.}
\label{model5}
\end{table}

%%%%%%%%%%%%%%%%%%%%%%%%%%%%%%%%% Model 6 %%%%%%%%%%%%%%%%%%%%%%%%%%%%%%%%%%%%%%%%%%%%%%%%%%%%%%%

\begin{table}[ht]\footnotesize\centering 
$\begin{array}{|c|c|c|c|c|c|c|c|c|c|c|c|}
\hline\multicolumn{2}{|c|}{\text{Model~\hyperref[spec6]{6}}} & \multicolumn{10}{c|}{ {\U(4)_C\times \U(2)_L \times \U(2)_R\times \USp(2)_2\times \USp(2)_3\times \USp(2)_4} }\\
\hline\rm{stack} & N & (n^1,l^1)\times(n^2,l^2)\times(n^3,l^3) & n_{\yng(2)}& n_{\yng(1,1)_{}} & b & b' & c & c' & 2 &  3 &  4\\
\hline
a & 8 & (1, -1)\times (1, 0)\times (1, 1) & 0 & 0  & -2 & -1 & 3 & 0 & -1 &  0 &  1\\
b & 4 &  (-1, 0)\times (-2, -1)\times (1, -3) & 5 & -5  & \text{-} & \text{-} & -7 & -10 & 0 &  1 &  -6\\
c & 4 &  (0, 1)\times (-1, 3)\times (-1, 1) & 2 & -2  & \text{-} & \text{-} & \text{-} & \text{-} & 1 &  0 &  0\\
\hline
 &   &  & \multicolumn{9}{c|}{ x_A = \frac{1}{3}x_B = \frac{1}{18}x_C = \frac{1}{3}x_D }\\
2 & 2 &  (1, 0)\times (0, -1)\times (0, 2) & \multicolumn{9}{c|}{ ~\beta^g_2=-3,~\beta^g_3=-5,~\beta^g_4=2, }\\
3 & 2  &  (0, -1)\times (1, 0)\times (0, 2)  & \multicolumn{9}{c|}{ \chi_1=\frac{1}{3 \sqrt{2}} , ~ \chi_2=\sqrt{2} , ~ \chi_3=\frac{\sqrt{2}}{3} }\\
4 & 2  &   (0, -1)\times (0, 1)\times (2, 0) & \multicolumn{9}{c|}{}\\
\hline  
\end{array}$
\caption{D6-brane configurations and intersection numbers of Model~\hyperref[spec6]{6}, and its MSSM gauge coupling relation is $g_a^2=\frac{54}{19}g_b^2=g_c^2=\frac{5 g_Y^2}{3}=\frac{48}{19} \sqrt[4]{2} \, \pi \,  e^{\phi _4}$.}
\label{model6}
\end{table}

%%%%%%%%%%%%%%%%%%%%%%%%%%%%%%%%% Model 7 %%%%%%%%%%%%%%%%%%%%%%%%%%%%%%%%%%%%%%%%%%%%%%%%%%%%%%%

\begin{table}[ht]\footnotesize\centering 
$\begin{array}{|c|c|c|c|c|c|c|c|c|c|c|}
\hline\multicolumn{2}{|c|}{\text{Model~\hyperref[spec7]{7}}} & \multicolumn{9}{c|}{ {\U(4)_C\times \U(2)_L \times \U(2)_R\times \USp(2)_2\times \USp(2)_4} }\\
\hline\rm{stack} & N & (n^1,l^1)\times(n^2,l^2)\times(n^3,l^3) & n_{\yng(2)}& n_{\yng(1,1)_{}} & b & b' & c & c' & 2 &  4\\
\hline
a & 8 & (1, -1)\times (1, 0)\times (1, 1) & 0 & 0  & -2 & -1 & 3 & 0 & -1 &  1\\
b & 4 &  (-1, 0)\times (-2, -1)\times (1, -3) & 5 & -5  & \text{-} & \text{-} & -8 & -8 & 0 &  -6\\
c & 4 &  (0, 1)\times (-2, 3)\times (-1, 1) & 1 & -1  & \text{-} & \text{-} & \text{-} & \text{-} & 2 &  0\\
\hline
 &   &  & \multicolumn{8}{c|}{ x_A = \frac{2}{3}x_B = \frac{1}{9}x_C = \frac{2}{3}x_D }\\
2 & 2 &  (1, 0)\times (0, -1)\times (0, 2) & \multicolumn{8}{c|}{ ~\beta^g_2=-2,~\beta^g_4=2, }\\
 &   &   & \multicolumn{8}{c|}{ \chi_1=\frac{1}{3} , ~ \chi_2=2 , ~ \chi_3=\frac{2}{3} }\\
4 & 2  &   (0, -1)\times (0, 1)\times (2, 0) & \multicolumn{8}{c|}{}\\
\hline  
\end{array}$
\caption{D6-brane configurations and intersection numbers of Model~\hyperref[spec7]{7}, and its MSSM gauge coupling relation is $g_a^2=\frac{18}{5}g_b^2=2g_c^2=\frac{10}{7}\frac{5 g_Y^2}{3}=\frac{24 \, \pi \,  e^{\phi _4}}{5}$.}
\label{model7}
\end{table}

%%%%%%%%%%%%%%%%%%%%%%%%%%%%%%%%% Model 8 %%%%%%%%%%%%%%%%%%%%%%%%%%%%%%%%%%%%%%%%%%%%%%%%%%%%%%%

\begin{table}[ht]\footnotesize\centering 
$\begin{array}{|c|c|c|c|c|c|c|c|c|c|c|}
\hline\multicolumn{2}{|c|}{\text{Model~\hyperref[spec8]{8}}} & \multicolumn{9}{c|}{ {\U(4)_C\times \U(2)_L \times \U(2)_R\times \USp(2)_1\times \USp(4)_3} }\\
\hline\rm{stack} & N & (n^1,l^1)\times(n^2,l^2)\times(n^3,l^3) & n_{\yng(2)}& n_{\yng(1,1)_{}} & b & b' & c & c' & 1 &  3\\
\hline
a & 8 & (1, -1)\times (-1, 0)\times (-1, -1) & 0 & 0  & -2 & -1 & 3 & 0 & 0 &  0\\
b & 4 &  (1, 0)\times (1, 1)\times (1, -3) & 2 & -2  & \text{-} & \text{-} & -4 & -8 & 0 &  1\\
c & 4 &  (-1, 4)\times (0, 1)\times (-1, 1) & 3 & -3  & \text{-} & \text{-} & \text{-} & \text{-} & -4 &  1\\
\hline
1 & 2  &  (1, 0)\times (1, 0)\times (2, 0) & \multicolumn{8}{c|}{ x_A = \frac{3}{4}x_B = \frac{1}{4}x_C = \frac{3}{4}x_D }\\
  &   &   & \multicolumn{8}{c|}{ \beta^g_1=-2,~\beta^g_3=-4, }\\
3 & 4  &  (0, -1)\times (1, 0)\times (0, 2)  & \multicolumn{8}{c|}{ \chi_1=\frac{1}{2} , ~ \chi_2=\frac{3}{2} , ~ \chi_3=1 }\\
 &   &    & \multicolumn{8}{c|}{}\\
\hline  
\end{array}$
\caption{D6-brane configurations and intersection numbers of Model~\hyperref[spec8]{8}, and its MSSM gauge coupling relation is $g_a^2=\frac{13}{5}g_b^2=3g_c^2=\frac{5}{3}\frac{5 g_Y^2}{3}=\frac{16}{5} \sqrt{3} \, \pi \,  e^{\phi _4}$.}
\label{model8}
\end{table}

%%%%%%%%%%%%%%%%%%%%%%%%%%%%%%%%% Model 9 %%%%%%%%%%%%%%%%%%%%%%%%%%%%%%%%%%%%%%%%%%%%%%%%%%%%%%%

\begin{table}[ht]\footnotesize\centering 
$\begin{array}{|c|c|c|c|c|c|c|c|c|c|c|}
\hline\multicolumn{2}{|c|}{\text{Model~\hyperref[spec9]{9}}} & \multicolumn{9}{c|}{ {\U(4)_C\times \U(2)_L \times \U(2)_R\times \USp(2)_1\times \USp(4)_3} }\\
\hline\rm{stack} & N & (n^1,l^1)\times(n^2,l^2)\times(n^3,l^3) & n_{\yng(2)}& n_{\yng(1,1)_{}} & b & b' & c & c' & 1 &  3\\
\hline
a & 8 & (-1, 1)\times (-1, 0)\times (1, 1) & 0 & 0  & 3 & 0 & -3 & 0 & 0 &  0\\
b & 4 &  (-1, 4)\times (0, 1)\times (-1, 1) & 3 & -3  & \text{-} & \text{-} & 0 & -4 & -4 &  1\\
c & 4 &  (1, 0)\times (1, 3)\times (1, -1) & -2 & 2  & \text{-} & \text{-} & \text{-} & \text{-} & 0 &  3\\
\hline
1 & 2  &  (1, 0)\times (1, 0)\times (2, 0) & \multicolumn{8}{c|}{ x_A = \frac{1}{12}x_B = \frac{1}{4}x_C = \frac{1}{12}x_D }\\
  &   &   & \multicolumn{8}{c|}{ \beta^g_1=-2,~\beta^g_3=-2, }\\
3 & 4  &  (0, -1)\times (1, 0)\times (0, 2)  & \multicolumn{8}{c|}{ \chi_1=\frac{1}{2} , ~ \chi_2=\frac{1}{6} , ~ \chi_3=1 }\\
 &   &    & \multicolumn{8}{c|}{}\\
\hline  
\end{array}$
\caption{D6-brane configurations and intersection numbers of Model~\hyperref[spec9]{9}, and its MSSM gauge coupling relation is $g_a^2=\frac{1}{3}g_b^2=g_c^2=\frac{5 g_Y^2}{3}=\frac{16 \, \pi \,  e^{\phi _4}}{5 \sqrt{3}}$.}
\label{model9}
\end{table}

%%%%%%%%%%%%%%%%%%%%%%%%%%%%%%%%% Model 9-dual %%%%%%%%%%%%%%%%%%%%%%%%%%%%%%%%%%%%%%%%%%%%%%%%%%%%%%%

\begin{table}[ht]\footnotesize\centering 
$\begin{array}{|c|c|c|c|c|c|c|c|c|c|c|}
\hline\multicolumn{2}{|c|}{\text{Model~\hyperref[spec9.5]{9-dual}}} & \multicolumn{9}{c|}{ {\U(4)_C\times \U(2)_L \times \U(2)_R\times \USp(2)_1\times \USp(4)_3} }\\
\hline\rm{stack} & N & (n^1,l^1)\times(n^2,l^2)\times(n^3,l^3) & n_{\yng(2)}& n_{\yng(1,1)_{}} & b & b' & c & c' & 1 &  3\\
\hline
a & 8 & (-1, 1)\times (-1, 0)\times (1, 1) & 0 & 0  & -3 & 0 & 3 & 0 & 0 &  0\\
b & 4 &  (1, 0)\times (1, 3)\times (1, -1) & -2 & 2  & \text{-} & \text{-} & 0 & -4 & 0 &  3\\
c & 4 &  (-1, 4)\times (0, 1)\times (-1, 1) & 3 & -3  & \text{-} & \text{-} & \text{-} & \text{-} & -4 &  1\\
\hline
1 & 2  &  (1, 0)\times (1, 0)\times (2, 0) & \multicolumn{8}{c|}{ x_A = \frac{1}{12}x_B = \frac{1}{4}x_C = \frac{1}{12}x_D }\\
  &   &   & \multicolumn{8}{c|}{ \beta^g_1=-2,~\beta^g_3=-2, }\\
3 & 4  &  (0, -1)\times (1, 0)\times (0, 2)  & \multicolumn{8}{c|}{ \chi_1=\frac{1}{2} , ~ \chi_2=\frac{1}{6} , ~ \chi_3=1 }\\
 &   &    & \multicolumn{8}{c|}{}\\
\hline  
\end{array}$
\caption{D6-brane configurations and intersection numbers of Model~\hyperref[spec9.5]{9-dual}, and its MSSM gauge coupling relation is $g_a^2=g_b^2=\frac{1}{3}g_c^2=\frac{5}{11}\frac{5 g_Y^2}{3}=\frac{16 \, \pi \,  e^{\phi _4}}{5 \sqrt{3}}$.}
\label{model9.5}
\end{table}

%%%%%%%%%%%%%%%%%%%%%%%%%%%%%%%%% Model 10 %%%%%%%%%%%%%%%%%%%%%%%%%%%%%%%%%%%%%%%%%%%%%%%%%%%%%%%

\begin{table}[ht]\footnotesize\centering 
$\begin{array}{|c|c|c|c|c|c|c|c|c|c|}
\hline\multicolumn{2}{|c|}{\text{Model~\hyperref[spec10]{10}}} & \multicolumn{8}{c|}{ {\U(4)_C\times \U(2)_L \times \U(2)_R\times \USp(4)_3} }\\
\hline\rm{stack} & N & (n^1,l^1)\times(n^2,l^2)\times(n^3,l^3) & n_{\yng(2)}& n_{\yng(1,1)_{}} & b & b' & c & c' & 3\\
\hline
a & 8 & (1, -1)\times (-1, 0)\times (-1, -1) & 0 & 0  & -2 & -1 & 3 & 0 & 0\\
b & 4 &  (1, 0)\times (2, 1)\times (1, -3) & 5 & -5  & \text{-} & \text{-} & -8 & -16 & 1\\
c & 4 &  (-1, 4)\times (0, 1)\times (-1, 1) & 3 & -3  & \text{-} & \text{-} & \text{-} & \text{-} & 1\\
\hline
 &   &  & \multicolumn{7}{c|}{ x_A = \frac{3}{2}x_B = \frac{1}{4}x_C = \frac{3}{2}x_D }\\
  &   &   & \multicolumn{7}{c|}{ ~\beta^g_3=-4, }\\
3 & 4  &  (0, -1)\times (1, 0)\times (0, 2)  & \multicolumn{7}{c|}{ \chi_1=\frac{1}{2} , ~ \chi_2=3 , ~ \chi_3=1 }\\
 &   &    & \multicolumn{7}{c|}{}\\
\hline  
\end{array}$
\caption{D6-brane configurations and intersection numbers of Model~\hyperref[spec10]{10}, and its MSSM gauge coupling relation is $g_a^2=\frac{26}{5}g_b^2=6g_c^2=2\frac{5 g_Y^2}{3}=\frac{16}{5} \sqrt{6} \, \pi \,  e^{\phi _4}$.}
\label{model10}
\end{table}

%%%%%%%%%%%%%%%%%%%%%%%%%%%%%%%%% Model 11 %%%%%%%%%%%%%%%%%%%%%%%%%%%%%%%%%%%%%%%%%%%%%%%%%%%%%%%

\begin{table}[ht]\footnotesize\centering 
$\begin{array}{|c|c|c|c|c|c|c|c|c|c|}
\hline\multicolumn{2}{|c|}{\text{Model~\hyperref[spec11]{11}}} & \multicolumn{8}{c|}{ {\U(4)_C\times \U(2)_L \times \U(2)_R\times \USp(4)_3} }\\
\hline\rm{stack} & N & (n^1,l^1)\times(n^2,l^2)\times(n^3,l^3) & n_{\yng(2)}& n_{\yng(1,1)_{}} & b & b' & c & c' & 3\\
\hline
a & 8 & (1, -1)\times (-1, 0)\times (-1, -1) & 0 & 0  & 3 & 0 & -3 & 0 & 0\\
b & 4 &  (-1, 4)\times (0, 1)\times (-1, 1) & 3 & -3  & \text{-} & \text{-} & 0 & -8 & 1\\
c & 4 &  (1, 0)\times (2, 3)\times (1, -1) & -1 & 1  & \text{-} & \text{-} & \text{-} & \text{-} & 3\\
\hline
 &   &  & \multicolumn{7}{c|}{ x_A = \frac{1}{6}x_B = \frac{1}{4}x_C = \frac{1}{6}x_D }\\
  &   &   & \multicolumn{7}{c|}{ ~\beta^g_3=-2, }\\
3 & 4  &  (0, -1)\times (1, 0)\times (0, 2)  & \multicolumn{7}{c|}{ \chi_1=\frac{1}{2} , ~ \chi_2=\frac{1}{3} , ~ \chi_3=1 }\\
 &   &    & \multicolumn{7}{c|}{}\\
\hline  
\end{array}$
\caption{D6-brane configurations and intersection numbers of Model~\hyperref[spec11]{11}, and its MSSM gauge coupling relation is $g_a^2=\frac{2}{3}g_b^2=2g_c^2=\frac{10}{7}\frac{5 g_Y^2}{3}=\frac{16}{5} \sqrt{\frac{2}{3}} \, \pi \,  e^{\phi _4}$.}
\label{model11}
\end{table}

%%%%%%%%%%%%%%%%%%%%%%%%%%%%%%%%% Model 11-dual %%%%%%%%%%%%%%%%%%%%%%%%%%%%%%%%%%%%%%%%%%%%%%%%%%%%%%%

\begin{table}[ht]\footnotesize\centering 
$\begin{array}{|c|c|c|c|c|c|c|c|c|c|}
\hline\multicolumn{2}{|c|}{\text{Model~\hyperref[spec11.5]{11-dual}}} & \multicolumn{8}{c|}{ {\U(4)_C\times \U(2)_L \times \U(2)_R\times \USp(4)_3} }\\
\hline\rm{stack} & N & (n^1,l^1)\times(n^2,l^2)\times(n^3,l^3) & n_{\yng(2)}& n_{\yng(1,1)_{}} & b & b' & c & c' & 3\\
\hline
a & 8 & (1, -1)\times (-1, 0)\times (-1, -1) & 0 & 0  & -3 & 0 & 3 & 0 & 0\\
b & 4 &  (1, 0)\times (2, 3)\times (1, -1) & -1 & 1  & \text{-} & \text{-} & 0 & -8 & 3\\
c & 4 &  (-1, 4)\times (0, 1)\times (-1, 1) & 3 & -3  & \text{-} & \text{-} & \text{-} & \text{-} & 1\\
\hline
 &   &  & \multicolumn{7}{c|}{ x_A = \frac{1}{6}x_B = \frac{1}{4}x_C = \frac{1}{6}x_D }\\
  &   &   & \multicolumn{7}{c|}{ ~\beta^g_3=-2, }\\
3 & 4  &  (0, -1)\times (1, 0)\times (0, 2)  & \multicolumn{7}{c|}{ \chi_1=\frac{1}{2} , ~ \chi_2=\frac{1}{3} , ~ \chi_3=1 }\\
 &   &    & \multicolumn{7}{c|}{}\\
\hline  
\end{array}$
\caption{D6-brane configurations and intersection numbers of Model~\hyperref[spec11.5]{11-dual}, and its MSSM gauge coupling relation is $g_a^2=2g_b^2=\frac{2}{3}g_c^2=\frac{10}{13}\frac{5 g_Y^2}{3}=\frac{16}{5} \sqrt{\frac{2}{3}} \, \pi \,  e^{\phi _4}$.}
\label{model11.5}
\end{table}

%%%%%%%%%%%%%%%%%%%%%%%%%%%%%%%%% Model 12 %%%%%%%%%%%%%%%%%%%%%%%%%%%%%%%%%%%%%%%%%%%%%%%%%%%%%%%

\begin{table}[ht]\footnotesize\centering 
$\begin{array}{|c|c|c|c|c|c|c|c|c|c|}
\hline\multicolumn{2}{|c|}{\text{Model~\hyperref[spec12]{12}}} & \multicolumn{8}{c|}{ {\U(4)_C\times \U(2)_L \times \U(2)_R\times \USp(2)_3} }\\
\hline\rm{stack} & N & (n^1,l^1)\times(n^2,l^2)\times(n^3,l^3) & n_{\yng(2)}& n_{\yng(1,1)_{}} & b & b' & c & c' & 3\\
\hline
a & 8 & (1, 1)\times (1, 0)\times (1, -1) & 0 & 0  & 3 & -6 & 3 & 0 & 0\\
b & 4 &  (-2, -1)\times (0, -1)\times (-5, -1) & 9 & -9  & \text{-} & \text{-} & -8 & 0 & -10\\
c & 4 &  (-2, 1)\times (-1, 1)\times (1, 1) & -2 & -6  & \text{-} & \text{-} & \text{-} & \text{-} & -2\\
\hline
 &   &  & \multicolumn{7}{c|}{ x_A = \frac{5}{6}x_B = 10x_C = \frac{5}{6}x_D }\\
  &   &   & \multicolumn{7}{c|}{ ~\beta^g_3=6, }\\
3 & 2  &  (0, -1)\times (1, 0)\times (0, 2)  & \multicolumn{7}{c|}{ \chi_1=\sqrt{10} , ~ \chi_2=\frac{\sqrt{\frac{5}{2}}}{6} , ~ \chi_3=2 \sqrt{10} }\\
 &   &    & \multicolumn{7}{c|}{}\\
\hline  
\end{array}$
\caption{D6-brane configurations and intersection numbers of Model~\hyperref[spec12]{12}, and its MSSM gauge coupling relation is $g_a^2=\frac{35}{66}g_b^2=\frac{7}{6}g_c^2=\frac{35}{32}\frac{5 g_Y^2}{3}=\frac{8 \sqrt[4]{2} 5^{3/4} \, \pi \,  e^{\phi _4}}{11 \sqrt{3}}$.}
\label{model12}
\end{table}

%%%%%%%%%%%%%%%%%%%%%%%%%%%%%%%%% Model 13 %%%%%%%%%%%%%%%%%%%%%%%%%%%%%%%%%%%%%%%%%%%%%%%%%%%%%%%

\begin{table}[ht]\footnotesize\centering 
$\begin{array}{|c|c|c|c|c|c|c|c|c|c|c|}
\hline\multicolumn{2}{|c|}{\text{Model~\hyperref[spec13]{13}}} & \multicolumn{9}{c|}{ {\U(4)_C\times \U(2)_L \times \U(2)_R\times \USp(2)_1\times \USp(2)_4} }\\
\hline\rm{stack} & N & (n^1,l^1)\times(n^2,l^2)\times(n^3,l^3) & n_{\yng(2)}& n_{\yng(1,1)_{}} & b & b' & c & c' & 1 &  4\\
\hline
a & 8 & (-1, -1)\times (1, 1)\times (1, 1) & 0 & -4  & 6 & -3 & -3 & 0 & 1 &  -1\\
b & 4 &  (-1, 2)\times (-1, 0)\times (5, 1) & 9 & -9  & \text{-} & \text{-} & -9 & -10 & 0 &  1\\
c & 4 &  (-2, 1)\times (0, -1)\times (1, -1) & -1 & 1  & \text{-} & \text{-} & \text{-} & \text{-} & -1 &  0\\
\hline
1 & 2  &  (1, 0)\times (1, 0)\times (2, 0) & \multicolumn{8}{c|}{ x_A = 22x_B = 2x_C = \frac{11}{5}x_D }\\
  &   &   & \multicolumn{8}{c|}{ \beta^g_1=-3,~\beta^g_4=-3, }\\
 &   &   & \multicolumn{8}{c|}{ \chi_1=\frac{1}{\sqrt{5}} , ~ \chi_2=\frac{11}{\sqrt{5}} , ~ \chi_3=4 \sqrt{5} }\\
4 & 2  &   (0, -1)\times (0, 1)\times (2, 0) & \multicolumn{8}{c|}{}\\
\hline  
\end{array}$
\caption{D6-brane configurations and intersection numbers of Model~\hyperref[spec13]{13}, and its MSSM gauge coupling relation is $g_a^2=\frac{5}{14}g_b^2=\frac{11}{6}g_c^2=\frac{11}{8}\frac{5 g_Y^2}{3}=\frac{8}{63} 5^{3/4} \sqrt{11} \, \pi \,  e^{\phi _4}$.}
\label{model13}
\end{table}

%%%%%%%%%%%%%%%%%%%%%%%%%%%%%%%%% Model 14 %%%%%%%%%%%%%%%%%%%%%%%%%%%%%%%%%%%%%%%%%%%%%%%%%%%%%%%

\begin{table}[ht]\footnotesize\centering 
$\begin{array}{|c|c|c|c|c|c|c|c|c|c|c|c|c|}
\hline\multicolumn{2}{|c|}{\text{Model~\hyperref[spec14]{14}}} & \multicolumn{11}{c|}{ {\U(4)_C\times \U(2)_L \times \U(2)_R\times \USp(2)_1\times \USp(2)_2\times \USp(2)_3\times \USp(2)_4} }\\
\hline\rm{stack} & N & (n^1,l^1)\times(n^2,l^2)\times(n^3,l^3) & n_{\yng(2)}& n_{\yng(1,1)_{}} & b & b' & c & c' & 1 &  2 &  3 &  4\\
\hline
a & 8 & (1, -1)\times (1, 0)\times (1, 1) & 0 & 0  & -3 & 0 & 3 & 0 & 0 &  -1 &  0 &  1\\
b & 4 &  (-1, 0)\times (-1, -3)\times (1, -1) & -2 & 2  & \text{-} & \text{-} & 0 & 0 & 0 &  0 &  3 &  -1\\
c & 4 &  (0, 1)\times (-1, 3)\times (-1, 1) & 2 & -2  & \text{-} & \text{-} & \text{-} & \text{-} & -3 &  1 &  0 &  0\\
\hline
1 & 2  &  (1, 0)\times (1, 0)\times (2, 0) & \multicolumn{10}{c|}{ x_A = \frac{1}{3}x_B = x_C = \frac{1}{3}x_D }\\
2 & 2 &  (1, 0)\times (0, -1)\times (0, 2) & \multicolumn{10}{c|}{ \beta^g_1=-3,~\beta^g_2=-3,~\beta^g_3=-3,~\beta^g_4=-3, }\\
3 & 2  &  (0, -1)\times (1, 0)\times (0, 2)  & \multicolumn{10}{c|}{ \chi_1=1 , ~ \chi_2=\frac{1}{3} , ~ \chi_3=2 }\\
4 & 2  &   (0, -1)\times (0, 1)\times (2, 0) & \multicolumn{10}{c|}{}\\
\hline  
\end{array}$
\caption{D6-brane configurations and intersection numbers of Model~\hyperref[spec14]{14}, and its MSSM gauge coupling relation is $g_a^2=g_b^2=g_c^2=\frac{5 g_Y^2}{3}=4 \sqrt{\frac{2}{3}} \, \pi \,  e^{\phi _4}$.}
\label{model14}
\end{table}

%%%%%%%%%%%%%%%%%%%%%%%%%%%%%%%%% Model 15 %%%%%%%%%%%%%%%%%%%%%%%%%%%%%%%%%%%%%%%%%%%%%%%%%%%%%%%

\begin{table}[ht]\footnotesize\centering 
$\begin{array}{|c|c|c|c|c|c|c|c|c|c|c|c|}
\hline\multicolumn{2}{|c|}{\text{Model~\hyperref[spec15]{15}}} & \multicolumn{10}{c|}{ {\U(4)_C\times \U(2)_L \times \U(2)_R\times \USp(4)_1\times \USp(2)_3\times \USp(2)_4} }\\
\hline\rm{stack} & N & (n^1,l^1)\times(n^2,l^2)\times(n^3,l^3) & n_{\yng(2)}& n_{\yng(1,1)_{}} & b & b' & c & c' & 1 &  3 &  4\\
\hline
a & 8 & (-1, 1)\times (1, -1)\times (1, -1) & 0 & 4  & -3 & 0 & 3 & 0 & -1 &  1 &  1\\
b & 4 &  (-4, 1)\times (-1, 0)\times (1, 1) & -3 & 3  & \text{-} & \text{-} & 0 & 2 & 0 &  0 &  4\\
c & 4 &  (-2, -1)\times (0, 1)\times (1, 1) & 1 & -1  & \text{-} & \text{-} & \text{-} & \text{-} & 1 &  -2 &  0\\
\hline
1 & 4  &  (1, 0)\times (1, 0)\times (2, 0) & \multicolumn{9}{c|}{ x_A = \frac{5}{2}x_B = 2x_C = 10x_D }\\
  &   &   & \multicolumn{9}{c|}{ \beta^g_1=-3,~\beta^g_3=-2,~\beta^g_4=0, }\\
3 & 2  &  (0, -1)\times (1, 0)\times (0, 2)  & \multicolumn{9}{c|}{ \chi_1=2 \sqrt{2} , ~ \chi_2=\frac{5}{\sqrt{2}} , ~ \chi_3=\sqrt{2} }\\
4 & 2  &   (0, -1)\times (0, 1)\times (2, 0) & \multicolumn{9}{c|}{}\\
\hline  
\end{array}$
\caption{D6-brane configurations and intersection numbers of Model~\hyperref[spec15]{15}, and its MSSM gauge coupling relation is $g_a^2=\frac{4}{9}g_b^2=\frac{10}{9}g_c^2=\frac{50}{47}\frac{5 g_Y^2}{3}=\frac{16}{27} 2^{3/4} \sqrt{5} \, \pi \,  e^{\phi _4}$.}
\label{model15}
\end{table}

%%%%%%%%%%%%%%%%%%%%%%%%%%%%%%%%% Model 15-dual %%%%%%%%%%%%%%%%%%%%%%%%%%%%%%%%%%%%%%%%%%%%%%%%%%%%%%%

\begin{table}[ht]\footnotesize\centering 
$\begin{array}{|c|c|c|c|c|c|c|c|c|c|c|c|}
\hline\multicolumn{2}{|c|}{\text{Model~\hyperref[spec15.5]{15-dual}}} & \multicolumn{10}{c|}{ {\U(4)_C\times \U(2)_L \times \U(2)_R\times \USp(4)_1\times \USp(2)_2\times \USp(2)_4} }\\
\hline\rm{stack} & N & (n^1,l^1)\times(n^2,l^2)\times(n^3,l^3) & n_{\yng(2)}& n_{\yng(1,1)_{}} & b & b' & c & c' & 1 &  2 &  4\\
\hline
a & 8 & (1, 1)\times (-1, -1)\times (1, 1) & 0 & -4  & -3 & 0 & 3 & 0 & 1 &  -1 &  -1\\
b & 4 &  (0, 1)\times (-2, 1)\times (-1, 1) & -1 & 1  & \text{-} & \text{-} & 0 & -2 & -1 &  2 &  0\\
c & 4 &  (-1, 0)\times (4, 1)\times (-1, 1) & 3 & -3  & \text{-} & \text{-} & \text{-} & \text{-} & 0 &  0 &  -4\\
\hline
1 & 4  &  (1, 0)\times (1, 0)\times (2, 0) & \multicolumn{9}{c|}{ x_A = 2x_B = \frac{5}{2}x_C = 10x_D }\\
2 & 2 &  (1, 0)\times (0, -1)\times (0, 2) & \multicolumn{9}{c|}{ \beta^g_1=-3,~\beta^g_2=-2,~\beta^g_4=0, }\\
 &   &   & \multicolumn{9}{c|}{ \chi_1=\frac{5}{\sqrt{2}} , ~ \chi_2=2 \sqrt{2} , ~ \chi_3=\sqrt{2} }\\
4 & 2  &   (0, -1)\times (0, 1)\times (2, 0) & \multicolumn{9}{c|}{}\\
\hline  
\end{array}$
\caption{D6-brane configurations and intersection numbers of Model~\hyperref[spec15.5]{15-dual}, and its MSSM gauge coupling relation is $g_a^2=\frac{10}{9}g_b^2=\frac{4}{9}g_c^2=\frac{4}{7}\frac{5 g_Y^2}{3}=\frac{16}{27} 2^{3/4} \sqrt{5} \, \pi \,  e^{\phi _4}$.}
\label{model15.5}
\end{table}

%%%%%%%%%%%%%%%%%%%%%%%%%%%%%%%%% Model 16 %%%%%%%%%%%%%%%%%%%%%%%%%%%%%%%%%%%%%%%%%%%%%%%%%%%%%%%

\begin{table}[ht]\footnotesize\centering 
$\begin{array}{|c|c|c|c|c|c|c|c|c|c|c|}
\hline\multicolumn{2}{|c|}{\text{Model~\hyperref[spec16]{16}}} & \multicolumn{9}{c|}{ {\U(4)_C\times \U(2)_L \times \U(2)_R\times \USp(4)_1\times \USp(2)_4} }\\
\hline\rm{stack} & N & (n^1,l^1)\times(n^2,l^2)\times(n^3,l^3) & n_{\yng(2)}& n_{\yng(1,1)_{}} & b & b' & c & c' & 1 &  4\\
\hline
a & 8 & (1, 1)\times (-2, -1)\times (1, 1) & 0 & -8  & 3 & 0 & -3 & 0 & 1 &  -2\\
b & 4 &  (-4, -1)\times (1, 0)\times (-1, 1) & 3 & -3  & \text{-} & \text{-} & 0 & -2 & 0 &  -4\\
c & 4 &  (-2, 1)\times (1, 1)\times (-1, 1) & -2 & -6  & \text{-} & \text{-} & \text{-} & \text{-} & -1 &  -2\\
\hline
1 & 4  &  (1, 0)\times (1, 0)\times (2, 0) & \multicolumn{8}{c|}{ x_A = \frac{13}{2}x_B = \frac{13}{8}x_C = 26x_D }\\
  &   &   & \multicolumn{8}{c|}{ \beta^g_1=-3,~\beta^g_4=4, }\\
 &   &   & \multicolumn{8}{c|}{ \chi_1=\sqrt{\frac{13}{2}} , ~ \chi_2=2 \sqrt{26} , ~ \chi_3=\frac{\sqrt{\frac{13}{2}}}{2} }\\
4 & 2  &   (0, -1)\times (0, 1)\times (2, 0) & \multicolumn{8}{c|}{}\\
\hline  
\end{array}$
\caption{D6-brane configurations and intersection numbers of Model~\hyperref[spec16]{16}, and its MSSM gauge coupling relation is $g_a^2=\frac{1}{6}g_b^2=\frac{7}{6}g_c^2=\frac{35}{32}\frac{5 g_Y^2}{3}=\frac{16}{135} 26^{3/4} \, \pi \,  e^{\phi _4}$.}
\label{model16}
\end{table}

%%%%%%%%%%%%%%%%%%%%%%%%%%%%%%%%% Model 16-dual %%%%%%%%%%%%%%%%%%%%%%%%%%%%%%%%%%%%%%%%%%%%%%%%%%%%%%%

\begin{table}[ht]\footnotesize\centering 
$\begin{array}{|c|c|c|c|c|c|c|c|c|c|c|}
\hline\multicolumn{2}{|c|}{\text{Model~\hyperref[spec16.5]{16-dual}}} & \multicolumn{9}{c|}{ {\U(4)_C\times \U(2)_L \times \U(2)_R\times \USp(4)_1\times \USp(2)_4} }\\
\hline\rm{stack} & N & (n^1,l^1)\times(n^2,l^2)\times(n^3,l^3) & n_{\yng(2)}& n_{\yng(1,1)_{}} & b & b' & c & c' & 1 &  4\\
\hline
a & 8 & (1, 1)\times (-2, -1)\times (1, 1) & 0 & -8  & -3 & 0 & 3 & 0 & 1 &  -2\\
b & 4 &  (-2, 1)\times (1, 1)\times (-1, 1) & -2 & -6  & \text{-} & \text{-} & 0 & -2 & -1 &  -2\\
c & 4 &  (-4, -1)\times (1, 0)\times (-1, 1) & 3 & -3  & \text{-} & \text{-} & \text{-} & \text{-} & 0 &  -4\\
\hline
1 & 4  &  (1, 0)\times (1, 0)\times (2, 0) & \multicolumn{8}{c|}{ x_A = \frac{13}{2}x_B = \frac{13}{8}x_C = 26x_D }\\
  &   &   & \multicolumn{8}{c|}{ \beta^g_1=-3,~\beta^g_4=4, }\\
 &   &   & \multicolumn{8}{c|}{ \chi_1=\sqrt{\frac{13}{2}} , ~ \chi_2=2 \sqrt{26} , ~ \chi_3=\frac{\sqrt{\frac{13}{2}}}{2} }\\
4 & 2  &   (0, -1)\times (0, 1)\times (2, 0) & \multicolumn{8}{c|}{}\\
\hline  
\end{array}$
\caption{D6-brane configurations and intersection numbers of Model~\hyperref[spec16.5]{16-dual}, and its MSSM gauge coupling relation is $g_a^2=\frac{7}{6}g_b^2=\frac{1}{6}g_c^2=\frac{1}{4}\frac{5 g_Y^2}{3}=\frac{16}{135} 26^{3/4} \, \pi \,  e^{\phi _4}$.}
\label{model16.5}
\end{table}

%%%%%%%%%%%%%%%%%%%%%%%%%%%%%%%%% Model 17 %%%%%%%%%%%%%%%%%%%%%%%%%%%%%%%%%%%%%%%%%%%%%%%%%%%%%%%

\begin{table}[ht]\footnotesize\centering 
$\begin{array}{|c|c|c|c|c|c|c|c|c|c|c|c|}
\hline\multicolumn{2}{|c|}{\text{Model~\hyperref[spec17]{17}}} & \multicolumn{10}{c|}{ {\U(4)_C\times \U(2)_L \times \U(2)_R\times \USp(2)_2\times \USp(2)_3\times \USp(2)_4} }\\
\hline\rm{stack} & N & (n^1,l^1)\times(n^2,l^2)\times(n^3,l^3) & n_{\yng(2)}& n_{\yng(1,1)_{}} & b & b' & c & c' & 2 &  3 &  4\\
\hline
a & 8 & (1, -1)\times (1, 0)\times (1, 1) & 0 & 0  & -3 & 0 & 3 & 0 & -1 &  0 &  1\\
b & 4 &  (-1, 0)\times (-2, -3)\times (1, -1) & -1 & 1  & \text{-} & \text{-} & 0 & -3 & 0 &  3 &  -2\\
c & 4 &  (0, 1)\times (-1, 3)\times (-1, 1) & 2 & -2  & \text{-} & \text{-} & \text{-} & \text{-} & 1 &  0 &  0\\
\hline
 &   &  & \multicolumn{9}{c|}{ x_A = \frac{1}{3}x_B = \frac{1}{2}x_C = \frac{1}{3}x_D }\\
2 & 2 &  (1, 0)\times (0, -1)\times (0, 2) & \multicolumn{9}{c|}{ ~\beta^g_2=-3,~\beta^g_3=-3,~\beta^g_4=-2, }\\
3 & 2  &  (0, -1)\times (1, 0)\times (0, 2)  & \multicolumn{9}{c|}{ \chi_1=\frac{1}{\sqrt{2}} , ~ \chi_2=\frac{\sqrt{2}}{3} , ~ \chi_3=\sqrt{2} }\\
4 & 2  &   (0, -1)\times (0, 1)\times (2, 0) & \multicolumn{9}{c|}{}\\
\hline  
\end{array}$
\caption{D6-brane configurations and intersection numbers of Model~\hyperref[spec17]{17}, and its MSSM gauge coupling relation is $g_a^2=2g_b^2=g_c^2=\frac{5 g_Y^2}{3}=\frac{16 \sqrt[4]{2} \, \pi \,  e^{\phi _4}}{3 \sqrt{3}}$.}
\label{model17}
\end{table}

%%%%%%%%%%%%%%%%%%%%%%%%%%%%%%%%% Model 17-dual %%%%%%%%%%%%%%%%%%%%%%%%%%%%%%%%%%%%%%%%%%%%%%%%%%%%%%%

\begin{table}[ht]\footnotesize\centering 
$\begin{array}{|c|c|c|c|c|c|c|c|c|c|c|c|}
\hline\multicolumn{2}{|c|}{\text{Model~\hyperref[spec17.5]{17-dual}}} & \multicolumn{10}{c|}{ {\U(4)_C\times \U(2)_L \times \U(2)_R\times \USp(2)_2\times \USp(2)_3\times \USp(2)_4} }\\
\hline\rm{stack} & N & (n^1,l^1)\times(n^2,l^2)\times(n^3,l^3) & n_{\yng(2)}& n_{\yng(1,1)_{}} & b & b' & c & c' & 2 &  3 &  4\\
\hline
a & 8 & (1, -1)\times (1, 0)\times (1, 1) & 0 & 0  & 3 & 0 & -3 & 0 & -1 &  0 &  1\\
b & 4 &  (0, 1)\times (-1, 3)\times (-1, 1) & 2 & -2  & \text{-} & \text{-} & 0 & -3 & 1 &  0 &  0\\
c & 4 &  (-1, 0)\times (-2, -3)\times (1, -1) & -1 & 1  & \text{-} & \text{-} & \text{-} & \text{-} & 0 &  3 &  -2\\
\hline
 &   &  & \multicolumn{9}{c|}{ x_A = \frac{1}{3}x_B = \frac{1}{2}x_C = \frac{1}{3}x_D }\\
2 & 2 &  (1, 0)\times (0, -1)\times (0, 2) & \multicolumn{9}{c|}{ ~\beta^g_2=-3,~\beta^g_3=-3,~\beta^g_4=-2, }\\
3 & 2  &  (0, -1)\times (1, 0)\times (0, 2)  & \multicolumn{9}{c|}{ \chi_1=\frac{1}{\sqrt{2}} , ~ \chi_2=\frac{\sqrt{2}}{3} , ~ \chi_3=\sqrt{2} }\\
4 & 2  &   (0, -1)\times (0, 1)\times (2, 0) & \multicolumn{9}{c|}{}\\
\hline  
\end{array}$
\caption{D6-brane configurations and intersection numbers of Model~\hyperref[spec17.5]{17-dual}, and its MSSM gauge coupling relation is $g_a^2=g_b^2=2g_c^2=\frac{10}{7}\frac{5 g_Y^2}{3}=\frac{16 \sqrt[4]{2} \, \pi \,  e^{\phi _4}}{3 \sqrt{3}}$.}
\label{model17.5}
\end{table}

%%%%%%%%%%%%%%%%%%%%%%%%%%%%%%%%% Model 18 %%%%%%%%%%%%%%%%%%%%%%%%%%%%%%%%%%%%%%%%%%%%%%%%%%%%%%%

\begin{table}[ht]\footnotesize\centering 
$\begin{array}{|c|c|c|c|c|c|c|c|c|c|c|c|}
\hline\multicolumn{2}{|c|}{\text{Model~\hyperref[spec18]{18}}} & \multicolumn{10}{c|}{ {\U(4)_C\times \U(2)_L \times \U(2)_R\times \USp(2)_2\times \USp(2)_3\times \USp(2)_4} }\\
\hline\rm{stack} & N & (n^1,l^1)\times(n^2,l^2)\times(n^3,l^3) & n_{\yng(2)}& n_{\yng(1,1)_{}} & b & b' & c & c' & 2 &  3 &  4\\
\hline
a & 8 & (1, -1)\times (1, 0)\times (1, 1) & 0 & 0  & 3 & 0 & -3 & 0 & -1 &  0 &  1\\
b & 4 &  (-1, 4)\times (0, 1)\times (-1, 1) & 3 & -3  & \text{-} & \text{-} & 0 & -7 & 0 &  1 &  0\\
c & 4 &  (2, 1)\times (1, 1)\times (1, -1) & 2 & 6  & \text{-} & \text{-} & \text{-} & \text{-} & 1 &  2 &  -2\\
\hline
 &   &  & \multicolumn{9}{c|}{ x_A = \frac{1}{9}x_B = \frac{1}{4}x_C = \frac{1}{9}x_D }\\
2 & 2 &  (1, 0)\times (0, -1)\times (0, 2) & \multicolumn{9}{c|}{ ~\beta^g_2=-3,~\beta^g_3=-3,~\beta^g_4=-2, }\\
3 & 2  &  (0, -1)\times (1, 0)\times (0, 2)  & \multicolumn{9}{c|}{ \chi_1=\frac{1}{2} , ~ \chi_2=\frac{2}{9} , ~ \chi_3=1 }\\
4 & 2  &   (0, -1)\times (0, 1)\times (2, 0) & \multicolumn{9}{c|}{}\\
\hline  
\end{array}$
\caption{D6-brane configurations and intersection numbers of Model~\hyperref[spec18]{18}, and its MSSM gauge coupling relation is $g_a^2=\frac{4}{9}g_b^2=\frac{17}{9}g_c^2=\frac{85}{61}\frac{5 g_Y^2}{3}=\frac{32 \, \pi \,  e^{\phi _4}}{15}$.}
\label{model18}
\end{table}

%%%%%%%%%%%%%%%%%%%%%%%%%%%%%%%%% Model 18-dual %%%%%%%%%%%%%%%%%%%%%%%%%%%%%%%%%%%%%%%%%%%%%%%%%%%%%%%

\begin{table}[ht]\footnotesize\centering 
$\begin{array}{|c|c|c|c|c|c|c|c|c|c|c|c|}
\hline\multicolumn{2}{|c|}{\text{Model~\hyperref[spec18.5]{18-dual}}} & \multicolumn{10}{c|}{ {\U(4)_C\times \U(2)_L \times \U(2)_R\times \USp(2)_2\times \USp(2)_3\times \USp(2)_4} }\\
\hline\rm{stack} & N & (n^1,l^1)\times(n^2,l^2)\times(n^3,l^3) & n_{\yng(2)}& n_{\yng(1,1)_{}} & b & b' & c & c' & 2 &  3 &  4\\
\hline
a & 8 & (1, -1)\times (1, 0)\times (1, 1) & 0 & 0  & -3 & 0 & 3 & 0 & -1 &  0 &  1\\
b & 4 &  (2, 1)\times (1, 1)\times (1, -1) & 2 & 6  & \text{-} & \text{-} & 0 & -7 & 1 &  2 &  -2\\
c & 4 &  (-1, 4)\times (0, 1)\times (-1, 1) & 3 & -3  & \text{-} & \text{-} & \text{-} & \text{-} & 0 &  1 &  0\\
\hline
 &   &  & \multicolumn{9}{c|}{ x_A = \frac{1}{9}x_B = \frac{1}{4}x_C = \frac{1}{9}x_D }\\
2 & 2 &  (1, 0)\times (0, -1)\times (0, 2) & \multicolumn{9}{c|}{ ~\beta^g_2=-3,~\beta^g_3=-3,~\beta^g_4=-2, }\\
3 & 2  &  (0, -1)\times (1, 0)\times (0, 2)  & \multicolumn{9}{c|}{ \chi_1=\frac{1}{2} , ~ \chi_2=\frac{2}{9} , ~ \chi_3=1 }\\
4 & 2  &   (0, -1)\times (0, 1)\times (2, 0) & \multicolumn{9}{c|}{}\\
\hline  
\end{array}$
\caption{D6-brane configurations and intersection numbers of Model~\hyperref[spec18.5]{18-dual}, and its MSSM gauge coupling relation is $g_a^2=\frac{17}{9}g_b^2=\frac{4}{9}g_c^2=\frac{4}{7}\frac{5 g_Y^2}{3}=\frac{32 \, \pi \,  e^{\phi _4}}{15}$.}
\label{model18.5}
\end{table}

%%%%%%%%%%%%%%%%%%%%%%%%%%%%%%%%% Model 19 %%%%%%%%%%%%%%%%%%%%%%%%%%%%%%%%%%%%%%%%%%%%%%%%%%%%%%%

\begin{table}[ht]\footnotesize\centering 
$\begin{array}{|c|c|c|c|c|c|c|c|c|c|c|}
\hline\multicolumn{2}{|c|}{\text{Model~\hyperref[spec19]{19}}} & \multicolumn{9}{c|}{ {\U(4)_C\times \U(2)_L \times \U(2)_R\times \USp(2)_1\times \USp(2)_4} }\\
\hline\rm{stack} & N & (n^1,l^1)\times(n^2,l^2)\times(n^3,l^3) & n_{\yng(2)}& n_{\yng(1,1)_{}} & b & b' & c & c' & 1 &  4\\
\hline
a & 8 & (-1, -1)\times (1, 1)\times (1, 1) & 0 & -4  & -3 & 0 & 3 & 0 & 1 &  -1\\
b & 4 &  (1, 4)\times (1, 0)\times (1, -1) & -3 & 3  & \text{-} & \text{-} & 0 & 7 & 0 &  -1\\
c & 4 &  (-2, 1)\times (2, 1)\times (-1, 1) & -3 & -13  & \text{-} & \text{-} & \text{-} & \text{-} & -1 &  -4\\
\hline
1 & 2  &  (1, 0)\times (1, 0)\times (2, 0) & \multicolumn{8}{c|}{ x_A = 28x_B = \frac{28}{23}x_C = 7x_D }\\
  &   &   & \multicolumn{8}{c|}{ \beta^g_1=-3,~\beta^g_4=1, }\\
 &   &   & \multicolumn{8}{c|}{ \chi_1=\sqrt{\frac{7}{23}} , ~ \chi_2=\sqrt{161} , ~ \chi_3=8 \sqrt{\frac{7}{23}} }\\
4 & 2  &   (0, -1)\times (0, 1)\times (2, 0) & \multicolumn{8}{c|}{}\\
\hline  
\end{array}$
\caption{D6-brane configurations and intersection numbers of Model~\hyperref[spec19]{19}, and its MSSM gauge coupling relation is $g_a^2=\frac{1}{6}g_b^2=\frac{11}{6}g_c^2=\frac{11}{8}\frac{5 g_Y^2}{3}=\frac{8}{405} \sqrt{2} 161^{3/4} \, \pi \,  e^{\phi _4}$.}
\label{model19}
\end{table}

%%%%%%%%%%%%%%%%%%%%%%%%%%%%%%%%% Model 19-dual %%%%%%%%%%%%%%%%%%%%%%%%%%%%%%%%%%%%%%%%%%%%%%%%%%%%%%%

\begin{table}[ht]\footnotesize\centering 
$\begin{array}{|c|c|c|c|c|c|c|c|c|c|c|}
\hline\multicolumn{2}{|c|}{\text{Model~\hyperref[spec19.5]{19-dual}}} & \multicolumn{9}{c|}{ {\U(4)_C\times \U(2)_L \times \U(2)_R\times \USp(2)_1\times \USp(2)_4} }\\
\hline\rm{stack} & N & (n^1,l^1)\times(n^2,l^2)\times(n^3,l^3) & n_{\yng(2)}& n_{\yng(1,1)_{}} & b & b' & c & c' & 1 &  4\\
\hline
a & 8 & (-1, -1)\times (1, 1)\times (1, 1) & 0 & -4  & 3 & 0 & -3 & 0 & 1 &  -1\\
b & 4 &  (-2, 1)\times (2, 1)\times (-1, 1) & -3 & -13  & \text{-} & \text{-} & 0 & 7 & -1 &  -4\\
c & 4 &  (1, 4)\times (1, 0)\times (1, -1) & -3 & 3  & \text{-} & \text{-} & \text{-} & \text{-} & 0 &  -1\\
\hline
1 & 2  &  (1, 0)\times (1, 0)\times (2, 0) & \multicolumn{8}{c|}{ x_A = 28x_B = \frac{28}{23}x_C = 7x_D }\\
  &   &   & \multicolumn{8}{c|}{ \beta^g_1=-3,~\beta^g_4=1, }\\
 &   &   & \multicolumn{8}{c|}{ \chi_1=\sqrt{\frac{7}{23}} , ~ \chi_2=\sqrt{161} , ~ \chi_3=8 \sqrt{\frac{7}{23}} }\\
4 & 2  &   (0, -1)\times (0, 1)\times (2, 0) & \multicolumn{8}{c|}{}\\
\hline  
\end{array}$
\caption{D6-brane configurations and intersection numbers of Model~\hyperref[spec19.5]{19-dual}, and its MSSM gauge coupling relation is $g_a^2=\frac{11}{6}g_b^2=\frac{1}{6}g_c^2=\frac{1}{4}\frac{5 g_Y^2}{3}=\frac{8}{405} \sqrt{2} 161^{3/4} \, \pi \,  e^{\phi _4}$.}
\label{model19.5}
\end{table}

%%%%%%%%%%%%%%%%%%%%%%%%%%%%%%%%% Model 20 %%%%%%%%%%%%%%%%%%%%%%%%%%%%%%%%%%%%%%%%%%%%%%%%%%%%%%%

\begin{table}[ht]\footnotesize\centering 
$\begin{array}{|c|c|c|c|c|c|c|c|c|c|c|}
\hline\multicolumn{2}{|c|}{\text{Model~\hyperref[spec20]{20}}} & \multicolumn{9}{c|}{ {\U(4)_C\times \U(2)_L \times \U(2)_R\times \USp(2)_1\times \USp(2)_4} }\\
\hline\rm{stack} & N & (n^1,l^1)\times(n^2,l^2)\times(n^3,l^3) & n_{\yng(2)}& n_{\yng(1,1)_{}} & b & b' & c & c' & 1 &  4\\
\hline
a & 8 & (1, 1)\times (-1, -1)\times (1, 1) & 0 & -4  & 3 & 0 & -3 & 0 & 1 &  -1\\
b & 4 &  (-1, 0)\times (5, 2)\times (-1, 1) & 3 & -3  & \text{-} & \text{-} & 0 & -1 & 0 &  -5\\
c & 4 &  (0, 1)\times (-2, 1)\times (-1, 1) & -1 & 1  & \text{-} & \text{-} & \text{-} & \text{-} & -1 &  0\\
\hline
1 & 2  &  (1, 0)\times (1, 0)\times (2, 0) & \multicolumn{8}{c|}{ x_A = 2x_B = \frac{14}{5}x_C = 7x_D }\\
  &   &   & \multicolumn{8}{c|}{ \beta^g_1=-3,~\beta^g_4=1, }\\
 &   &   & \multicolumn{8}{c|}{ \chi_1=\frac{7}{\sqrt{5}} , ~ \chi_2=\sqrt{5} , ~ \chi_3=\frac{4}{\sqrt{5}} }\\
4 & 2  &   (0, -1)\times (0, 1)\times (2, 0) & \multicolumn{8}{c|}{}\\
\hline  
\end{array}$
\caption{D6-brane configurations and intersection numbers of Model~\hyperref[spec20]{20}, and its MSSM gauge coupling relation is $g_a^2=\frac{5}{6}g_b^2=\frac{7}{6}g_c^2=\frac{35}{32}\frac{5 g_Y^2}{3}=\frac{8}{27} 5^{3/4} \sqrt{7} \, \pi \,  e^{\phi _4}$.}
\label{model20}
\end{table}

%%%%%%%%%%%%%%%%%%%%%%%%%%%%%%%%% Model 20-dual %%%%%%%%%%%%%%%%%%%%%%%%%%%%%%%%%%%%%%%%%%%%%%%%%%%%%%%

\begin{table}[ht]\footnotesize\centering 
$\begin{array}{|c|c|c|c|c|c|c|c|c|c|c|}
\hline\multicolumn{2}{|c|}{\text{Model~\hyperref[spec20.5]{20-dual}}} & \multicolumn{9}{c|}{ {\U(4)_C\times \U(2)_L \times \U(2)_R\times \USp(2)_1\times \USp(2)_4} }\\
\hline\rm{stack} & N & (n^1,l^1)\times(n^2,l^2)\times(n^3,l^3) & n_{\yng(2)}& n_{\yng(1,1)_{}} & b & b' & c & c' & 1 &  4\\
\hline
a & 8 & (1, 1)\times (-1, -1)\times (1, 1) & 0 & -4  & -3 & 0 & 3 & 0 & 1 &  -1\\
b & 4 &  (0, 1)\times (-2, 1)\times (-1, 1) & -1 & 1  & \text{-} & \text{-} & 0 & -1 & -1 &  0\\
c & 4 &  (-1, 0)\times (5, 2)\times (-1, 1) & 3 & -3  & \text{-} & \text{-} & \text{-} & \text{-} & 0 &  -5\\
\hline
1 & 2  &  (1, 0)\times (1, 0)\times (2, 0) & \multicolumn{8}{c|}{ x_A = 2x_B = \frac{14}{5}x_C = 7x_D }\\
  &   &   & \multicolumn{8}{c|}{ \beta^g_1=-3,~\beta^g_4=1, }\\
 &   &   & \multicolumn{8}{c|}{ \chi_1=\frac{7}{\sqrt{5}} , ~ \chi_2=\sqrt{5} , ~ \chi_3=\frac{4}{\sqrt{5}} }\\
4 & 2  &   (0, -1)\times (0, 1)\times (2, 0) & \multicolumn{8}{c|}{}\\
\hline  
\end{array}$
\caption{D6-brane configurations and intersection numbers of Model~\hyperref[spec20.5]{20-dual}, and its MSSM gauge coupling relation is $g_a^2=\frac{7}{6}g_b^2=\frac{5}{6}g_c^2=\frac{25}{28}\frac{5 g_Y^2}{3}=\frac{8}{27} 5^{3/4} \sqrt{7} \, \pi \,  e^{\phi _4}$.}
\label{model20.5}
\end{table}

%%%%%%%%%%%%%%%%%%%%%%%%%%%%%%%%% Model 21 %%%%%%%%%%%%%%%%%%%%%%%%%%%%%%%%%%%%%%%%%%%%%%%%%%%%%%%

\begin{table}[ht]\footnotesize\centering 
$\begin{array}{|c|c|c|c|c|c|c|c|c|c|c|}
\hline\multicolumn{2}{|c|}{\text{Model~\hyperref[spec21]{21}}} & \multicolumn{9}{c|}{ {\U(4)_C\times \U(2)_L \times \U(2)_R\times \USp(2)_2\times \USp(2)_4} }\\
\hline\rm{stack} & N & (n^1,l^1)\times(n^2,l^2)\times(n^3,l^3) & n_{\yng(2)}& n_{\yng(1,1)_{}} & b & b' & c & c' & 2 &  4\\
\hline
a & 8 & (1, -1)\times (1, 0)\times (1, 1) & 0 & 0  & -3 & 0 & 3 & 0 & -1 &  1\\
b & 4 &  (-1, 0)\times (-2, -3)\times (1, -1) & -1 & 1  & \text{-} & \text{-} & 0 & 0 & 0 &  -2\\
c & 4 &  (0, 1)\times (-2, 3)\times (-1, 1) & 1 & -1  & \text{-} & \text{-} & \text{-} & \text{-} & 2 &  0\\
\hline
 &   &  & \multicolumn{8}{c|}{ x_A = \frac{2}{3}x_B = x_C = \frac{2}{3}x_D }\\
2 & 2 &  (1, 0)\times (0, -1)\times (0, 2) & \multicolumn{8}{c|}{ ~\beta^g_2=-2,~\beta^g_4=-2, }\\
 &   &   & \multicolumn{8}{c|}{ \chi_1=1 , ~ \chi_2=\frac{2}{3} , ~ \chi_3=2 }\\
4 & 2  &   (0, -1)\times (0, 1)\times (2, 0) & \multicolumn{8}{c|}{}\\
\hline  
\end{array}$
\caption{D6-brane configurations and intersection numbers of Model~\hyperref[spec21]{21}, and its MSSM gauge coupling relation is $g_a^2=2g_b^2=2g_c^2=\frac{10}{7}\frac{5 g_Y^2}{3}=\frac{8 \, \pi \,  e^{\phi _4}}{\sqrt{3}}$.}
\label{model21}
\end{table}

%%%%%%%%%%%%%%%%%%%%%%%%%%%%%%%%% Model 22 %%%%%%%%%%%%%%%%%%%%%%%%%%%%%%%%%%%%%%%%%%%%%%%%%%%%%%%

\begin{table}[ht]\footnotesize\centering 
$\begin{array}{|c|c|c|c|c|c|c|c|c|c|}
\hline\multicolumn{2}{|c|}{\text{Model~\hyperref[spec22]{22}}} & \multicolumn{8}{c|}{ {\U(4)_C\times \U(2)_L \times \U(2)_R\times \USp(2)_3} }\\
\hline\rm{stack} & N & (n^1,l^1)\times(n^2,l^2)\times(n^3,l^3) & n_{\yng(2)}& n_{\yng(1,1)_{}} & b & b' & c & c' & 3\\
\hline
a & 8 & (1, -1)\times (1, 0)\times (1, 1) & 0 & 0  & 3 & 0 & -3 & 0 & 0\\
b & 4 &  (-2, 5)\times (0, 1)\times (-1, 1) & 3 & -3  & \text{-} & \text{-} & 0 & -8 & 2\\
c & 4 &  (2, 1)\times (1, 1)\times (1, -1) & 2 & 6  & \text{-} & \text{-} & \text{-} & \text{-} & 2\\
\hline
 &   &  & \multicolumn{7}{c|}{ x_A = \frac{1}{6}x_B = \frac{2}{5}x_C = \frac{1}{6}x_D }\\
  &   &   & \multicolumn{7}{c|}{ ~\beta^g_3=-2, }\\
3 & 2  &  (0, -1)\times (1, 0)\times (0, 2)  & \multicolumn{7}{c|}{ \chi_1=\sqrt{\frac{2}{5}} , ~ \chi_2=\frac{\sqrt{\frac{5}{2}}}{6} , ~ \chi_3=2 \sqrt{\frac{2}{5}} }\\
 &   &    & \multicolumn{7}{c|}{}\\
\hline  
\end{array}$
\caption{D6-brane configurations and intersection numbers of Model~\hyperref[spec22]{22}, and its MSSM gauge coupling relation is $g_a^2=\frac{5}{6}g_b^2=\frac{11}{6}g_c^2=\frac{11}{8}\frac{5 g_Y^2}{3}=\frac{8 \sqrt[4]{2} 5^{3/4} \, \pi \,  e^{\phi _4}}{7 \sqrt{3}}$.}
\label{model22}
\end{table}

%%%%%%%%%%%%%%%%%%%%%%%%%%%%%%%%% Model 22-dual %%%%%%%%%%%%%%%%%%%%%%%%%%%%%%%%%%%%%%%%%%%%%%%%%%%%%%%

\begin{table}[ht]\footnotesize\centering 
$\begin{array}{|c|c|c|c|c|c|c|c|c|c|}
\hline\multicolumn{2}{|c|}{\text{Model~\hyperref[spec22.5]{22-dual}}} & \multicolumn{8}{c|}{ {\U(4)_C\times \U(2)_L \times \U(2)_R\times \USp(2)_3} }\\
\hline\rm{stack} & N & (n^1,l^1)\times(n^2,l^2)\times(n^3,l^3) & n_{\yng(2)}& n_{\yng(1,1)_{}} & b & b' & c & c' & 3\\
\hline
a & 8 & (1, -1)\times (1, 0)\times (1, 1) & 0 & 0  & -3 & 0 & 3 & 0 & 0\\
b & 4 &  (2, 1)\times (1, 1)\times (1, -1) & 2 & 6  & \text{-} & \text{-} & 0 & -8 & 2\\
c & 4 &  (-2, 5)\times (0, 1)\times (-1, 1) & 3 & -3  & \text{-} & \text{-} & \text{-} & \text{-} & 2\\
\hline
 &   &  & \multicolumn{7}{c|}{ x_A = \frac{1}{6}x_B = \frac{2}{5}x_C = \frac{1}{6}x_D }\\
  &   &   & \multicolumn{7}{c|}{ ~\beta^g_3=-2, }\\
3 & 2  &  (0, -1)\times (1, 0)\times (0, 2)  & \multicolumn{7}{c|}{ \chi_1=\sqrt{\frac{2}{5}} , ~ \chi_2=\frac{\sqrt{\frac{5}{2}}}{6} , ~ \chi_3=2 \sqrt{\frac{2}{5}} }\\
 &   &    & \multicolumn{7}{c|}{}\\
\hline  
\end{array}$
\caption{D6-brane configurations and intersection numbers of Model~\hyperref[spec22.5]{22-dual}, and its MSSM gauge coupling relation is $g_a^2=\frac{11}{6}g_b^2=\frac{5}{6}g_c^2=\frac{25}{28}\frac{5 g_Y^2}{3}=\frac{8 \sqrt[4]{2} 5^{3/4} \, \pi \,  e^{\phi _4}}{7 \sqrt{3}}$.}
\label{model22.5}
\end{table}

\FloatBarrier 
  
%\bibliographystyle{JHEP}
%\bibliography{References}

\begin{thebibliography}{10}

\bibitem{Berkooz:1996km}
M.~Berkooz, M.~R. Douglas and R.~G. Leigh, \emph{{Branes intersecting at
  angles}}, \href{https://doi.org/10.1016/S0550-3213(96)00452-X}{\emph{Nucl.
  Phys. B} {\bfseries 480} (1996) 265}
  [\href{https://arxiv.org/abs/hep-th/9606139}{{\ttfamily hep-th/9606139}}].

\bibitem{Aldazabal:2000cn}
G.~Aldazabal, S.~Franco, L.~E. Ibanez, R.~Rabadan and A.~M. Uranga,
  \emph{{Intersecting brane worlds}},
  \href{https://doi.org/10.1088/1126-6708/2001/02/047}{\emph{JHEP} {\bfseries
  02} (2001) 047} [\href{https://arxiv.org/abs/hep-ph/0011132}{{\ttfamily
  hep-ph/0011132}}].

\bibitem{Witten:1998cd}
E.~Witten, \emph{{D-branes and K-theory}},
  \href{https://doi.org/10.1088/1126-6708/1998/12/019}{\emph{JHEP} {\bfseries
  12} (1998) 019} [\href{https://arxiv.org/abs/hep-th/9810188}{{\ttfamily
  hep-th/9810188}}].

\bibitem{Uranga:2000xp}
A.~M. Uranga, \emph{{D-brane probes, RR tadpole cancellation and K-theory
  charge}}, \href{https://doi.org/10.1016/S0550-3213(00)00787-2}{\emph{Nucl.
  Phys. B} {\bfseries 598} (2001) 225}
  [\href{https://arxiv.org/abs/hep-th/0011048}{{\ttfamily hep-th/0011048}}].

\bibitem{Mansha:2024yqz}
A.~Mansha, T.~Li and M.~Sabir, \emph{{Revisiting the supersymmetric
  trinification models from intersecting D6-branes}},
  \href{https://doi.org/10.1088/1572-9494/ad565f}{\emph{Commun. Theor. Phys.}
  {\bfseries 76} (2024) 095201}
  [\href{https://arxiv.org/abs/2406.07586}{{\ttfamily 2406.07586}}].

\bibitem{Cvetic:2004ui}
M.~Cvetic, T.~Li and T.~Liu, \emph{{Supersymmetric Pati-Salam models from
  intersecting D6-branes: A Road to the standard model}},
  \href{https://doi.org/10.1016/j.nuclphysb.2004.07.036}{\emph{Nucl. Phys. B}
  {\bfseries 698} (2004) 163}
  [\href{https://arxiv.org/abs/hep-th/0403061}{{\ttfamily hep-th/0403061}}].

\bibitem{Blumenhagen:2006ci}
R.~Blumenhagen, B.~Kors, D.~Lust and S.~Stieberger, \emph{{Four-dimensional
  String Compactifications with D-Branes, Orientifolds and Fluxes}},
  \href{https://doi.org/10.1016/j.physrep.2007.04.003}{\emph{Phys. Rept.}
  {\bfseries 445} (2007) 1}
  [\href{https://arxiv.org/abs/hep-th/0610327}{{\ttfamily hep-th/0610327}}].

\bibitem{Blumenhagen:2005mu}
R.~Blumenhagen, M.~Cvetic, P.~Langacker and G.~Shiu, \emph{{Toward realistic
  intersecting D-brane models}},
  \href{https://doi.org/10.1146/annurev.nucl.55.090704.151541}{\emph{Ann. Rev.
  Nucl. Part. Sci.} {\bfseries 55} (2005) 71}
  [\href{https://arxiv.org/abs/hep-th/0502005}{{\ttfamily hep-th/0502005}}].

\bibitem{Li:2019nvi}
T.~Li, A.~Mansha and R.~Sun, \emph{{Revisiting the supersymmetric
  Pati\textendash{}Salam models from intersecting D6-branes}},
  \href{https://doi.org/10.1140/epjc/s10052-021-08839-w}{\emph{Eur. Phys. J. C}
  {\bfseries 81} (2021) 82} [\href{https://arxiv.org/abs/1910.04530}{{\ttfamily
  1910.04530}}].

\bibitem{Li:2021pxo}
T.~Li, A.~Mansha, R.~Sun, L.~Wu and W.~He, \emph{{N=1 supersymmetric $SU(12)_C
  \times SU (2)_L \times SU(2)_R$ models,~$SU(4)_C \times SU(6)_L \times
  SU(2)_R$~models, and $SU(4)_C \times SU(2)_L \times SU(6)_R$~ models from
  intersecting D6-branes}},
  \href{https://doi.org/10.1103/PhysRevD.104.046018}{\emph{Phys. Rev. D}
  {\bfseries 104} (2021) 046018}
  [\href{https://arxiv.org/abs/1912.11633}{{\ttfamily 1912.11633}}].

\bibitem{Mansha:2022pnd}
A.~Mansha, T.~Li, M.~Sabir and L.~Wu, \emph{{Three-family supersymmetric
  Pati\textendash{}Salam models with symplectic groups from intersecting
  D6-branes}},
  \href{https://doi.org/10.1140/epjc/s10052-024-12411-7}{\emph{Eur. Phys. J. C}
  {\bfseries 84} (2024) 151}
  [\href{https://arxiv.org/abs/2212.09644}{{\ttfamily 2212.09644}}].

\bibitem{Sabir:2022hko}
M.~Sabir, T.~Li, A.~Mansha and X.-C. Wang, \emph{{The supersymmetry breaking
  soft terms, and fermion masses and mixings in the supersymmetric Pati-Salam
  model from intersecting D6-branes}},
  \href{https://doi.org/10.1007/JHEP04(2022)089}{\emph{JHEP} {\bfseries 04}
  (2022) 089} [\href{https://arxiv.org/abs/2202.07048}{{\ttfamily
  2202.07048}}].

\bibitem{Mansha:2023kwq}
A.~Mansha, T.~Li and L.~Wu, \emph{{The hidden sector variations in the
  $\mathcal{N}=1$ supersymmetric three-family Pati\textendash{}Salam models
  from intersecting D6-branes}},
  \href{https://doi.org/10.1140/epjc/s10052-023-12167-6}{\emph{Eur. Phys. J. C}
  {\bfseries 83} (2023) 1067}
  [\href{https://arxiv.org/abs/2303.02864}{{\ttfamily 2303.02864}}].

\bibitem{He:2021gug}
W.~He, T.~Li and R.~Sun, \emph{{The complete search for the supersymmetric
  Pati-Salam models from intersecting D6-branes}},
  \href{https://doi.org/10.1007/JHEP08(2022)044}{\emph{JHEP} {\bfseries 08}
  (2022) 044} [\href{https://arxiv.org/abs/2112.09632}{{\ttfamily
  2112.09632}}].

\bibitem{Sabir:2024cgt}
M.~Sabir, A.~Mansha, T.~Li and Z.-W. Wang, \emph{{Fermion masses and mixings in
  the supersymmetric Pati-Salam landscape from Intersecting D6-Branes}},
  \href{https://doi.org/10.1007/JHEP10(2024)252}{\emph{JHEP} {\bfseries 10}
  (2024) 252} [\href{https://arxiv.org/abs/2409.09110}{{\ttfamily
  2409.09110}}].

\bibitem{Sabir:2024mfv}
M.~Sabir, T.~Li, A.~Mansha and Z.-W. Wang, \emph{{Fermion Masses and Mixings in
  String Theory with Dirac Neutrinos}},
  \href{https://arxiv.org/abs/2407.19458}{{\ttfamily 2407.19458}}.

\bibitem{Gimon:1996rq}
E.~G. Gimon and J.~Polchinski, \emph{{Consistency conditions for orientifolds
  and D-manifolds}},
  \href{https://doi.org/10.1103/PhysRevD.54.1667}{\emph{Phys. Rev. D}
  {\bfseries 54} (1996) 1667}
  [\href{https://arxiv.org/abs/hep-th/9601038}{{\ttfamily hep-th/9601038}}].

\bibitem{Green:1984sg}
M.~B. Green and J.~H. Schwarz, \emph{{Anomaly Cancellation in Supersymmetric
  D=10 Gauge Theory and Superstring Theory}},
  \href{https://doi.org/10.1016/0370-2693(84)91565-X}{\emph{Phys. Lett. B}
  {\bfseries 149} (1984) 117}.

\bibitem{Cascales:2003zp}
J.~F.~G. Cascales and A.~M. Uranga, \emph{{Chiral 4d string vacua with D branes
  and NSNS and RR fluxes}},
  \href{https://doi.org/10.1088/1126-6708/2003/05/011}{\emph{JHEP} {\bfseries
  05} (2003) 011} [\href{https://arxiv.org/abs/hep-th/0303024}{{\ttfamily
  hep-th/0303024}}].

\bibitem{Marchesano:2004yq}
F.~Marchesano and G.~Shiu, \emph{{MSSM vacua from flux compactifications}},
  \href{https://doi.org/10.1103/PhysRevD.71.011701}{\emph{Phys. Rev. D}
  {\bfseries 71} (2005) 011701}
  [\href{https://arxiv.org/abs/hep-th/0408059}{{\ttfamily hep-th/0408059}}].

\bibitem{Marchesano:2004xz}
F.~Marchesano and G.~Shiu, \emph{{Building MSSM flux vacua}},
  \href{https://doi.org/10.1088/1126-6708/2004/11/041}{\emph{JHEP} {\bfseries
  11} (2004) 041} [\href{https://arxiv.org/abs/hep-th/0409132}{{\ttfamily
  hep-th/0409132}}].

\bibitem{Chen:2006gd}
C.-M. Chen, T.~Li and D.~V. Nanopoulos, \emph{{Type IIA Pati-Salam flux
  vacua}}, \href{https://doi.org/10.1016/j.nuclphysb.2006.01.039}{\emph{Nucl.
  Phys. B} {\bfseries 740} (2006) 79}
  [\href{https://arxiv.org/abs/hep-th/0601064}{{\ttfamily hep-th/0601064}}].

\bibitem{Cvetic:2004nk}
M.~Cvetic, P.~Langacker, T. Li and T.~Liu, \emph{{D6-brane splitting on type
  IIA orientifolds}},
  \href{https://doi.org/10.1016/j.nuclphysb.2004.12.028}{\emph{Nucl. Phys. B}
  {\bfseries 709} (2005) 241}
  [\href{https://arxiv.org/abs/hep-th/0407178}{{\ttfamily hep-th/0407178}}].

\bibitem{Cvetic:2007ku}
M.~Cvetic, R.~Richter and T.~Weigand, \emph{{Computation of D-brane instanton
  induced superpotential couplings: Majorana masses from string theory}},
  \href{https://doi.org/10.1103/PhysRevD.76.086002}{\emph{Phys. Rev. D}
  {\bfseries 76} (2007) 086002}
  [\href{https://arxiv.org/abs/hep-th/0703028}{{\ttfamily hep-th/0703028}}].

\bibitem{Chen:2007zu}
C.-M. Chen, T.~Li, V.~E. Mayes and D.~V. Nanopoulos, \emph{{Towards realistic
  supersymmetric spectra and Yukawa textures from intersecting branes}},
  \href{https://doi.org/10.1103/PhysRevD.77.125023}{\emph{Phys. Rev. D}
  {\bfseries 77} (2008) 125023}
  [\href{https://arxiv.org/abs/0711.0396}{{\ttfamily 0711.0396}}].

\bibitem{Blumenhagen:2006xt}
R.~Blumenhagen, M.~Cvetic and T.~Weigand, \emph{{Spacetime instanton
  corrections in 4D string vacua: The Seesaw mechanism for D-Brane models}},
  \href{https://doi.org/10.1016/j.nuclphysb.2007.02.016}{\emph{Nucl. Phys. B}
  {\bfseries 771} (2007) 113}
  [\href{https://arxiv.org/abs/hep-th/0609191}{{\ttfamily hep-th/0609191}}].

\bibitem{Haack:2006cy}
M.~Haack, D.~Krefl, D.~Lust, A.~Van~Proeyen and M.~Zagermann, \emph{{Gaugino
  Condensates and D-terms from D7-branes}},
  \href{https://doi.org/10.1088/1126-6708/2007/01/078}{\emph{JHEP} {\bfseries
  01} (2007) 078} [\href{https://arxiv.org/abs/hep-th/0609211}{{\ttfamily
  hep-th/0609211}}].

\bibitem{Florea:2006si}
B.~Florea, S.~Kachru, J.~McGreevy and N.~Saulina, \emph{{Stringy Instantons and
  Quiver Gauge Theories}},
  \href{https://doi.org/10.1088/1126-6708/2007/05/024}{\emph{JHEP} {\bfseries
  05} (2007) 024} [\href{https://arxiv.org/abs/hep-th/0610003}{{\ttfamily
  hep-th/0610003}}].

\bibitem{Cremmer:1982en}
E.~Cremmer, S.~Ferrara, L.~Girardello and A.~Van~Proeyen, \emph{{Yang-Mills
  Theories with Local Supersymmetry: Lagrangian, Transformation Laws and
  SuperHiggs Effect}},
  \href{https://doi.org/10.1016/0550-3213(83)90679-X}{\emph{Nucl. Phys. B}
  {\bfseries 212} (1983) 413}.

\bibitem{Lust:2004cx}
D.~Lust, P.~Mayr, R.~Richter and S.~Stieberger, \emph{{Scattering of gauge,
  matter, and moduli fields from intersecting branes}},
  \href{https://doi.org/10.1016/j.nuclphysb.2004.06.052}{\emph{Nucl. Phys. B}
  {\bfseries 696} (2004) 205}
  [\href{https://arxiv.org/abs/hep-th/0404134}{{\ttfamily hep-th/0404134}}].

\bibitem{Klebanov:2003my}
I.~R. Klebanov and E.~Witten, \emph{{Proton decay in intersecting D-brane
  models}}, \href{https://doi.org/10.1016/S0550-3213(03)00410-3}{\emph{Nucl.
  Phys. B} {\bfseries 664} (2003) 3}
  [\href{https://arxiv.org/abs/hep-th/0304079}{{\ttfamily hep-th/0304079}}].

\bibitem{Blumenhagen:2003jy}
R.~Blumenhagen, D.~Lust and S.~Stieberger, \emph{{Gauge unification in
  supersymmetric intersecting brane worlds}},
  \href{https://doi.org/10.1088/1126-6708/2003/07/036}{\emph{JHEP} {\bfseries
  07} (2003) 036} [\href{https://arxiv.org/abs/hep-th/0305146}{{\ttfamily
  hep-th/0305146}}].

\bibitem{Ibanez:2001nd}
L.~E. Ibanez, F.~Marchesano and R.~Rabadan, \emph{{Getting just the standard
  model at intersecting branes}},
  \href{https://doi.org/10.1088/1126-6708/2001/11/002}{\emph{JHEP} {\bfseries
  11} (2001) 002} [\href{https://arxiv.org/abs/hep-th/0105155}{{\ttfamily
  hep-th/0105155}}].

\bibitem{Blumenhagen:2005tn}
R.~Blumenhagen, M.~Cvetic, F.~Marchesano and G.~Shiu, \emph{{Chiral D-brane
  models with frozen open string moduli}},
  \href{https://doi.org/10.1088/1126-6708/2005/03/050}{\emph{JHEP} {\bfseries
  03} (2005) 050} [\href{https://arxiv.org/abs/hep-th/0502095}{{\ttfamily
  hep-th/0502095}}].

\bibitem{Font:2004cx}
A.~Font and L.~E. Ibanez, \emph{{SUSY-breaking soft terms in a MSSM magnetized
  D7-brane model}},
  \href{https://doi.org/10.1088/1126-6708/2005/03/040}{\emph{JHEP} {\bfseries
  03} (2005) 040} [\href{https://arxiv.org/abs/hep-th/0412150}{{\ttfamily
  hep-th/0412150}}].

\bibitem{Cvetic:2003ch}
M.~Cvetic and I.~Papadimitriou, \emph{{Conformal field theory couplings for
  intersecting D-branes on orientifolds}},
  \href{https://doi.org/10.1103/PhysRevD.70.029903}{\emph{Phys. Rev. D}
  {\bfseries 68} (2003) 046001}
  [\href{https://arxiv.org/abs/hep-th/0303083}{{\ttfamily hep-th/0303083}}].

\bibitem{Kawamura:1996ex}
Y.~Kawamura, T.~Kobayashi and T.~Komatsu, \emph{{Specific scalar mass relations
  in $\SU(3) \times \SU(2) \times \U(1)$ orbifold model}},
  \href{https://doi.org/10.1016/S0370-2693(97)00364-X}{\emph{Phys. Lett. B}
  {\bfseries 400} (1997) 284}
  [\href{https://arxiv.org/abs/hep-ph/9609462}{{\ttfamily hep-ph/9609462}}].

\bibitem{Komargodski:2009pc}
Z.~Komargodski and N.~Seiberg, \emph{{Comments on the Fayet-Iliopoulos Term in
  Field Theory and Supergravity}},
  \href{https://doi.org/10.1088/1126-6708/2009/06/007}{\emph{JHEP} {\bfseries
  06} (2009) 007} [\href{https://arxiv.org/abs/0904.1159}{{\ttfamily
  0904.1159}}].

\bibitem{Kane:2004hm}
G.~L. Kane, P.~Kumar, J.~D. Lykken and T.~T. Wang, \emph{{Some phenomenology of
  intersecting D-brane models}},
  \href{https://doi.org/10.1103/PhysRevD.71.115017}{\emph{Phys. Rev. D}
  {\bfseries 71} (2005) 115017}
  [\href{https://arxiv.org/abs/hep-ph/0411125}{{\ttfamily hep-ph/0411125}}].

\bibitem{Brignole:1997dp}
A.~Brignole, L.~E. Ibanez and C.~Munoz, \emph{{Soft supersymmetry breaking
  terms from supergravity and superstring models}},
  \href{https://doi.org/10.1142/9789812839657_0003}{\emph{Adv. Ser. Direct.
  High Energy Phys.} {\bfseries 18} (1998) 125}
  [\href{https://arxiv.org/abs/hep-ph/9707209}{{\ttfamily hep-ph/9707209}}].

\bibitem{Brignole:1993dj}
A.~Brignole, L.~E. Ibanez and C.~Munoz, \emph{{Towards a theory of soft terms
  for the supersymmetric Standard Model}},
  \href{https://doi.org/10.1016/0550-3213(94)00068-9}{\emph{Nucl. Phys. B}
  {\bfseries 422} (1994) 125}
  [\href{https://arxiv.org/abs/hep-ph/9308271}{{\ttfamily hep-ph/9308271}}].

\bibitem{Lee:2019wij}
S.-J. Lee, W.~Lerche and T.~Weigand, \emph{{Emergent strings from infinite
  distance limits}}, \href{https://doi.org/10.1007/JHEP02(2022)190}{\emph{JHEP}
  {\bfseries 02} (2022) 190}
  [\href{https://arxiv.org/abs/1910.01135}{{\ttfamily 1910.01135}}].

\bibitem{Casas:2024ttx}
G.~F. Casas, L.~E. Ib\'a\~nez and F.~Marchesano, \emph{{Yukawa couplings at
  infinite distance and swampland towers in chiral theories}},
  \href{https://doi.org/10.1007/JHEP09(2024)170}{\emph{JHEP} {\bfseries 09}
  (2024) 170} [\href{https://arxiv.org/abs/2403.09775}{{\ttfamily
  2403.09775}}].

\end{thebibliography}

\providecommand{\href}[2]{#2}\begingroup\raggedright\endgroup

\end{document}